\documentclass[a4paper,11pt]{article}
\pdfoutput=1 
\usepackage{jheppub} 
\usepackage[T1]{fontenc} 
\usepackage[english]{babel}
\usepackage{graphicx}			    
\usepackage{slashed}
\usepackage{caption}
\usepackage{amsmath}
\usepackage{amsthm}
\usepackage{amsfonts}
\usepackage{mathtools}
\usepackage{subcaption}
\usepackage{tikz} 
\usepackage{tikz-feynman}
\usepackage{tikz-3dplot}
\tikzfeynmanset{compat=1.1.0}
\usepackage{placeins}
\usepackage{scalerel}
\usepackage{verbatim}
\usetikzlibrary{shapes.misc}
\tikzset{cross/.style={cross out, draw=black, minimum size=2*(#1-\pgflinewidth), inner sep=0pt, outer sep=0pt},
	cross/.default={5pt}}

\DeclareMathOperator{\sDet}{sDet}
\DeclareMathOperator{\Det}{Det}
\DeclareMathOperator{\Tr}{Tr}

\newcommand*\Laplace{\mathop{}\!\mathbin\bigtriangleup}
\theoremstyle{plain}
\theoremstyle{plain}

\theoremstyle{definition}

\makeatletter
\def\@fpheader{\relax}
\makeatother
\usepackage{dcolumn}
\usepackage{bm}
\newcommand{\VarPrimeSup}[3]{#1^{\prime #2 #3}} 

\makeatletter
\def\@fpheader{\relax}
\makeatother

\title{Generating functional of correlators of twist-$2$ operators in $\mathcal{N} = 1$ SUSY Yang-Mills theory, I}

\author[a]{Marco Bochicchio,}
\author[b,a]{Mauro Papinutto, }
\author[b,a]{Francesco Scardino}

\affiliation[a]{Physics Department, INFN Roma1, 
Piazzale A. Moro 2, Roma, I-00185, Italy}
\affiliation[b]{Physics Department, Sapienza University,Piazzale A. Moro 2, Roma, I-00185, Italy}

\emailAdd{marco.bochicchio@roma1.infn.it}
\emailAdd{mauro.papinutto@roma1.infn.it}
\emailAdd{francesco.scardino@roma1.infn.it}

\abstract{The present paper is the first installment where, extending our previous work in pure Yang-Mills (YM) theory, we compute the generating functional of correlators of collinear twist-$2$ operators that enter the components of balanced superfields -- i.e., superfields with an equal number of dotted and undotted indices in their spinor representation -- in $\mathcal{N} = 1$ SUSY SU($N$) YM theory in Minkowskian and Euclidean space-time, in the conformal limit and renormalization-group (RG) improved form, and to the leading and next-to-leading order in the large-$N$ expansion. Moreover, we compare our asymptotic RG-improved generating functional to the next-to-leading large-$N$ order with the corresponding nonperturbative object arising from the glueball/gluinoball one-loop effective action, which it should be asymptotic to at short distances because of the asymptotic freedom. Remarkably, we find that both have the structure of the logarithm of a functional superdeterminant. Hence, our large-$N$ computation sets strong ultraviolet asymptotic constraints on the nonperturbative solution of large-$N$ $\mathcal{N} = 1$ SUSY YM theory that may be a pivotal guide for the search of such a solution.}

\begin{document} 

\definecolor{c969696}{RGB}{150,150,150}
\maketitle 
\flushbottom
  \section{Introduction and physics motivations} \label{s1}

Recently, we have computed \cite{BPSpaper2} the short-distance (i.e., ultraviolet (UV)) asymptotics of the generating functional of correlators of collinear twist-$2$ operators in pure SU($N$) Yang-Mills (YM) theory. \par
The present paper is the first of two instalments, where we extend the above computation to $\mathcal{N} = 1$ supersymmetric (SUSY) SU($N$) YM theory for twist-2 operators that are the components of balanced superfields -- in the present paper -- and the components of unbalanced ones -- in the second installment. \par
We refer to operators -- and superfields as well -- as either balanced or unbalanced \cite{BPSpaper2,BPS1} if in their spinor representation they carry either an equal or a different number of dotted and undotted indices, respectively.\par
To summarize our main result, we compute the UV asymptotics of the renormalization-group (RG) improved generating functional of correlators of twist-$2$ operators to the leading and next-to-leading -- i.e., leading nonplanar -- order in 't Hooft large-$N$ expansion \cite{tHooft:1973alw}. \par
Remarkably, we find that our asymptotic generating functional to the leading nonplanar order has the structure of the logarithm of a functional superdeterminant that matches the structure of the corresponding nonperturbative object arising from the glueball/gluinoball one-loop effective action. \par
Hence, our computation sets strong UV constraints on the yet-to-come nonperturbative solution of large-$N$ $\mathcal{N} = 1$ SUSY YM theory that may be a pivotal guide for the search of such a solution. \par
Actually, our calculation involves twist-$2$ operators that have maximal-spin component along the light-cone direction $p_+$ in Minkowskian space-time -- which we refer to as collinear twist-$2$ operators \cite{BPSpaper2,BPS1} -- and their analytic continuation to Euclidean space-time, and it is based on a number of innovations \cite{BPSpaper2,BPS1} described below. \par
Indeed, we have computed in pure YM theory by Feynman diagrams the Minkowskian conformal correlators to the lowest order of perturbation theory and reconstructed from them the corresponding generating functional in the balanced and unbalanced sectors separately \cite{BPS1}.
We have also computed in a much simpler way the complete generating functional of conformal correlators by functional-integral methods \cite{BPSpaper2} and found perfect agreement with our previous computation \cite{BPS1}, the equivalence occurring thanks to a dictionary established by means of a tedious but straightforward calculation \cite{BPSpaper2}. \par
Moreover, we have lifted the generating functional of the Euclidean conformal correlators to the generating functional of the UV asymptotic correlators \cite{BPSpaper2} by means of the RG improvement and a careful choice of the renormalization scheme -- dubbed nonresonant diagonal according to the Poincar\'e-Dulac theorem \cite{MB1} -- that involves a novel geometric interpretation of operator mixing \cite{MB1}, which reduces the mixing of collinear twist-$2$ operators to the multiplicatively renormalizable case to all orders of perturbation theory \cite{BPSpaper2}. Remarkably, the RG-improved correlators should be asymptotic to the exact ones thanks to the asymptotic freedom (AF) \cite{BPSpaper2,BPSL,QCD24}.\par
Finally, we have expanded the latter generating functional to the leading and next-to-leading order in 't Hooft large-$N$ expansion \cite{tHooft:1973alw}. \par
All the above techniques and results extend straightforwardly to $\mathcal{N} = 1$ SUSY SU($N$) YM theory, the only difference being that the final result for the Euclidean asymptotic RG-improved generating functional of connected correlators in the SUSY YM theory has the structure of the logarithm of a functional superdeterminant, instead of a determinant as in the pure YM case.\par
Perturbatively, the logarithm of a superdeterminant occurs because, in the light-cone gauge, collinear twist-$2$ operators are quadratic in the fundamental fields and thus their generating functional to the lowest order arises from a Gaussian integral \cite{BPSpaper2} -- also involving anticommuting variables in the SUSY case. \par
We demonstrate that the above structure of the logarithm of a superdeterminant is then inherited by the Euclidean RG-improved asymptotic generating functional and its leading and next-to-leading large-$N$ expansion \cite{tHooft:1973alw} -- a nontrivial consequence \cite{BPSpaper2} of AF of the $\mathcal{N}=1$ SUSY SU($N$) YM theory that also applies to its large-$N$ limit. \par
Indeed, since the early days of large-$N$ QCD \cite{tHooft:1973alw} it has been known at a qualitative level that AF applies to large-$N$ correlators, and specifically to the nonperturbative $2$-point correlators \cite{Witten:1979kh,1987gauge}.
Accordingly, the UV constraints on the large-$N$ spectral representation of $2$-point correlators that follow from AF have been investigated at a quantitative level both for multiplicatively renormalizable operators \cite{Bochicchio:2013eda} and twist-$2$ operators that mix by renormalization \cite{Bochicchio:2023ols}. Obviously, the above considerations also apply to large-$N$ $\mathcal{N}=1$ SUSY YM theory for $2$-point correlators \cite{Bochicchio:2013eda,Bochicchio:2023ols} and $n$-point correlators of twist-$2$ operators in the present paper. \par
Nonperturbatively, the logarithm of a (super)determinant should arise to the next-to-leading order in the large-$N$ expansion from the glueball/gluinoball one-loop effective action, according to the general principles of quantum field theory as originally predicted in \cite{Bochicchio:2016toi}, and asymptotically verified in detail in the pure YM case \cite{BPSpaper2,BPSL,QCD24} and, in the present paper, in the SUSY YM case.\par
Hence, the UV asymptotics of the generating functional to the leading-nonplanar order in the present paper -- that is only consequence of lowest-order perturbation theory, renormalization group and AF -- matches the nonperturbative structure of the large-$N$ theory to the leading-nonplanar order -- a remarkable fact in itself.\par
Therefore, the above computation sets strong UV constraints on the nonperturbative solution of large-$N$ $\mathcal{N} = 1$ SUSY YM theory and it may be a pivotal guide in the search for such a solution.

\section{Plan of the paper} \label{qq}

In Sec. \ref{NP} -- as a nonperturbative detour to motivate our perturbative and RG-improved computations -- we work out the general structure of the generating functional of large-$N$ correlators of balanced twist-$2$ operators arising from the corresponding nonperturbative glueball/gluinoball one-loop effective action in $\mathcal{N}=1$ SUSY YM theory. \par
Secs. \ref{defSYM}-\ref{IX} instead concern our perturbative and RG-improved computations involving gluons and gluinos:\par
In Sec. \ref{defSYM} we recall the definition of $\mathcal{N}=1$ SUSY YM theory in the light-cone gauge.\par	
In Sec. \ref{twist2op} we define the twist-$2$ operators that enter the components of balanced superfields.\par
In Sec. \ref{functionalInt} we evaluate to the lowest perturbative order the corresponding Minkowskian conformal generating functional as a superdeterminant of a quadratic form. We also rewrite it as a Fredholm superdeterminant employing the aforementioned dictionary in \cite{BPSpaper2}.\par
In Sec. \ref{euclgen} we analytic continue the Minkowskian conformal generating functional to Euclidean space-time.\par
In Sec. \ref{s0} we work out the conformal generating functional of supermultiplet balanced operators. \par
In Sec. \ref{IX} we work out the generating functional of Euclidean RG-improved asymptotic correlators and its large-$N$ expansion.\par
Finally, Sec. \ref{Conc} contains our main conclusions regarding the matching of the structure of the logarithm of a superdeterminant arising from the nonperturbative large-$N$ one-loop effective action -- in terms of glueballs and gluinoballs in Sec. \ref{NP} -- with our asymptotic computation -- in terms of gluons and gluinos in Sec. \ref{IX} -- to the leading nonplanar order.\par
In the Appendices we report basic definitions and several relevant ancillary computations.  \par

\section{Nonperturbative effective action in large-$N$ $\mathcal{N}=1$ SUSY YM theory}\label{NP}

It has been known for almost fifty years that $\mathcal{N}=1$ SUSY SU($N$) YM theory admits 't Hooft large-$N$ topological expansion \cite{tHooft:1973alw} for the $n$-point connected correlators of gauge-invariant single-trace operators. The corresponding Feynman diagrams in 't Hooft double-line representation -- after a suitable gluing of reversely oriented lines -- are topologically classified \cite{tHooft:1973alw,Veneziano:1976wm} by the sum on the genus $g$ of $n$-punctured closed Riemann surfaces, where each topology is weighted by a factor $N^{\chi}$, with $\chi=2-2g-n$ the Euler characteristic of the Riemann surface.\par
Consequently, the corresponding nonperturbative large-$N$ effective theory involves an infinite number of weakly interacting glueballs and gluinoballs with coupling of order $\frac{1}{N}$ \cite{tHooft:1973alw,Migdal:1977nu,Witten:1979kh} and masses proportional to the RG-invariant scale $\Lambda_{SYM}$. \par
By assuming 't Hooft topological expansion in $\mathcal{N}=1$ SUSY SU($N$) YM theory, the generating functional $\mathcal{W}^E[J_O,J_M]=\log \mathcal{Z}^E[J_O,J_M]$ of Euclidean connected correlators of single-trace bosonic and fermionic balanced operators $O$, $M$, respectively,  with
\begin{align}
	\mathcal{Z}^E[J_O,J_M]=\frac{1}{\mathcal{Z}^E}  \int \mathcal{D}A\mathcal{D}\chi\, e^{-  S_{SYM}+ \sum_s\int J_{O_s}O_s+  J_{M_s}M_s}
\end{align}
reads
\begin{align}
	\label{thooft}
	\mathcal{W}^E[J_O,J_M]=\mathcal{W}^E_{\text{sphere}}[J_O,J_M]+\mathcal{W}^E_{\text{torus}}[J_O,J_M]+ \cdots \,.
\end{align}
Nonperturbatively, $\mathcal{W}^E_{\text{sphere}}[J_O,J_M]$, which perturbatively is the ('t Hooft-)planar contribution \cite{tHooft:1973alw}, is a sum of tree diagrams involving glueball/gluinoball propagators and vertices, while $\mathcal{W}^E_{\text{torus}}[J_O,J_M]$, which perturbatively is the leading-non('t Hooft-)planar contribution, is a sum of glueball/gluinoball one-loop diagrams. \par
Nonperturbatively, $\mathcal{W}^E_{\text{torus}}[J_O,J_M]$ should have the structure of the logarithm of a functional (super)determinant, as it has been originally predicted in the pure YM case \cite{Bochicchio:2016toi} on the basis of fundamental principles, and subsequently asymptotically verified \cite{BPSpaper2,BPSL,QCD24}. \par 
Indeed, in analogy with the pure YM case \cite{Bochicchio:2016toi,BPSL,QCD24}, in the yet-to-come nonperturbative solution of large-$N$ $\mathcal{N}=1$ SUSY YM theory, the very same correlators should be computed by the correlators of glueball $\Phi$ and gluinoball $\Psi$ fields with an infinite number of components, the corresponding generating functional being schematically \cite{Bochicchio:2016toi,BPSL,QCD24}
\begin{align}
	\mathcal{Z}^E_{\text{glueball/gluinoball}}[J_{\Phi},J_{\Psi}]= \mathcal{Z}^{E-1}_{\text{glueball/gluinoball}} 
	\int\mathcal{D}\Phi \mathcal{D}\Psi\,e^{-S_{\text{glueball/gluinoball}}(\Phi,\Psi)+\int \Phi \ast_1 J_{\Phi}+\Psi \ast'_1 J_{\Psi}} \,,
\end{align}
with \cite{Bochicchio:2016toi,BPSL,QCD24}
\begin{align}
	S_{\text{glueball/gluinoball}}(\Phi,\Psi)
	&=\frac{1}{2}\int\,\Phi\ast_2(-\Delta+M^2)\Phi + \Psi\ast'_2(-\Delta+M^2)\Psi \nonumber \\
	&\quad+\frac{1}{N}(\frac{1}{3}\Phi \ast_3\Phi\ast_3\Phi+\Psi \ast'_3\Phi\ast'_3\Psi)+\cdots \, ,
\end{align}
where $\ast_2$, $\ast_1$ and $\ast'_2$, $\ast'_1$ are fixed below, the ellipses and $\ast_3$, $\ast'_3$ respectively stand for $n$-glueball/gluinoball vertices with $n>3$ and some presently unknown operation on the glueball/gluinoball fields that, by assuming locality and Euclidean invariance, may involve derivatives. Hence, nonperturbatively the connected generating functional $\mathcal{W}^E_{\text{glueball/gluinoball}}[J_{\Phi},J_{\Psi}] = \log\mathcal{Z}^E_{\text{glueball/gluinoball}}[J_{\Phi},J_{\Psi}]$ reads to one loop of glueballs/gluinoballs \cite{Bochicchio:2016toi,BPSL,QCD24}
	\begin{equation}
		\label{glueballW}
				\resizebox{0.98\textwidth}{!}{%
			$\begin{aligned}
		\mathcal{W}^E_{\text{glueball/gluinoball}}[J_{\Phi},J_{\Psi}] =& -S_{\text{glueball/gluinoball}}(\Phi_J,\Psi_J)+\int \Phi_J \ast_1 J_{\Phi}+\int \Psi_J \ast'_1 J_{\Psi} + \cdots \\
		&+\frac{1}{2}\log\text{sDet}
		\begin{pmatrix}\ast'_2(-\Delta+M^2)+\frac{1}{N}\ast'_3\Phi_J\ast'_3& \frac{1}{N}\ast'_3\ast'_3\Psi_J\\ 
			\frac{1}{N}\ast'_3\ast'_3\Psi_J&\ast_2(-\Delta+M^2)+\frac{1}{N}\ast_3\Phi_J\ast_3 \end{pmatrix} \,,
				\end{aligned}$
	} 
	\end{equation}
where $\Phi_J$, $\Psi_J$ are determined by 
\begin{equation}
	\frac{\delta S_{\text{glueball/gluinoball}}}{\delta\Phi}\Big\rvert_{\Phi_J}=\ast_1 J_{\Phi} 
\end{equation}
and
\begin{equation}
	\frac{\delta S_{\text{glueball/gluinoball}}}{\delta\Psi}\Big\rvert_{\Psi_J}=\ast'_1 J_{\Psi}\,,
\end{equation}
with the superdeterminant defined by \cite{zinn1993quantum}
\begin{align}\label{eq:sdet}
	\text{sDet}\begin{pmatrix}A & B \\ C & D\end{pmatrix}=\Det(A-BD^{-1}C)\Det(D)^{-1}= \Det(A)\Det(D-CA^{-1}B)^{-1}\,,
\end{align}
where $A$, $D$ are bosonic entries and $B$, $C$ fermionic ones.\par
The dictionary between $\mathcal{W}^E[J_O,J_M]$ and $\mathcal{W}^E_{\text{glueball/gluinoball}}[J_{\Phi},J_{\Psi}]$ is obtained by matching the corresponding
spectral representations -- as a sum of free propagators with residues $R_{sm}$, $R'_{sm}$ -- for the $2$-point correlators at $N =\infty$ \cite{Migdal:1977nu} of $O_s$, $M_s$, respectively, that, by fixing $\ast_2$, $\ast'_2$ according to the canonical normalization of the glueball/gluinoball kinetic term, uniquely determines the coupling of $J_{\Phi}$, $J_{\Psi}$ to the tower of glueball/gluinoball fields $\Phi \ast_1 J_{\Phi} = \sum_{sm} \Phi_{sm} \sqrt{R_{sm}} J_{\Phi_s}$, $\Psi \ast'_1 J_{\Psi} = \sum_{sm} \Psi_{sm} \sqrt{R'_{sm}} J_{\Psi_s}$, respectively.\par
Until the present paper nothing has been known quantitatively on $\mathcal{W}^E[J_O,J_M]$ and $\mathcal{W}^E_{\text{glueball/gluinoball}}[J_{\Phi},J_{\Psi}]$. \par Indeed, in the next Secs. we will compute the UV asymptotics of $\mathcal{W}^E[J_O,J_M]$ and $\mathcal{W}^E_{\text{glueball/gluinoball}}[J_{\Phi},J_{\Psi}]$ for balanced twist-$2$ operators to the planar and leading-nonplanar order in the large-$N$ expansion.

\section{$\mathcal{N} = 1$ SUSY YM theory in the light-cone gauge \label{defSYM}} 

The action of $\mathcal{N} = 1$ SUSY SU($N$) YM theory in Mikowskian space-time reads \cite{Brink:1976bc,Belitsky:2004sc}
\begin{equation}\label{eq:fundamentaltheory}
	S= \int -\frac{1}{4} F^a_{\mu\nu}F^{a\,\mu\nu} +\frac{i}{2}\bar{\chi}^a\gamma^\mu (D_\mu\chi)^a \,d^4x
\end{equation}
with $a=1, \ldots, N^2-1$, where $\chi$ is Majorana spinor $\chi=C\bar{\chi}^T$ in the adjoint representation of the Lie algebra of SU($N$), $C$ the charge conjugation, and
\begin{align}
	&F_{\mu\nu} = \partial_\mu A_\nu-\partial_\nu A_\mu +i\frac{g}{\sqrt{N}}\left[A_\mu,A_\nu\right]\nonumber\\
	&D_\mu \chi= \partial_\mu\chi+i\frac{g}{\sqrt{N}}\left[A_\mu,\chi\right]
\end{align}  
in matrix notation, with $g^2= g^2_{YM}N$ 't Hooft coupling \cite{tHooft:1973alw}. The action is invariant under the SUSY transformations \cite{Brink:1976bc,Belitsky:2003sh}
\begin{align}
	&\delta A^a_\mu =-i \bar{\zeta}\gamma_\mu \chi^a\nonumber\\
	&\delta\chi^a=\frac{i}{2}\sigma^{\mu\nu}F_{\mu\nu}^a\zeta
\end{align}
with $\sigma^{\mu\nu} = \frac{i}{2}\left[\gamma^\mu,\gamma^\nu\right]$ and $\zeta$ a Majorana spinor. A Majorana spinor can be decomposed into two complex-conjugate Weyl spinors, $\lambda_\alpha$ and $\bar{\lambda}_{\dot{\alpha}}$,
\begin{equation}
	\chi = \begin{pmatrix}
		\lambda_\alpha \\
		\bar{\lambda}^{\dot{\alpha}}
	\end{pmatrix}\hspace{1cm}\bar{\chi} = \chi^\dagger \gamma^0 =  \begin{pmatrix}
		\lambda^\alpha & \hspace{-0.1cm}
		\bar{\lambda}_{\dot{\alpha}}
	\end{pmatrix}\,.
\end{equation}
Correspondingly, the action reads
\begin{equation}
	S= - \int \frac{1}{4} F^a_{\mu\nu}F^{a\,\mu\nu} +i\bar{\lambda}^{a\,\dot{\alpha}}\sigma^\mu_{\alpha \dot{\alpha}} (D_\mu\lambda^\alpha)^a \,d^4x\,.
\end{equation}
The fundamental fields, $A_\mu$ and $\chi$, respectively interpolate for the gluon and gluino.
In the light-cone gauge $A_+ = 0$,
\begin{equation}
	\chi_-=\Pi_{-} \chi=\begin{pmatrix}
		0 \\
		\lambda_2 \\
		\bar{\lambda}_{\dot{2}} \\
		0
	\end{pmatrix}
\end{equation}
and $A_-$ can be integrated out, with the projectors $\Pi_{\pm} = \frac{1}{2} \gamma^{\mp}\gamma^{\pm}$. The corresponding action is expressed in terms of the physical fields only -- the transverse components of the gluon, $A$ and $\bar{A}$,
\begin{align}
	& A = \frac{1}{\sqrt{2}} (A_1 + i A_2)\nonumber \\
	& \bar{A} = \frac{1}{\sqrt{2}} (A_1 - i A_2) 
\end{align}
and the plus component of the gluino, $\chi_+$,
\begin{equation}
	\chi_+ = \Pi_+ \chi= \begin{pmatrix}
		\lambda_1 \\
		0 \\
		0 \\
		-\bar{\lambda}_{\dot{1}}
	\end{pmatrix}\,.
\end{equation}
For simplicity we introduce two anticommuting fields, $\lambda$ and $\bar \lambda$, so that their normalization
\begin{align}
	\lambda = \frac{1}{2^{\frac{1}{4}}}\lambda_1 \nonumber \\
	\bar{\lambda} = \frac{1}{2^{\frac{1}{4}}}\bar{\lambda}_{\dot{1}}
\end{align}
matches the canonical normalization of the kinetic term of the action in the light-cone gauge \cite{Belitsky:2004sc}
\begin{align}
	S(A, \bar A,\lambda,\bar{\lambda}) 
	=& \int -\bar{A}^a \square A^a-i\bar{\lambda}^a\square\partial_+^{-1}\lambda^a-2\frac{g}{\sqrt{N}} f^{abc}(A^a\partial_+\bar{A}^b\bar{\partial}\partial_+^{-1}A^c+\bar{A}^a\partial_+A^b\partial\partial_+^{-1}\bar{A}^c)\nonumber\\ &-2\frac{g^2}{N}f^{abc}f^{ade}\partial_+^{-1}(A^b\partial_+ \bar{A}^c)\partial_+^{-1}(\bar{A}^d\partial_+A^e) -2i\frac{g}{\sqrt{N}}f^{abc}\bar{\lambda}^a\lambda^b(\bar{\partial}A^c+\partial\bar{A}^c)\nonumber\\
	&-2i\frac{g^2}{N}f^{abe}f^{cde}\partial_+^{-1}(\bar{A}^a\partial_+ A^b+A^a\partial_+ \bar{A}^b)\partial_{+}^{-1}(\bar{\lambda}^c\lambda^d)\nonumber\\
	&+2\frac{g^2}{N}f^{abe}f^{cde}\partial_{+}^{-1}(\bar{\lambda}^a\lambda^b)\partial_{+}^{-1}(\bar{\lambda}^c\lambda^d)\nonumber\\
	&+2i\frac{g}{\sqrt{N}}f^{abc}\bar{A}^a\bar{\lambda}^b\partial_+^{-1}\partial\lambda^c+2i\frac{g}{\sqrt{N}}f^{abc}A^a\lambda^b\partial_+^{-1}\bar{\partial}\bar{\lambda}^c\nonumber\\
	&-2i\frac{g^2}{N}f^{abe}f^{cde}\bar{A}^a\bar{\lambda}^b\partial_{+}^{-1}(A^c\lambda^d) \,\,d^4x
\end{align} 
with
\begin{equation}
	\square= g^{\mu\nu}\partial_{\mu}\partial_{\nu}=\partial_0^2-\sum_{i=1}^{3}\partial_i^2\,,
\end{equation}
where we employ the mostly minus metric $g_{\mu\nu}$ in Minkowskian space-time (appendix \ref{appN}). 
The v.e.v. of a product of local gauge-invariant operators $\mathcal{O}_i$ that do not depend on $A_{-}$ and $\chi_-$ reads
\begin{align}
	\langle\mathcal{O}_1(x_1)\ldots \mathcal{O}_n(x_n)\rangle 
	=\frac{1}{Z}\int   \mathcal{D} A \mathcal{D} \bar{A} \mathcal{D} \lambda \mathcal{D} \bar{\lambda}\, e^{iS(A,\bar A,\lambda,\bar{\lambda})} \mathcal{O}_1(x_1)\ldots \mathcal{O}_n(x_n)\,.
\end{align}
To the leading perturbative order the above v.e.v. reduces to
\begin{align}
	\langle \mathcal{O}_1(x_1)\ldots \mathcal{O}_n(x_n)\rangle
	=\frac{1}{Z}\int \mathcal{D} A \mathcal{D} \bar{A} \mathcal{D} \lambda \mathcal{D} \bar{\lambda}\,  e^{\int -i \bar{A}^a \square A^a+\bar{\lambda}^a\Box \partial_+^{-1} \lambda^a  \, d^4x}\mathcal{O}_1(x_1)\ldots \mathcal{O}_n(x_n)\,.
\end{align}
We obtain for the corresponding conformal generating functional to the leading order
\begin{align}
	\mathcal{Z}_{\text{conf}}[J_{\mathcal{O}}]
	= \frac{1}{Z} \int \mathcal{D} A \mathcal{D} \bar{A} \mathcal{D} \lambda \mathcal{D} \bar{\lambda} 
	e^{\int -i\bar{A}^a \square A^a+\bar{\lambda}^a\Box \partial_+^{-1} \lambda^a \,  d^4x} 
	\exp\left(\int d^4x\sum_{i}\, J_{\mathcal{O}_i}\mathcal{O}_i\right)
\end{align}
that is the supersymmetric generalization of the conformal generating functional in YM theory \cite{BPSpaper2}.

\section{Twist-2 balanced superfields in $\mathcal{N} = 1$ SUSY YM theory \label{twist2op}}
\label{twist2}

\subsection{Collinear twist-2 operators}\label{twist22}

We list the gauge-invariant collinear twist-2 operators that enter the components of balanced superfields -- to be defined below -- in $\mathcal{N} = 1$ SUSY YM theory \cite{Belitsky:2004sc, Belitsky:2003sh}:
	\begin{itemize}
		\item \underline{gluon-gluon operators}
		\subitem 
		\begin{align}
			O^A_s &= \frac{1}{2} \partial_+ \bar{A}^a(i\overrightarrow{\partial}_++ i\overleftarrow{\partial}_+)^{s-2}C^{\frac{5}{2}}_{s-2}\Bigg(\frac{\overrightarrow{\partial}_+- \overleftarrow{\partial}_+}{\overrightarrow{\partial_+}+\overleftarrow{\partial}_+}\Bigg)\partial_+ A^a \hspace{1cm} \text{ $s =2,4,6,\ldots$}\nonumber\\
			&\equiv\frac{1}{2}\bar{A}^a\mathcal{Y}^{\frac{5}{2}}_{s-2}\left(\overrightarrow{\partial_+},\overleftarrow{\partial}_+\right) A^a
		\end{align}
		\subitem 
		\begin{align}
			\tilde{O}^A_s &= \frac{1}{2} \partial_+ \bar{A}^a(i\overrightarrow{\partial}_++ i\overleftarrow{\partial}_+)^{s-2}C^{\frac{5}{2}}_{s-2}\Bigg(\frac{\overrightarrow{\partial}_+- \overleftarrow{\partial}_+}{\overrightarrow{\partial_+}+\overleftarrow{\partial}_+}\Bigg)\partial_+ A^a \hspace{1cm} \text{ $s =3,5,7,\ldots$}\nonumber\\
			&\equiv\frac{1}{2}\bar{A}^a\mathcal{H}^{\frac{5}{2}}_{s-2}\left(\overrightarrow{\partial_+},\overleftarrow{\partial}_+\right) A^a
		\end{align}
		with $PT=(-1)^s$ and $C=+1$, where $P$ is parity, $T$ time-reversal and $C$ charge conjugation.
		\item \underline{gluino-gluino operators}
		\subitem \begin{align}
			O^\lambda_s &=  \frac{1}{2} \bar{\lambda}^a(i\overrightarrow{\partial}_++ i\overleftarrow{\partial}_+)^{s-1}C^{\frac{3}{2}}_{s-1}\Bigg(\frac{\overrightarrow{\partial}_+- \overleftarrow{\partial}_+}{\overrightarrow{\partial_+}+\overleftarrow{\partial}_+}\Bigg) \lambda^a \hspace{1cm} \text{ $s = 2,4,6,\ldots$}\nonumber\\
			&\equiv\frac{1}{2}\bar{\lambda}^a\mathcal{Y}^{\frac{3}{2}}_{s-1}\left(\overrightarrow{\partial_+},\overleftarrow{\partial}_+\right) \lambda^a
		\end{align}
		\subitem \begin{align}
			\tilde{O}^\lambda_s &=  \frac{1}{2} \bar{\lambda}^a(i\overrightarrow{\partial}_++ i\overleftarrow{\partial}_+)^{s-1}C^{\frac{3}{2}}_{s-1}\Bigg(\frac{\overrightarrow{\partial}_+- \overleftarrow{\partial}_+}{\overrightarrow{\partial_+}+\overleftarrow{\partial}_+}\Bigg) \lambda^a \hspace{1cm} \text{ $s =1,3,5,\ldots$}\nonumber\\
			&\equiv\frac{1}{2}\bar{\lambda}^a\mathcal{H}^{\frac{3}{2}}_{s-1}\left(\overrightarrow{\partial_+},\overleftarrow{\partial}_+\right) \lambda^a
		\end{align}
		with $PT=(-1)^{s}$ and $C=+1$.
		\item \underline{gluon-gluino operators}
		\begin{align}
			M_s &= \frac{1}{2}\hspace{0.08cm} \partial_+A^a (i\overrightarrow{\partial}_++ i\overleftarrow{\partial}_+)^{s-1}P^{(2,1)}_{s-1}\Bigg(\frac{\overrightarrow{\partial}_+- \overleftarrow{\partial}_+}{\overrightarrow{\partial_+}+\overleftarrow{\partial}_+}\Bigg) \lambda^a \hspace{1cm} \text{ $s = 1,2,3,\ldots$}\nonumber\\
			\nonumber\\
			& \equiv \frac{1}{2}A^a \mathcal{G}_{s-1}^{(2,1)}\left(\overrightarrow{\partial_+},\overleftarrow{\partial}_+\right) \lambda^a = \frac{1}{2}(-1)^{s-1}\lambda^a \mathcal{G}_{s-1}^{(1,2)}\left(\overrightarrow{\partial_+},\overleftarrow{\partial}_+\right)  A^a
		\end{align}
		\subitem
		\begin{align}
			\bar{M}_s &=  \hspace{0.08cm}\frac{1}{2}\bar{\lambda}^a(i\overrightarrow{\partial}_++ i\overleftarrow{\partial}_+)^{s-1}P^{(1,2)}_{s-1}\Bigg(\frac{\overrightarrow{\partial}_+- \overleftarrow{\partial}_+}{\overrightarrow{\partial_+}+\overleftarrow{\partial}_+}\Bigg) \partial_+\bar{A}^a \hspace{1cm} \text{ $s = 1,2,3,\ldots$}\nonumber\\
			\nonumber\\
			& \equiv\frac{1}{2}\bar{\lambda}^a  \mathcal{G}_{s-1}^{(1,2)}\left(\overrightarrow{\partial_+},\overleftarrow{\partial}_+\right) \bar{A}^a=\frac{1}{2} (-1)^{s-1}\bar{A}^a\mathcal{G}_{s-1}^{(2,1)}\left(\overrightarrow{\partial_+},\overleftarrow{\partial}_+\right) \bar{\lambda}^a 
		\end{align}
	\end{itemize}

where $C^{\alpha}_n$ are Gegenbauer polynomials, $P^{(\alpha,\beta)}_n$ Jacobi polynomials (appendix \ref{appB}), and
we have defined
\begin{align}
	\label{defY1}
	\mathcal{Y}_{s-2}^{\frac{5}{2}}(\overrightarrow{\partial}_+,\overleftarrow{\partial}_+) 
	&= \overleftarrow{\partial}_+ (i\overrightarrow{\partial}_++i\overleftarrow{\partial}_+)^{s-2}C^{\frac{5}{2}}_{s-2}\left(\frac{\overrightarrow{\partial}_+-\overleftarrow{\partial}_+}{\overrightarrow{\partial}_++\overleftarrow{\partial}_+}\right)\overrightarrow{\partial}_+\nonumber\\
	&=\frac{\Gamma(3)\Gamma(s+3)}{\Gamma(5)\Gamma(s+1)}i^{s-2}\sum_{k=0}^{s-2} {s\choose k}{s\choose k+2}(-1)^{s-k} \overleftarrow{\partial}_{+}^{s-k-1} \overrightarrow{\partial}_{+}^{k+1}
\end{align}
and
\begin{align}	\label{defY2}
	\mathcal{Y}_{s-1}^{\frac{3}{2}}(\overrightarrow{\partial}_+,\overleftarrow{\partial}_+) &=  (i\overrightarrow{\partial}_++i\overleftarrow{\partial}_+)^{s-2}C^{\frac{3}{2}}_{s-1}\left(\frac{\overrightarrow{\partial}_+-\overleftarrow{\partial}_+}{\overrightarrow{\partial}_++\overleftarrow{\partial}_+}\right)\nonumber\\
	&=\frac{(s+1)}{2}i^{s-1}\sum_{k = 0}^{s-1}{s\choose k}{s\choose k+1}(-1)^{s-k-1} \overleftarrow{\partial_{+}}^{s-k-1}\overrightarrow{\partial_{+}}^{k} 
\end{align}
for even $s$, and
\begin{align}
	\label{defH1}
	\mathcal{H}_{s-2}^{\frac{5}{2}}(\overrightarrow{\partial}_+,\overleftarrow{\partial}_+)
	&= \overleftarrow{\partial}_+ (i\overrightarrow{\partial}_++i\overleftarrow{\partial}_+)^{s-2}C^{\frac{5}{2}}_{s-2}\left(\frac{\overrightarrow{\partial}_+-\overleftarrow{\partial}_+}{\overrightarrow{\partial}_++\overleftarrow{\partial}_+}\right)\overrightarrow{\partial}_+\nonumber\\
	&=\frac{\Gamma(3)\Gamma(s+3)}{\Gamma(5)\Gamma(s+1)}i^{s-2}\sum_{k=0}^{s-2} {s\choose k}{s\choose k+2}(-1)^{s-k} \overleftarrow{\partial}_{+}^{s-k-1} \overrightarrow{\partial}_{+}^{k+1}
\end{align}
and
\begin{align}
	\label{defH2}
	\mathcal{H}_{s-1}^{\frac{3}{2}}(\overrightarrow{\partial}_+,\overleftarrow{\partial}_+) 
	&=(i\overrightarrow{\partial}_++i\overleftarrow{\partial}_+)^{s-2}C^{\frac{3}{2}}_{s-1}\left(\frac{\overrightarrow{\partial}_+-\overleftarrow{\partial}_+}{\overrightarrow{\partial}_++\overleftarrow{\partial}_+}\right)\nonumber\\
	&=\frac{(s+1)}{2}i^{s-1} \sum_{k = 0}^{s-1}{s\choose k}{s\choose k+1}(-1)^{s-k-1} \overleftarrow{\partial_{+}}^{s-k-1}\overrightarrow{\partial_{+}}^{k} 
\end{align}
for odd $s$, with
\begin{align}
	\label{gdef1}
	\mathcal{G}_{s-1}^{(1,2)} (\overrightarrow{\partial}_+,\overleftarrow{\partial}_+)
	&=(i\overrightarrow{\partial}_++ i\overleftarrow{\partial}_+)^{s-1}P^{(1,2)}_{s-1}\Bigg(\frac{\overrightarrow{\partial}_+- \overleftarrow{\partial}_+}{\overrightarrow{\partial_+}+\overleftarrow{\partial}_+}\Bigg) \overrightarrow{\partial}_+ \nonumber\\
	&=  i^{s-1} \sum_{k = 0}^{s-1}{s\choose k}{s+1\choose k+2} (-1)^{s-k-1} \overleftarrow{\partial_{+}}^{s-k-1} \overrightarrow{\partial_{+}}^{k+1}
\end{align}
and
\begin{align}
	\label{gdef2}
	\mathcal{G}_{s-1}^{(2,1)} (\overrightarrow{\partial}_+,\overleftarrow{\partial}_+)
	&=\overleftarrow{\partial}_+(i\overrightarrow{\partial}_++ i\overleftarrow{\partial}_+)^{s-1}P^{(2,1)}_{s-1}\Bigg(\frac{\overrightarrow{\partial}_+- \overleftarrow{\partial}_+}{\overrightarrow{\partial_+}+\overleftarrow{\partial}_+}\Bigg)\nonumber\\
	&=   i^{s-1} \sum_{k = 0}^{s-1}{s+1\choose k}{s\choose k+1} (-1)^{s-k-1} \overleftarrow{\partial_{+}}^{s-k} \overrightarrow{\partial_{+}}^{k}
\end{align}
for odd $s$. The spin projection on the $p_+$ direction is $s$ for $O^A_s, \tilde{O}^A_s$ and $O^\lambda_s,\tilde{O}^\lambda_s$, while it is $s + \frac{1}{2}$ for $M_s$ and $\bar{M}_s$.
The bosonic operators $O^A_s, \tilde{O}^A_s$ and $O^\lambda_s,\tilde{O}^\lambda_s$ are Hermitian. The fermionic operators $M_s$, $\bar{M}_s$ are Hermitian conjugate of each other.

\subsection{Balanced superfields}

Suitable linear combinations of the above operators form irreducible representations of the superalgebra of SUSY transformations restricted to the light-cone \cite{Belitsky:2003sh,Belitsky:1998gu}
\begin{align}
	&\delta A = -2i\bar{\lambda}\zeta\nonumber\\
	&\delta \bar{A} = 2i\bar{\zeta}\lambda\nonumber\\
	&\delta\lambda=2\zeta\partial_+\bar{A}\nonumber\\
	&\delta\bar{\lambda} = 2\bar{\zeta}\partial_+A\, ,
\end{align}
where $\zeta$ is Majorana spinor satisfying $\Pi_+\zeta = 0$. Indeed, since the restricted SUSY transformations are linear, the set of the above operators is closed under their action. 
The operators that form supermultiplets -- i.e., the aforementioned representations -- read as follows \cite{Belitsky:2004sc,Belitsky:1998gu}. \par
For even spin with $s \geq 2$
\begin{align}
	\label{S}
	& S^{(1)}_s = \frac{6}{s-1}O^A_s -  O^\lambda_s \nonumber\\
	& S^{(2)}_s = \frac{6}{s+2}\, O^A_s +O^\lambda_s
\end{align}
and for odd spin with $s\geq 3$
\begin{align}
	\label{Stilde}
	& \tilde{S}^{(1)}_s =   -\frac{6}{s-1}\tilde{O}^A_s -  \tilde{O}^\lambda_s \nonumber\\
	& \tilde{S}^{(2)}_s =  -\frac{6}{s+2}\,\tilde{O}^A_s + \tilde{O}^\lambda_s\, ,
\end{align}
where $ \tilde{O}^A_s$ is not defined for $s = 1$, while $\tilde{O}^\lambda_1$ it is, and $\tilde{S}^{(2)}_1 =\tilde{O}^\lambda_1$.
The above operators also diagonalize the matrix of anomalous dimension to order $g^2$, where  $\mathcal{N} = 1$ SUSY YM theory is actually conformal invariant in the conformal scheme (appendix \ref{B}). The components of the above supermultiplets enter the balanced superfields (Sec. 6.3 in \cite{SS1})
\begin{align}
	\mathbb{W}_s(x,\theta,\bar{\theta}) \sim \mathbb{S}^{(2)}_{s+1}+\theta \bar{M}_{s+1}+\bar{\theta}M_{s+1}+\theta\bar{\theta}\mathbb{S}^{(1)}_{s+2}\, ,
\end{align}
where $\mathbb{S}^{(i)}= \{S^{(i)},\tilde{S}^{(i)}\}$ include both even- and odd-spin operators.

\section{Generating functional of Minkowskian conformal correlators \label{functionalInt}}	

The corresponding generating functional reads to the lowest perturbative order
\begin{align}\label{eq:generatingfunctionalfirst}
	& \mathcal{Z}_{\text{conf}}\left[J_{O^A},J_{\tilde{O}^A},J_{O^\lambda},J_{\tilde{O}^\lambda},\bar{J}_{M},J_{\bar M}\right] =\frac{1}{Z} \int \mathcal{D} A \mathcal{D} \bar{A} \mathcal{D} \lambda \mathcal{D} \bar{\lambda} \, e^{-\int i\bar{A}^a \square A^a-\bar{\lambda}^a\Box \partial_+^{-1} \lambda^a \, d^4x}  \nonumber\\
	&\quad\exp \Big( \int \sum_s O^A_s J_{O^A_s}+\tilde{O}^A_s J_{\tilde{O}^A_s} + O^\lambda_s J_{O^\lambda_s} +\tilde{O}^\lambda_s J_{\tilde{O}^\lambda_s} + \bar{J}_{M_s} M_s + \bar{M}_s J_{\bar M_s} \,d^4 x \Big) \,.
\end{align}
The sources $J_{O^A_s}$, $J_{\tilde{O}^A_s}$, $J_{O^\lambda_s}$ and $J_{\tilde{O}^\lambda_s}$ are bosonic, while $J_{\bar M_s}$ and $\bar{J}_{M_s}$ are fermionic, so that we make the formal association
\begin{align}
	\frac{\delta}{\delta J_{O^A_{s}}(x)} \longleftrightarrow O^A_{s}(x) \nonumber\\
	\frac{\delta}{\delta J_{O^\lambda_{s}}(x)} \longleftrightarrow O^\lambda_{s}(x) 
\end{align}
and
\begin{align}
	\frac{\delta}{\delta \bar{J}_{M_{s}}(x)} \longleftrightarrow M_{s}(x) \nonumber\\
	-\frac{\delta}{\delta J_{\bar M_{s}}(x)} \longleftrightarrow \bar{M}_{s}(x)\,.
\end{align}
The generating functional of connected correlators $\mathcal{W}_{\text{conf}} = \log \mathcal{Z}_{\text{conf}}$ follows.
For example,
\begin{equation}
	\langle O^A_{s_1}(x)O^A_{s_2}(y) \rangle = \, \frac{\delta}{\delta J_{O^A_{s_1}}(x)} \frac{\delta}{\delta J_{O^A_{s_2}}(y)} \mathcal{W}_{\text{conf}}\,\Bigg|_{J = 0}
\end{equation}
and
\begin{align}
	\langle M_{s_1}(x)\bar{M}_{s_2}(y) \rangle = \, \frac{\delta}{\delta \bar{J}_{M_{s_1}}(x)} \Big(-\frac{\delta}{\delta J_{\bar M_{s_2}}(y)}\Big) \mathcal{W}_{\text{conf}}\,\Bigg|_{J = 0}\,.
\end{align}
For brevity we will omit from now on the specification $\big|_{J = 0}$.

\subsection{Connected generating functional $\mathcal{W}_{\text{conf}}$ as the $\log$ of a superdeterminant of a quadratic form}

The above functional integral is quadratic in the elementary fields and, therefore, it may be computed exactly \cite{BPSpaper2}. Employing the symmetry properties \cite{BPSpaper2} of the Gegenbauer polynomials (appendix \ref{appB}) we get
\begin{align}
	&\mathcal{Z}_{\text{conf}}\left[J_{O^A},J_{\tilde{O}^A},J_{O^\lambda},J_{\tilde{O}^\lambda},\bar{J}_{M},J_{\bar M}\right] = \frac{1}{Z} \int \mathcal{D}A\mathcal{D}\bar{A} \mathcal{D} \lambda \mathcal{D} \bar{\lambda}\nonumber\\
	&\quad \exp\Bigg(-\frac{1}{2}\int d^4x\, \begin{pmatrix}
		\bar{A}^a(x) &A^a(x)
	\end{pmatrix} \mathcal{M}_{AA}^{ab}\begin{pmatrix}
		&A^b(x) & \\
		&\bar{A}^b(x) &
	\end{pmatrix}\nonumber\\
	&\quad-\frac{1}{2}\int d^4x\, \begin{pmatrix}
		\bar{\lambda}^a(x)&\lambda^a(x)
	\end{pmatrix} \mathcal{M}_{\lambda\lambda}^{ab} \begin{pmatrix}
		&\lambda^b(x) & \\
		&\bar{\lambda}^b(x) &
	\end{pmatrix}+\frac{1}{4}\int d^4x\,\begin{pmatrix}
		&\bar{\lambda}^a(x) &\lambda^a(x)
	\end{pmatrix} \mathcal{M}^{ab}_{\lambda A}\begin{pmatrix}
		&A^b(x) & \\
		&\bar{A}^b(x)  &
	\end{pmatrix}\nonumber\\
	&\quad+\frac{1}{4}\int d^4x\,\begin{pmatrix}
		&\bar{A}^a(x)&A^a(x) 
	\end{pmatrix} \mathcal{M}_{A\lambda}^{ab} \begin{pmatrix}
		&\lambda^b(x) & \\
		&\bar{\lambda}^b(x)  &
	\end{pmatrix}\Bigg)\,,
\end{align}
with
	\begin{equation}
		\label{matrixM}
		\resizebox{0.98\textwidth}{!}{%
			$\begin{aligned}
				& \mathcal{M}_{AA}^{ab}=\delta^{ab}
				\begin{pmatrix}i\square-\frac{1}{2}\sum_sJ_{O^A_{s}}\otimes\mathcal{Y}_{s-2}^{\frac{5}{2}}-\frac{1}{2}\sum_sJ_{\tilde{O}^A_{s}}\otimes\mathcal{H}_{s-2}^{\frac{5}{2}} & 0\\ 0&i\square-\frac{1}{2}\sum_sJ_{O^A_{s}}\otimes\mathcal{Y}_{s-2}^{\frac{5}{2}}+\frac{1}{2}\sum_sJ_{\tilde{O}^A_{s}}\otimes\mathcal{H}_{s-2}^{\frac{5}{2}}  \end{pmatrix}\\\\
				& \mathcal{M}_{\lambda\lambda}^{ab}=\delta^{ab}
				\begin{pmatrix}-\square\partial_+^{-1}-\frac{1}{2}\sum_sJ_{O^\lambda_{s}}\otimes\mathcal{Y}_{s-1}^{\frac{3}{2}}-\frac{1}{2}\sum_sJ_{\tilde{O}^\lambda_{s}}\otimes\mathcal{H}_{s-1}^{\frac{3}{2}}& 0\\ 0&-\square\partial_+^{-1}-\frac{1}{2}\sum_sJ_{O^\lambda_{s}}\otimes\mathcal{Y}_{s-1}^{\frac{3}{2}}+\frac{1}{2}\sum_sJ_{\tilde{O}^\lambda_{s}}\otimes\mathcal{H}_{s-1}^{\frac{3}{2}}   \end{pmatrix}\\
				& \mathcal{M}_{A\lambda}^{ab}=\frac{\delta^{ab}}{2}
				\begin{pmatrix}0 & -\sum_sJ_{\bar M_s}\otimes\mathcal{G}_{s-1}^{(2,1)}(-1)^{s-1}\\ \sum_s\bar{J}_{M_s}\otimes  \mathcal{G}_{s-1}^{(2,1)}&0 \end{pmatrix}\\\\
				& \mathcal{M}_{\lambda A}^{ab}=\frac{\delta^{ab}}{2}
				\begin{pmatrix}0& \sum_sJ_{\bar M_s} \otimes\mathcal{G}_{s-1}^{(1,2)} \\ -\sum_s\bar{J}_{M_s}\otimes  \mathcal{G}_{s-1}^{(1,2)}(-1)^{s-1}&0 \end{pmatrix}\,,
			\end{aligned}$
		} 
	\end{equation}
where the symbol $\otimes$ implies that the right and left derivatives do not act on the sources $J$.
We group the above matrices into a single supermatrix
\begin{align}
	\mathcal{X}^{ab} = 	\begin{pmatrix}\mathcal{M}^{ab}_{\lambda\lambda} & \mathcal{M}^{ab}_{\lambda A}\\ \mathcal{M}^{ab}_{A\lambda}&\mathcal{M}^{ab}_{AA} \end{pmatrix}\,,
\end{align}
so that the generating functional reads
	\begin{align}
		&\mathcal{Z}_{\text{conf}}\left[J_{O^A},J_{\tilde{O}^A},J_{O^\lambda},J_{\tilde{O}^\lambda},\bar{J}_{M},J_{\bar M}\right] \nonumber\\
		&= \frac{1}{Z} \int \mathcal{D}A\mathcal{D}\bar{A} \mathcal{D} \lambda \mathcal{D} \bar{\lambda} 
		\exp\Bigg(-\frac{1}{2}\int d^4x\, \begin{pmatrix}
			\bar{\lambda}^a(x)&\lambda^a(x)&	\bar{A}^a(x) &A^a(x)
		\end{pmatrix} \mathcal{X}^{ab}\begin{pmatrix}
			&\lambda^b(x) &\\
			&\bar{\lambda}^b(x) & \\
			&A^b(x) &\\
			&\bar{A}^b(x) & 
		\end{pmatrix}\Bigg)
	\end{align}
that reduces to the superdeterminant \cite{zinn1993quantum}
\begin{align}
	\mathcal{Z}_{\text{conf}}\left[J_{O^A},J_{\tilde{O}^A},J_{O^\lambda},J_{\tilde{O}^\lambda},\bar{J}_{M},J_{\bar M}\right] = \sDet^{\frac{1}{2}}\left(\mathcal{X}\right)\,.
\end{align}
More explicitly,
\begin{align}
	\mathcal{Z}_{\text{conf}}\left[J_{O^A},J_{\tilde{O}^A},J_{O^\lambda},J_{\tilde{O}^\lambda},\bar{J}_{M},J_{\bar M}\right] 
	&= \Det^{\frac{1}{2}}(\mathcal{M}_{\lambda\lambda} )\Det^{-\frac{1}{2}}\left(\mathcal{M}_{AA}-\mathcal{M}_{A\lambda}\mathcal{M}_{\lambda\lambda}^{-1}\mathcal{M}_{\lambda A}\right)\nonumber\\
	&= \Det^{-\frac{1}{2}}(\mathcal{M}_{AA})\Det^{\frac{1}{2}}\left(\mathcal{M}_{\lambda\lambda}-\mathcal{M}_{\lambda A }\mathcal{M}_{AA}^{-1}\mathcal{M}_{A\lambda }\right)\, ,
\end{align}
where the two results depend on first integrating either on the fermionic or bosonic variables. \par
The entries of $\mathcal{M}_{A\lambda}\mathcal{M}_{\lambda\lambda}^{-1}\mathcal{M}_{\lambda A}$ read
	\begin{equation}
		\resizebox{0.98\textwidth}{!}{%
			$\begin{aligned}
				&\left[	\mathcal{M}_{\lambda A }\mathcal{M}_{AA}^{-1}\mathcal{M}_{A\lambda }\right]_{11}=\frac{1}{4}\mathcal{G}_{s-1}^{(1,2)}(-1)^{s-1}\otimes \bar{J}_{M_s}(i\square-\frac{1}{2}J_{O^A_{s_1}}\otimes\mathcal{Y}_{s_1-2}^{\frac{5}{2}}-\frac{1}{2}J_{\tilde{O}^A_{s_1}}\otimes\mathcal{H}_{s_1-2}^{\frac{5}{2}} )^{-1}   \mathcal{G}_{s_2-1}^{(2,1)}(-1)^{s_2-1}\otimes J_{\bar M_{s_2}}\\
				&\left[\mathcal{M}_{\lambda A }\mathcal{M}_{AA}^{-1}\mathcal{M}_{A\lambda }\right]_{12}=0\\
				&\left[	\mathcal{M}_{\lambda A }\mathcal{M}_{AA}^{-1}\mathcal{M}_{A\lambda }\right]_{21}=0\\
				&\left[\mathcal{M}_{\lambda A }\mathcal{M}_{AA}^{-1}\mathcal{M}_{A\lambda }\right]_{22}=\frac{1}{4}\mathcal{G}_{s-1}^{(1,2)}\otimes J_{\bar M_s}(i\square-\frac{1}{2}J_{O^A_{s_1}}\otimes\mathcal{Y}_{s_1-2}^{\frac{5}{2}}+\frac{1}{2}J_{\tilde{O}^A_{s_1}}\otimes\mathcal{H}_{s_1-2}^{\frac{5}{2}})^{-1} \mathcal{G}_{s_2-1}^{(2,1)}\otimes \bar{J}_{M_{s_2}}\,,
			\end{aligned}$
		} 
	\end{equation}
	where the sum over repeated spin indices is understood, and similarly
	\begin{equation}
		\resizebox{0.98\textwidth}{!}{%
			$\begin{aligned}
				&\left[\mathcal{M}_{A\lambda}\mathcal{M}_{\lambda\lambda}^{-1}\mathcal{M}_{\lambda A}\right]_{11}=\frac{1}{4}  \mathcal{G}_{s-1}^{(2,1)}(-1)^{s-1}\otimes J_{\bar M_s}(-\square\partial_+^{-1}-\frac{1}{2}J_{O^\lambda_{s_1}}\otimes\mathcal{Y}_{s_1-1}^{\frac{3}{2}}+\frac{1}{2}J_{\tilde{O}^\lambda_{s_1}}\otimes\mathcal{H}_{s_1-1}^{\frac{3}{2}} )^{-1} \mathcal{G}_{s_2-1}^{(1,2)}(-1)^{s_2-1}\otimes \bar{J}_{M_{s_2}}\\
				&\left[\mathcal{M}_{A\lambda}\mathcal{M}_{\lambda\lambda}^{-1}\mathcal{M}_{\lambda A}\right]_{12}=0\\
				&\left[\mathcal{M}_{A\lambda}\mathcal{M}_{\lambda\lambda}^{-1}\mathcal{M}_{\lambda A}\right]_{21}=0\\
				&\left[\mathcal{M}_{A\lambda}\mathcal{M}_{\lambda\lambda}^{-1}\mathcal{M}_{\lambda A}\right]_{22}=\frac{1}{4} \mathcal{G}_{s-1}^{(2,1)}\otimes \bar{J}_{M_s}( -\square\partial_+^{-1}-\frac{1}{2}J_{O^\lambda_{s_1}}\otimes\mathcal{Y}_{s_1-1}^{\frac{3}{2}}-\frac{1}{2}J_{\tilde{O}^\lambda_{s_1}}\otimes\mathcal{H}_{s_1-1}^{\frac{3}{2}})^{-1}   \mathcal{G}_{s_2-1}^{(1,2)}\otimes J_{\bar M_{s_2}}\,.
			\end{aligned}$
		} 
	\end{equation}
Evaluating the determinant of block matrices \cite{BPSpaper2} by means of
\begin{align}
	\label{ourdet}
	\Det 	\begin{pmatrix}A& B\\ C& D \end{pmatrix} 
	=\Det (A)\,\Det(D)\Det\left(1-D^{-1}C A^{-1}B\right)
\end{align}
provided that $A^{-1}$ and $D^{-1}$ exist, we get for the connected generating functional
	\begin{align}
		\label{wgen1}
		&\mathcal{W}_{\text{conf}}\left[J_{O^A},J_{\tilde{O}^A},J_{O^\lambda},J_{\tilde{O}^\lambda},\bar{J}_{M},J_{\bar M}\right] \nonumber\\
		&=-\frac{1}{2}\log\Det \Big(\mathcal{I} +\frac{1}{2}i\square^{-1}J_{O^A_{s}}\otimes\mathcal{Y}_{s-2}^{\frac{5}{2}}+\frac{1}{2}i\square^{-1}J_{\tilde{O}^A_{s}}\otimes\mathcal{H}_{s-2}^{\frac{5}{2}} \Big)  \nonumber\\
		&\quad-\frac{1}{2}\log\Det \Big( \mathcal{I}+\frac{1}{2}i\square^{-1}J_{O^A_{s}}\otimes\mathcal{Y}_{s-2}^{\frac{5}{2}}-\frac{1}{2}i\square^{-1}J_{\tilde{O}^A_{s}}\otimes\mathcal{H}_{s-2}^{\frac{5}{2}} \Big)  \nonumber\\
		&\quad +\frac{1}{2}\log\Det \Big(\mathcal{I}+\frac{1}{2}\partial_+\square^{-1}J_{O^\lambda_{s}}\otimes\mathcal{Y}_{s-1}^{\frac{3}{2}}+\frac{1}{2}\partial_+\square^{-1}J_{\tilde{O}^\lambda_{s}}\otimes\mathcal{H}_{s-1}^{\frac{3}{2}} \Big)\nonumber\\
		&\quad+\frac{1}{2}\log\Det \Big(\mathcal{I}+\frac{1}{2}\partial_+\square^{-1}J_{O^\lambda_{s}}\otimes\mathcal{Y}_{s-1}^{\frac{3}{2}}-\frac{1}{2}\partial_+\square^{-1}J_{\tilde{O}^\lambda_{s}}\otimes\mathcal{H}_{s-1}^{\frac{3}{2}}  \Big)\nonumber\\
		& \quad+\frac{1}{2}\log\Det \Big(\mathcal{I}-\frac{1}{4}(\mathcal{I}+\frac{1}{2}\partial_+\square^{-1}J_{O^\lambda_{s}}\otimes\mathcal{Y}_{s-1}^{\frac{3}{2}}-\frac{1}{2}\partial_+\square^{-1}J_{\tilde{O}^\lambda_{s}}\otimes\mathcal{H}_{s-1}^{\frac{3}{2}} )^{-1}\nonumber\\ &\qquad\partial_+\square^{-1}\bar{J}_{M_{s_1}}\otimes\mathcal{G}_{s_1-1}^{(1,2)}(-1)^{s_1-1} \nonumber\\
		&\qquad  (\mathcal{I}+\frac{1}{2}i\square^{-1}J_{O^A_{s_2}}\otimes\mathcal{Y}_{s_2-2}^{\frac{5}{2}}+\frac{1}{2}i\square^{-1}J_{\tilde{O}^A_{s_2}}\otimes\mathcal{H}_{s_2-2}^{\frac{5}{2}} )^{-1}i\square^{-1} J_{\bar M_{s_3}}\otimes\mathcal{G}_{s_3-1}^{(2,1)}(-1)^{s_3-1} \Big)\nonumber\\
		&\quad +\frac{1}{2}\log\Det \Big(\mathcal{I}  - \frac{1}{4}(\mathcal{I}+\frac{1}{2}\partial_+\square^{-1}J_{O^\lambda_{s}}\otimes\mathcal{Y}_{s-1}^{\frac{3}{2}}+\frac{1}{2}\partial_+\square^{-1}J_{\tilde{O}^\lambda_{s}}\otimes\mathcal{H}_{s-1}^{\frac{3}{2}} )^{-1}\nonumber\\
		&\qquad\partial_+\square^{-1}J_{\bar M_{s_1}}\otimes\mathcal{G}_{s_1-1}^{(1,2)} \nonumber\\
		& \qquad (\mathcal{I}+\frac{1}{2}i\square^{-1}J_{O^A_{s_2}}\otimes\mathcal{Y}_{s_2-2}^{\frac{5}{2}}-\frac{1}{2}i\square^{-1}J_{\tilde{O}^A_{s_2}}\otimes\mathcal{H}_{s_2-2}^{\frac{5}{2}})^{-1} i\square^{-1}\bar{J}_{M_{s_3}}\otimes\mathcal{G}_{s_3-1}^{(2,1)}\Big)
	\end{align}
	and
	\begin{align}
		\label{wgen2}
		&\mathcal{W}_{\text{conf}}\left[J_{O^A},J_{\tilde{O}^A},J_{O^\lambda},J_{\tilde{O}^\lambda},\bar{J}_{M},J_{\bar M}\right] \nonumber\\
		&=+\frac{1}{2}\log\Det \Big(\mathcal{I}+\frac{1}{2}\partial_+\square^{-1}J_{O^\lambda_{s}}\otimes\mathcal{Y}_{s-1}^{\frac{3}{2}}+\frac{1}{2}\partial_+\square^{-1}J_{\tilde{O}^\lambda_{s}}\otimes\mathcal{H}_{s-1}^{\frac{3}{2}} \Big)  \nonumber\\
		&\quad+\frac{1}{2}\log\Det \Big(\mathcal{I}+\frac{1}{2}\partial_+\square^{-1}J_{O^\lambda_{s}}\otimes\mathcal{Y}_{s-1}^{\frac{3}{2}}-\frac{1}{2}\partial_+\square^{-1}J_{\tilde{O}^\lambda_{s}}\otimes\mathcal{H}_{s-1}^{\frac{3}{2}}  \Big)  \nonumber\\
		& \quad-\frac{1}{2}\log\Det \Big(\mathcal{I} +\frac{1}{2}i\square^{-1}J_{O^A_{s}}\otimes\mathcal{Y}_{s-2}^{\frac{5}{2}}+\frac{1}{2}i\square^{-1}J_{\tilde{O}^A_{s}}\otimes\mathcal{H}_{s-2}^{\frac{5}{2}}  \Big)  \nonumber\\
		&\quad-\frac{1}{2}\log\Det \Big( \mathcal{I}+\frac{1}{2}i\square^{-1}J_{O^A_{s}}\otimes\mathcal{Y}_{s-2}^{\frac{5}{2}}-\frac{1}{2}i\square^{-1}J_{\tilde{O}^A_{s}}\otimes\mathcal{H}_{s-2}^{\frac{5}{2}} \Big)  \nonumber\\
		& \quad-\frac{1}{2}\log\Det \Big(\mathcal{I} -\frac{1}{4}(\mathcal{I} +\frac{1}{2}i\square^{-1}J_{O^A_{s}}\otimes\mathcal{Y}_{s-2}^{\frac{5}{2}}+\frac{1}{2}i\square^{-1}J_{\tilde{O}^A_{s}}\otimes\mathcal{H}_{s-2}^{\frac{5}{2}} )^{-1}\nonumber\\
		&\qquad  i\square^{-1}J_{\bar M_{s_1}}\otimes{\mathcal{G}}_{s_1-1}^{(2,1)}(-1)^{s_1-1}\nonumber\\
		& \qquad(\mathcal{I}+\frac{1}{2}\partial_+\square^{-1}J_{O^\lambda_{s_2}}\otimes\mathcal{Y}_{s_2-1}^{\frac{3}{2}}-\frac{1}{2}\partial_+\square^{-1}J_{\tilde{O}^\lambda_{s_2}}\otimes\mathcal{H}_{s_2-1}^{\frac{3}{2}} )^{-1}\partial_+\square^{-1} \bar{J}_{M_{s_3}}\otimes\mathcal{G}_{s_3-1}^{(1,2)}(-1)^{s_3-1} \Big)\nonumber\\
		&\quad -\frac{1}{2}\log\Det \Big(\mathcal{I} - \frac{1}{4}( \mathcal{I}+\frac{1}{2}i\square^{-1}J_{O^A_{s}}\otimes\mathcal{Y}_{s-2}^{\frac{5}{2}}-\frac{1}{2}i\square^{-1}J_{\tilde{O}^A_{s}}\otimes\mathcal{H}_{s-2}^{\frac{5}{2}})^{-1}\nonumber\\
		&\qquad  i\square^{-1}\bar{J}_{M_{s_1}}\otimes\mathcal{G}_{s_1-1}^{(2,1)}\nonumber\\
		&\qquad (\mathcal{I}+\frac{1}{2}\partial_+\square^{-1}J_{O^\lambda_{s_2}}\otimes\mathcal{Y}_{s_2-1}^{\frac{3}{2}}+\frac{1}{2}\partial_+\square^{-1}J_{\tilde{O}^\lambda_{s_2}}\otimes\mathcal{H}_{s_2-1}^{\frac{3}{2}}  )^{-1} \partial_+\square^{-1}J_{\bar M_{s_3}}\otimes  \mathcal{G}_{s_3-1}^{(1,2)}\Big)\,,
	\end{align}
with $\mathcal{I}$ the identity in both color and space-time, and the sum over repeated spin indices understood. 
After rescaling the operators
\begin{align}
	\label{rescale}
	&O^{'A}_s(x)=\frac{1}{N}\frac{2\Gamma(5)\Gamma(s+1)}{\Gamma(3)\Gamma(s+3)}O^A_s(x)\nonumber\\
	&\tilde{O}^{'A}_s(x)=\frac{1}{N}\frac{2\Gamma(5)\Gamma(s+1)}{\Gamma(3)\Gamma(s+3)}	\tilde{O}^A_s(x)
\end{align}
and
\begin{align}
	&O^{'\lambda}_s(x)=\frac{1}{N}\frac{4}{s+1}	O^\lambda_s(x)\nonumber\\
	&\tilde{O}^{'\lambda}_s(x)=\frac{1}{N}\frac{4}{s+1}	\tilde{O}^\lambda_s(x)\nonumber\\
	&M'_s = \frac{2}{N}M_s\,,
\end{align}
	so that their $2$-point correlators are of order $1$ for large $N$, we obtain more explicitly
	\begin{equation}
		\label{wexplicit1}
				\resizebox{0.98\textwidth}{!}{%
			$\begin{aligned}
		&{\mathcal{W}_{\text{conf}}}\left[J_{O^{'A}},J_{\tilde{O}^{'A}},J_{O^{'\lambda}},J_{\tilde{O}^{'\lambda}},\bar{J}_{M'},J_{\bar M'}\right] =\\
		& -\frac{N^2-1}{2}\log\Det\left(I+\frac{1}{N}\sum_{k=0}^{s-2}{s\choose k}{s\choose k+2}(i\overrightarrow{\partial}_+)^{s-k-1}i \square^{-1}(J_{O^{'A}_{s}}+J_{\tilde{O}^{'A}_{s}})(i\overrightarrow{\partial}_+)^{k+1} \right)\\
		&-\frac{N^2-1}{2}\log\Det\left(I+\frac{1}{N}\sum_{k=0}^{s-2}{s\choose k}{s\choose k+2}(i\overrightarrow{\partial}_+)^{s-k-1}i \square^{-1}(J_{O^{'A}_{s}}-J_{\tilde{O}^{'A}_{s}})(i\overrightarrow{\partial}_+)^{k+1} \right)\\
		&+\frac{N^2-1}{2}\log\Det\left(I-\frac{1}{N}\sum_{k=0}^{s-1}{s\choose k}{s\choose k+1}(i\overrightarrow{\partial}_+)^{s-k-1}(i\overrightarrow{\partial}_+)i \square^{-1}(J_{O^{'\lambda}_{s}}+J_{\tilde{O}^{'\lambda}_{s}})(i\overrightarrow{\partial}_+)^{k} \right)\\
		&+\frac{N^2-1}{2}\log\Det\left(I-\frac{1}{N}\sum_{k=0}^{s-1}{s\choose k}{s\choose k+1}(i\overrightarrow{\partial}_+)^{s-k-1}(i\overrightarrow{\partial}_+)i \square^{-1}(J_{O^{'\lambda}_{s}}-J_{\tilde{O}^{'\lambda}_{s}})(i\overrightarrow{\partial}_+)^{k} \right)\\
		&+\frac{N^2-1}{2}\log\Det\Bigg[I+\frac{1}{N^2}\left(I-\frac{1}{N}\sum_{k=0}^{s-1}{s\choose k}{s\choose k+1}(i\overrightarrow{\partial}_+)^{s-k-1}(i\overrightarrow{\partial}_+)i \square^{-1}(J_{O^{'\lambda}_{s}}-J_{\tilde{O}^{'\lambda}_{s}})(i\overrightarrow{\partial}_+)^{k}\right)^{-1}\\
		&\quad \sum_{k_1 = 0}^{s_1-1}{s_1\choose k_1}{s_1+1\choose k_1+2} (-1)^{s_1-1}
		(i\overrightarrow{\partial}_+)^{s_1-k_1-1}(i\overrightarrow{\partial}_+)i\square^{-1}\bar{J}_{M'_{s_1}} (i\overrightarrow{\partial}_+)^{k_1+1}  \\
		&\quad\left(I+\frac{1}{N}\sum_{k_2=0}^{s_2-2}{s_2\choose k_2}{s_2\choose k_2+2}(i\overrightarrow{\partial}_+)^{s_2-k_2-1}i \square^{-1}(J_{O^{'A}_{s_2}}+J_{\tilde{O}^{'A}_{s_2}})(i\overrightarrow{\partial}_+)^{k_2+1}\right)^{-1}\\
		&\quad \sum_{k_3 = 0}^{s_3-1}{s_3+1\choose k_3}{s_3\choose k_3+1} (-1)^{s_3-1}
		(i\overrightarrow{\partial}_+)^{s_3-k_3}i\square^{-1}J_{\bar M'_{s_3}} (i\overrightarrow{\partial}_+)^{k_3}\Bigg]\\
		&+\frac{N^2-1}{2}\log\Det\Bigg[I+\frac{1}{N^2}\left(I-\frac{1}{N}\sum_{k=0}^{s-1}{s\choose k}{s\choose k+1}(i\overrightarrow{\partial}_+)^{s-k-1}(i\overrightarrow{\partial}_+)i \square^{-1}(J_{O^{'\lambda}_{s}}+J_{\tilde{O}^{'\lambda}_{s}})(i\overrightarrow{\partial}_+)^{k}\right)^{-1}\\
		&\quad \sum_{k_1 = 0}^{s_1-1}{s_1\choose k_1}{s_1+1\choose k_1+2} 
		(i\overrightarrow{\partial}_+)^{s_1-k_1-1}(i\overrightarrow{\partial}_+)i\square^{-1}J_{\bar M'_{s_1}} (i\overrightarrow{\partial}_+)^{k_1+1}   \\
		&\quad\left(I+\frac{1}{N}\sum_{k_2=0}^{s_2-2}{s_2\choose k_2}{s_2\choose k_2+2}(i\overrightarrow{\partial}_+)^{s_2-k_2-1}i \square^{-1}(J_{O^{'A}_{s_2}}-J_{\tilde{O}^{'A}_{s_2}})(i\overrightarrow{\partial}_+)^{k_2+1}\right)^{-1}\\
		&\quad  \sum_{k_3 = 0}^{s_3-1}{s_3+1\choose k_3}{s_3\choose k_3+1} 
		(i\overrightarrow{\partial}_+)^{s_3-k_3}i\square^{-1}\bar{J}_{M'_{s_3}} (i\overrightarrow{\partial}_+)^{k_3}\Bigg]
					\end{aligned}$
	} 
	\end{equation}
	and
	\begin{equation}
		\label{wexplicit2}
						\resizebox{0.98\textwidth}{!}{%
			$\begin{aligned}
		&{\mathcal{W}_{\text{conf}}}\left[J_{O^{'A}},J_{\tilde{O}^{'A}},J_{O^{'\lambda}},J_{\tilde{O}^{'\lambda}},\bar{J}_{M'},J_{\bar M'}\right] =\\
		& -\frac{N^2-1}{2}\log\Det\left(I+\frac{1}{N}\sum_{k=0}^{s-2}{s\choose k}{s\choose k+2}(i\overrightarrow{\partial}_+)^{s-k-1}i \square^{-1}(J_{O^{'A}_{s}}+J_{\tilde{O}^{'A}_{s}})(i\overrightarrow{\partial}_+)^{k+1} \right)\\
		&-\frac{N^2-1}{2}\log\Det\left(I+\frac{1}{N}\sum_{k=0}^{s-2}{s\choose k}{s\choose k+2}(i\overrightarrow{\partial}_+)^{s-k-1}i \square^{-1}(J_{O^{'A}_{s}}-J_{\tilde{O}^{'A}_{s}})(i\overrightarrow{\partial}_+)^{k+1} \right)\\
		&+\frac{N^2-1}{2}\log\Det\left(I-\frac{1}{N}\sum_{k=0}^{s-1}{s\choose k}{s\choose k+1}(i\overrightarrow{\partial}_+)^{s-k-1}(i\overrightarrow{\partial}_+)i \square^{-1}(J_{O^{'\lambda}_{s}}+J_{\tilde{O}^{'\lambda}_{s}})(i\overrightarrow{\partial}_+)^{k} \right)\\
		&+\frac{N^2-1}{2}\log\Det\left(I-\frac{1}{N}\sum_{k=0}^{s-1}{s\choose k}{s\choose k+1}(i\overrightarrow{\partial}_+)^{s-k-1}(i\overrightarrow{\partial}_+)i \square^{-1}(J_{O^{'\lambda}_{s}}-J_{\tilde{O}^{'\lambda}_{s}})(i\overrightarrow{\partial}_+)^{k} \right)\\
		&-\frac{N^2-1}{2}\log\Det\Bigg[I+\frac{1}{N^2}\left(I+\frac{1}{N}\sum_{k=0}^{s-2}{s\choose k}{s\choose k+2}(i\overrightarrow{\partial}_+)^{s-k-1}i \square^{-1}(J_{O^{'A}_{s}}-J_{\tilde{O}^{'A}_{s}})(i\overrightarrow{\partial}_+)^{k+1}\right)^{-1}\\
		&\quad  \sum_{k_1 = 0}^{s_1-1}{s_1+1\choose k_1}{s_1\choose k_1+1} (-1)^{s_1-1}
		(i\overrightarrow{\partial}_+)^{s_1-k_1}i\square^{-1}J_{\bar M'_{s_1}} (i\overrightarrow{\partial}_+)^{k_1}\\
		&\quad\left(I-\frac{1}{N}\sum_{k_2=0}^{s_2-1}{s_2\choose k_2}{s_2\choose k_2+1}(i\overrightarrow{\partial}_+)^{s_2-k_2-1}(i\overrightarrow{\partial}_+)i \square^{-1}(J_{O^{'\lambda}_{s_2}}+J_{\tilde{O}^{'\lambda}_{s_2}})(i\overrightarrow{\partial}_+)^{k_2}\right)^{-1}\\
		&\quad \sum_{k_3 = 0}^{s_3-1}{s_3\choose k_3}{s_3+1\choose k_3+2} (-1)^{s_3-1}
		(i\overrightarrow{\partial}_+)^{s_3-k_3-1}(i\overrightarrow{\partial}_+)i\square^{-1}\bar{J}_{M'_{s_3}} (i\overrightarrow{\partial}_+)^{k_3+1} \Bigg]\\
		&-\frac{N^2-1}{2}\log\Det\Bigg[I+\frac{1}{N^2}\left(I+\frac{1}{N}\sum_{k=0}^{s-2}{s\choose k}{s\choose k+2}(i\overrightarrow{\partial}_+)^{s-k-1}i \square^{-1}(J_{O^{'A}_{s}}+J_{\tilde{O}^{'A}_{s}})(i\overrightarrow{\partial}_+)^{k+1}\right)^{-1}\\
		&\quad \sum_{k_1 = 0}^{s_1-1}{s_1+1\choose k_1}{s_1\choose k_1+1} 
		(i\overrightarrow{\partial}_+)^{s_1-k_1}i\square^{-1}\bar{J}_{M'_{s_1}} (i\overrightarrow{\partial}_+)^{k_1} \\
		&\quad\left(I-\frac{1}{N}\sum_{k_2=0}^{s_2-1}{s_2\choose k_2}{s_2\choose k_2+1}(i\overrightarrow{\partial}_+)^{s_2-k_2-1}(i\overrightarrow{\partial}_+)i \square^{-1}(J_{O^{'\lambda}_{s_2}}-J_{\tilde{O}^{'\lambda}_{s_2}})(i\overrightarrow{\partial}_+)^{k_2}\right)^{-1}\\
		&\quad \sum_{k_3 = 0}^{s_3-1}{s_3+1\choose k_3}{s_3\choose k_3+1} 
		(i\overrightarrow{\partial}_+)^{s_3-k_3-1}(i\overrightarrow{\partial}_+)i\square^{-1}J_{\bar M'_{s_3}} (i\overrightarrow{\partial}_+)^{k_3}\Bigg]\,,
							\end{aligned}$
	} 
	\end{equation}

where $I$ is the identity in space-time with kernel
\begin{align}
	I \rightarrow \delta^{(4)}(x-y)
\end{align}
and, after performing the color trace, we have employed the definitions in Eqs. \eqref{defY1}, \eqref{defY2}, \eqref{defH1}, \eqref{defH2}, \eqref{gdef1}, \eqref{gdef2} and \cite{BPSpaper2}
\begin{align}
	i\square^{-1}\overleftarrow{\partial}_{+}^{s-k-1} = (-1)^{s-k-1}\overrightarrow{\partial}_{+}^{s-k-1} i\square^{-1}
\end{align}
that follows from (minus) the propagator in the coordinate representation \cite{BPS1}
\begin{equation}
	\label{propagator}
	i \square^{-1}  	  \rightarrow \frac{1}{4\pi^2}\frac{1}{\rvert x-y \rvert^2-i\epsilon}\,.
\end{equation}

\subsection{Connected generating functional $\Gamma_{\text{conf}}$ as the $\log$ a Fredholm superdeterminant \label{fredgen}}

By means of the dictionary \cite{BPSpaper2} we rewrite the generating functional $\mathcal{W}_{\text{conf}}$ -- that is actually the logarithm of a superdeterminant of a quadratic form as remarked above -- as the logarithm of a superdeterminant $\Gamma_{\text{conf}}$ of integral operators formally of Fredholm type
	\begin{equation}
		\label{genMg1}
							\resizebox{0.98\textwidth}{!}{%
		$\begin{aligned}	&\Gamma_{\text{conf}}\left[j_{O^{A}},j_{\tilde{O}^{A}},j_{O^{\lambda}},j_{\tilde{O}^{\lambda}},\bar{j}_{M},j_{\bar{M}}\right] \\
		&= 
		-\frac{N^2-1}{2}\log\Det \left[\mathbb{I}+\mathcal{D}^{-1}_Aj_{O^A}+\mathcal{D}^{-1}_Aj_{\tilde{O}^A}\right]-\frac{N^2-1}{2}\log\Det\left[\mathbb{I}+\mathcal{D}^{-1}_Aj_{O^A}-\mathcal{D}^{-1}_Aj_{\tilde{O}^A}\right]\\
		&\quad+\frac{N^2-1}{2}\log\Det \left[\mathbb{I}-\mathcal{D}^{-1}_\lambda j_{O^\lambda}-\mathcal{D}^{-1}_\lambda j_{\tilde{O}^\lambda}\right]+\frac{N^2-1}{2}\log\Det\left[\mathbb{I}-\mathcal{D}^{-1}_\lambda j_{O^\lambda}+\mathcal{D}^{-1}_\lambda j_{\tilde{O}^\lambda}\right]\\
		&\quad+\frac{N^2-1}{2}\log\Det\left[\mathbb{I}+\left(\mathbb{I}-\mathcal{D}^{-1}_\lambda j_{{O}^\lambda}-\mathcal{D}^{-1}_\lambda j_{\tilde{{O}}^\lambda}\right)^{-1}\mathcal{D}_M^{-1}\bar{j}_{M}\left(\mathbb{I}+\mathcal{D}^{-1}_Aj_{O^A}+\mathcal{D}^{-1}_Aj_{\tilde{O}^A}\right)^{-1}\mathcal{D}^{-1}_{\bar{M}}j_{\bar{M}}\right]\\
		&\quad+\frac{N^2-1}{2}\log\Det\left[\mathbb{I}+\left(\mathbb{I}-\mathcal{D}^{-1}_\lambda j_{{O}^\lambda}+\mathcal{D}^{-1}_\lambda j_{\tilde{{O}}^\lambda}\right)^{-1}\mathcal{D}_M^{-1}j_{\bar{M}}\left(\mathbb{I}+\mathcal{D}^{-1}_Aj_{O^A}-\mathcal{D}^{-1}_Aj_{\tilde{O}^A}\right)^{-1}\mathcal{D}^{-1}_{\bar{M}}\bar{j}_{M}\right]
									\end{aligned}$
	} 
	\end{equation}
	and
	\begin{equation}
		\label{genMg2}
									\resizebox{0.98\textwidth}{!}{%
			$\begin{aligned}
		&\Gamma_{\text{conf}}\left[j_{O^{A}},j_{\tilde{O}^{A}},j_{O^{\lambda}},j_{\tilde{O}^{\lambda}},\bar{j}_{M},j_{\bar{M}}\right] \\
		&= 
		-\frac{N^2-1}{2}\log\Det \left[\mathbb{I}+\mathcal{D}^{-1}_Aj_{O^A}+\mathcal{D}^{-1}_Aj_{\tilde{O}^A}\right]-\frac{N^2-1}{2}\log\Det\left[\mathbb{I}+\mathcal{D}^{-1}_Aj_{O^A}-\mathcal{D}^{-1}_Aj_{\tilde{O}^A}\right]\\
		&\quad+\frac{N^2-1}{2}\log\Det \left[\mathbb{I}-\mathcal{D}^{-1}_\lambda j_{O^\lambda}-\mathcal{D}^{-1}_\lambda j_{\tilde{O}^\lambda}\right]+\frac{N^2-1}{2}\log\Det\left[\mathbb{I}-\mathcal{D}^{-1}_\lambda j_{O^\lambda}+\mathcal{D}^{-1}_\lambda j_{\tilde{O}^\lambda}\right]\\
		&\quad-\frac{N^2-1}{2}\log\Det\left[\mathbb{I}+\left(\mathbb{I}+\mathcal{D}^{-1}_Aj_{O^A}+\mathcal{D}^{-1}_Aj_{\tilde{O}^A}\right)^{-1}\mathcal{D}_{\bar{M}}^{-1}j_{\bar{M}}\left(\mathbb{I}-\mathcal{D}^{-1}_\lambda j_{{O}^\lambda}-\mathcal{D}^{-1}_\lambda j_{\tilde{{O}}^\lambda}\right)^{-1}\mathcal{D}_M^{-1}\bar{j}_{M}\right]\\
		&\quad-\frac{N^2-1}{2}\log\Det\left[\mathbb{I}+\left(\mathbb{I}+\mathcal{D}^{-1}_Aj_{O^A}-\mathcal{D}^{-1}_Aj_{\tilde{O}^A}\right)^{-1}\mathcal{D}_{\bar{M}}^{-1}\bar{j}_{M}\left(\mathbb{I}-\mathcal{D}^{-1}_\lambda j_{{O}^\lambda}+\mathcal{D}^{-1}_\lambda j_{\tilde{{O}}^\lambda}\right)^{-1}\mathcal{D}^{-1}_M j_{\bar{M}}\right]\,,
									\end{aligned}$
	} 
	\end{equation}
	where $\mathbb{I}$ is the identity in space-time and discrete indices defined below. The corresponding kernels are defined as follows:
	\begin{itemize}
		\item \underline{gluon-gluon  kernel}
		\subitem 
		\begin{align}
			\label{kernelG}
			&\mathcal{D}^{-1}_{A} =\frac{1}{2}\frac{\Gamma(3)\Gamma(s_1+3)}{\Gamma(5)\Gamma(s_1+1)}{s_1\choose k_1}{s_2\choose k_2+2}(-i\partial_{+})^{s_1-k_1+k_2}i\square^{-1}\nonumber\\ &\rightarrow\mathcal{D}^{-1}_{A\,s_1k_1,s_2k_2}(x-y)=\frac{1}{8\pi^2}\frac{\Gamma(3)\Gamma(s_1+3)}{\Gamma(5)\Gamma(s_1+1)}{s_1\choose k_1}{s_2\choose k_2+2}(-i\partial_{+})^{s_1-k_1+k_2}\frac{1}{\rvert x-y\rvert^2-i\epsilon}
		\end{align}
		\item \underline{gluino-gluino  kernel}
		\subitem
		\begin{align}
			\label{kernelL}
			&\mathcal{D}^{-1}_{\lambda}=\frac{1}{2}\frac{s_1+1}{2}{s_1\choose k_1}{s_2\choose k_2+1}(-i\partial_{+})^{s_1-k_1+k_2-1}(-i\partial_{+})i\square^{-1}\nonumber\\	&\rightarrow\mathcal{D}^{-1}_{\lambda\,s_1k_1,s_2k_2}(x-y)  =\frac{1}{8\pi^2}\frac{s_1+1}{2}{s_1\choose k_1}{s_2\choose k_2+1}(-i\partial_{+})^{s_1-k_1+k_2-1}(-i\partial_{+})\frac{1}{\rvert x-y\rvert^2-i\epsilon}
		\end{align}
		\item \underline{gluon-gluino kernels}
		\begin{align}
			&\mathcal{D}^{-1}_{M}=-\frac{i}{2}{s_1+1\choose k_1+2}{s_2+1\choose k_2}(i\partial_{+})^{s_1-k_1+k_2}(i\partial_{+})i\square^{-1}\nonumber\\ &\rightarrow	\mathcal{D}^{-1}_{M\,s_1k_1,s_2k_2}(x-y) =-\frac{i}{8\pi^2}{s_1+1\choose k_1+2}{s_2+1\choose k_2}(i\partial_{+})^{s_1-k_1+k_2}(i\partial_{+})\frac{1}{\rvert x-y\rvert^2-i\epsilon}\\
			&\vspace{5cm}\nonumber\\	&\mathcal{D}^{-1}_{\bar{M}}
			=-\frac{i}{2}{s_1\choose k_1}{s_2\choose k_2+1}(i\partial_{+})^{s_1-k_1+k_2}i\square^{-1}\nonumber\\ &\rightarrow\mathcal{D}^{-1}_{\bar{M}\,s_1k_1,s_2k_2}(x-y)=-\frac{i}{8\pi^2}{s_1\choose k_1}{s_2\choose k_2+1}(i\partial_{+})^{s_1-k_1+k_2}\frac{1}{\rvert x-y\rvert^2-i\epsilon}\,.
		\end{align}
	\end{itemize}
The kernels are coupled to the currents $j_{\mathcal{O}_{sk}}$ that are dual to the component operators $\mathcal{O}_{sk}$ employed to construct the conformal operators $\mathcal{O}_s$ \cite{BPS1}
\begin{equation}
	\mathcal{O}_s = \sum_{k=0}^{l}\mathcal{O}_{sk}\,.
\end{equation}
Consequently, the $n$-point correlators are defined \cite{BPSpaper2} by
\begin{align}
	\label{gammacorr}
	\langle \mathcal{O}_{s_1}(x_1)\ldots\mathcal{O}_{s_n}(x_n)\rangle=\sum_{k_1 = 0}^{l_1} \frac{\delta}{\delta j_{\mathcal{O}_{s_1k_1}(x_1)}}\cdots\sum_{k_n = 0}^{l_n}\frac{\delta}{\delta j_{\mathcal{O}_{s_nk_n}(x_n)}} {\Gamma_{\text{conf}}}[j_{\mathcal{O}}]
\end{align}
and the following identity holds
\begin{align}
\sum_{k_1 = 0}^{l_1} \frac{\delta}{\delta j_{\mathcal{O}_{s_1k_1}(x_1)}}\cdots\sum_{k_n = 0}^{l_n}\frac{\delta}{\delta j_{\mathcal{O}_{s_nk_n}(x_n)}} {\Gamma_{\text{conf}}}[j_{\mathcal{O}}] = \frac{\delta}{\delta J_{\mathcal{O}_{s_1}(x_1)}}\cdots\frac{\delta}{\delta J_{\mathcal{O}_{s_n}(x_n)}} {\mathcal{W}_{\text{conf}}}[J_{\mathcal{O}}]
\end{align}
according  to the dictionary \cite{BPSpaper2}. Equivalently, according to the above equation, $\Gamma_{\text{conf}}[j_{\mathcal{O}_{sk}}]$ coincides with $\mathcal{W}_{\text{conf}}[J_{\mathcal{O}_{s}}]$ by the identification
\begin{equation} \label{74}
	j_{\mathcal{O}_{sk}} = J_{\mathcal{O}_{s}}
\end{equation}
for every $k$.

\section{Generating functional of Euclidean conformal correlators \label{euclgen}}

\subsection{Analytic continuation to Euclidean space-time}

The Minkowskian correlators can be analytically continued \cite{BPS1} to Euclidean space-time by substituting (appendix \ref{appN})
\begin{equation} \label{ab}
	x_+\rightarrow -i x_{z}
\end{equation}
and
\begin{equation} \label{bc}
	\frac{1}{\rvert x \rvert^2-i\epsilon}\rightarrow-\frac{1}{x^2}\,.
\end{equation}
The analytically continued operators read:

	\begin{itemize}
		\item \underline{gluon-gluon operators}
		\subitem 
		\begin{equation}
			O^A_s \rightarrow (-1)^{s+1}\Tr \partial_z \bar{A}^{E}(\overrightarrow{\partial_z} + \overleftarrow{\partial_z})^{s-2}C^{\frac{5}{2}}_{s-2}\Bigg(\frac{\overrightarrow{\partial_z} - \overleftarrow{\partial_z}}{\overrightarrow{\partial_z}+\overleftarrow{\partial_z}}\Bigg)\partial_z A^{E} = O^{A\,E}_s 
		\end{equation}
		\item \underline{gluino-gluino operators}
		\subitem \begin{equation}
			O^\lambda_s \rightarrow  (-1)^{s-1}\Tr\bar{\lambda}^{E}(\overrightarrow{\partial_z} + \overleftarrow{\partial_z})^{s-1}C^{\frac{3}{2}}_{s-1}\Bigg(\frac{\overrightarrow{\partial_z} - \overleftarrow{\partial_z}}{\overrightarrow{\partial_z}+\overleftarrow{\partial_z}}\Bigg) \lambda^{E}=O^{\lambda\,E}_s
		\end{equation}
		\item \underline{gluon-gluino operators}
		\begin{equation}
			M_s \rightarrow i(-1)^{s-1}\hspace{0.08cm} \partial_z\Tr A^{E} (\overrightarrow{\partial_z} + i\overleftarrow{\partial_z})^{s-1}P^{(2,1)}_{s-1}\Bigg(\frac{\overrightarrow{\partial_z} - \overleftarrow{\partial_z}}{\overrightarrow{\partial_z}+\overleftarrow{\partial_z}}\Bigg) \bar{\lambda}^{E} = M^E_s
		\end{equation}
		\subitem
		\begin{equation}
			\bar{M}_s \rightarrow  \hspace{0.08cm}i(-1)^{s-1}\Tr\lambda^{E}(\overrightarrow{\partial_z} + \overleftarrow{\partial_z})^{s-1}P^{(1,2)}_{s-1}\Bigg(\frac{\overrightarrow{\partial_z} - \overleftarrow{\partial_z}}{\overrightarrow{\partial_z}+\overleftarrow{\partial_z}}\Bigg) \partial_z\bar{A}^{E}=\bar{M}^E_s  \,.
		\end{equation}
	\end{itemize}
	
	\subsection{Analytic continuation of $\mathcal{W}_{\text{conf}}$}	
	Therefore, the generating functional of Euclidean correlators reads
	\begin{equation}
		\label{wexplicit1E}
				\resizebox{0.98\textwidth}{!}{%
			$\begin{aligned}
		&{\mathcal{W}^E_{\text{conf}}}\left[J_{O^{'A\,E}},J_{\tilde{O}^{'A\,E}},J_{O^{'\lambda\,E}},J_{\tilde{O}^{'\lambda\,E}},\bar{J}_{{M^{'E}}},J_{\bar M^{'E}}\right] =\\
		& -\frac{N^2-1}{2}\log\Det\left(I+\frac{1}{N}\sum_{k=0}^{s-2}{s\choose k}{s\choose k+2}(-\overrightarrow{\partial}_z)^{s-k-1}\Laplace^{-1}(J_{O^{'A\,E}_{s}}+J_{\tilde{O}^{'A\,E}_{s}})(-\overrightarrow{\partial}_z)^{k+1} \right)\\
		&-\frac{N^2-1}{2}\log\Det\left(I+\frac{1}{N}\sum_{k=0}^{s-2}{s\choose k}{s\choose k+2}(-\overrightarrow{\partial}_z)^{s-k-1}\Laplace^{-1}(J_{O^{'A\,E}_{s}}-J_{\tilde{O}^{'A\,E}_{s}})(-\overrightarrow{\partial}_z)^{k+1} \right)\\
		&+\frac{N^2-1}{2}\log\Det\left(I-\frac{1}{N}\sum_{k=0}^{s-1}{s\choose k}{s\choose k+1}(-\overrightarrow{\partial}_z)^{s-k-1}(-\overrightarrow{\partial}_z)\Laplace^{-1}(J_{O^{'\lambda\,E}_{s}}+J_{\tilde{O}^{'\lambda\,E}_{s}})(-\overrightarrow{\partial}_z)^{k} \right)\\
		&+\frac{N^2-1}{2}\log\Det\left(I-\frac{1}{N}\sum_{k=0}^{s-1}{s\choose k}{s\choose k+1}(-\overrightarrow{\partial}_z)^{s-k-1}(-\overrightarrow{\partial}_z)\Laplace^{-1}(J_{O^{'\lambda\,E}_{s}}-J_{\tilde{O}^{'\lambda\,E}_{s}})(-\overrightarrow{\partial}_z)^{k} \right)\\
		&+\frac{N^2-1}{2}\log\Det\Bigg[I+\frac{1}{N^2}\left(I-\frac{1}{N}\sum_{k=0}^{s-1}{s\choose k}{s\choose k+1}(-\overrightarrow{\partial}_z)^{s-k-1}(-\overrightarrow{\partial}_z)\Laplace^{-1}(J_{O^{'\lambda\,E}_{s}}-J_{\tilde{O}^{'\lambda\,E}_{s}})(-\overrightarrow{\partial}_z)^{k}\right)^{-1}\\
		&\quad \sum_{k_1 = 0}^{s_1-1}{s_1\choose k_1}{s_1+1\choose k_1+2} (-1)^{s_1-1}
		(-\overrightarrow{\partial}_z)^{s_1-k_1-1}(-\overrightarrow{\partial}_z)\Laplace^{-1}\bar{J}_{{M^{'E}}_{s_1}} (-\overrightarrow{\partial}_z)^{k_1+1}  \\
		&\quad\left(I+\frac{1}{N}\sum_{k_2=0}^{s_2-2}{s_2\choose k_2}{s_2\choose k_2+2}(-\overrightarrow{\partial}_z)^{s_2-k_2-1}\Laplace^{-1}(J_{O^{'A\,E}_{s_2}}+J_{\tilde{O}^{'A\,E}_{s_2}})(-\overrightarrow{\partial}_z)^{k_2+1}\right)^{-1}\\
		&\quad \sum_{k_3 = 0}^{s_3-1}{s_3+1\choose k_3}{s_3\choose k_3+1} (-1)^{s_3-1}
		(-\overrightarrow{\partial}_z)^{s_3-k_3}\Laplace^{-1}J_{{\bar M^{'E}}_{s_3}} (-\overrightarrow{\partial}_z)^{k_3}\Bigg]\\
		&+\frac{N^2-1}{2}\log\Det\Bigg[I+\frac{1}{N^2}\left(I-\frac{1}{N}\sum_{k=0}^{s-1}{s\choose k}{s\choose k+1}(-\overrightarrow{\partial}_z)^{s-k-1}(-\overrightarrow{\partial}_z)\Laplace^{-1}(J_{O^{'\lambda\,E}_{s}}+J_{\tilde{O}^{'\lambda\,E}_{s}})(-\overrightarrow{\partial}_z)^{k}\right)^{-1}\\
		&\quad \sum_{k_1 = 0}^{s_1-1}{s_1\choose k_1}{s_1+1\choose k_1+2} 
		(-\overrightarrow{\partial}_z)^{s_1-k_1-1}(-\overrightarrow{\partial}_z)\Laplace^{-1}J_{{\bar M^{'E}}_{s_1}} (-\overrightarrow{\partial}_z)^{k_1+1}   \\
		&\quad\left(I+\frac{1}{N}\sum_{k_2=0}^{s_2-2}{s_2\choose k_2}{s_2\choose k_2+2}(-\overrightarrow{\partial}_z)^{s_2-k_2-1}\Laplace^{-1}(J_{O^{'A\,E}_{s_2}}-J_{\tilde{O}^{'A\,E}_{s_2}})(-\overrightarrow{\partial}_z)^{k_2+1}\right)^{-1}\\
		&\quad  \sum_{k_3 = 0}^{s_3-1}{s_3+1\choose k_3}{s_3\choose k_3+1} 
		(-\overrightarrow{\partial}_z)^{s_3-k_3}\Laplace^{-1}\bar{J}_{{M^{'E}}_{s_3}} (-\overrightarrow{\partial}_z)^{k_3}\Bigg]
					\end{aligned}$
	} 
	\end{equation}
	and
	\begin{equation}
		\label{wexplicit2E}
				\resizebox{0.98\textwidth}{!}{%
			$\begin{aligned}
		&{\mathcal{W}^E_{\text{conf}}}\left[J_{O^{'A\,E}},J_{\tilde{O}^{'A\,E}},J_{O^{'\lambda\,E}},J_{\tilde{O}^{'\lambda\,E}},\bar{J}_{{M^{'E}}},J_{\bar M^{'E}}\right] =\\
		& -\frac{N^2-1}{2}\log\Det\left(I+\frac{1}{N}\sum_{k=0}^{s-2}{s\choose k}{s\choose k+2}(-\overrightarrow{\partial}_z)^{s-k-1}\Laplace^{-1}(J_{O^{'A\,E}_{s}}+J_{\tilde{O}^{'A\,E}_{s}})(-\overrightarrow{\partial}_z)^{k+1} \right)\\
		&-\frac{N^2-1}{2}\log\Det\left(I+\frac{1}{N}\sum_{k=0}^{s-2}{s\choose k}{s\choose k+2}(-\overrightarrow{\partial}_z)^{s-k-1}\Laplace^{-1}(J_{O^{'A\,E}_{s}}-J_{\tilde{O}^{'A\,E}_{s}})(-\overrightarrow{\partial}_z)^{k+1} \right)\\
		&+\frac{N^2-1}{2}\log\Det\left(I-\frac{1}{N}\sum_{k=0}^{s-1}{s\choose k}{s\choose k+1}(-\overrightarrow{\partial}_z)^{s-k-1}(-\overrightarrow{\partial}_z)\Laplace^{-1}(J_{O^{'\lambda\,E}_{s}}+J_{\tilde{O}^{'\lambda\,E}_{s}})(-\overrightarrow{\partial}_z)^{k} \right)\\
		&+\frac{N^2-1}{2}\log\Det\left(I-\frac{1}{N}\sum_{k=0}^{s-1}{s\choose k}{s\choose k+1}(-\overrightarrow{\partial}_z)^{s-k-1}(-\overrightarrow{\partial}_z)\Laplace^{-1}(J_{O^{'\lambda\,E}_{s}}-J_{\tilde{O}^{'\lambda\,E}_{s}})(-\overrightarrow{\partial}_z)^{k} \right)\\
		&-\frac{N^2-1}{2}\log\Det\Bigg[I+\frac{1}{N^2}\left(I+\frac{1}{N}\sum_{k=0}^{s-2}{s\choose k}{s\choose k+2}(-\overrightarrow{\partial}_z)^{s-k-1}\Laplace^{-1}(J_{O^{'A\,E}_{s}}-J_{\tilde{O}^{'A\,E}_{s}})(-\overrightarrow{\partial}_z)^{k+1}\right)^{-1}\\
		&\quad  \sum_{k_1 = 0}^{s_1-1}{s_1+1\choose k_1}{s_1\choose k_1+1} (-1)^{s_1-1}
		(-\overrightarrow{\partial}_z)^{s_1-k_1}\Laplace^{-1}J_{{\bar M^{'E}}_{s_1}} (-\overrightarrow{\partial}_z)^{k_1}\\
		&\quad\left(I-\frac{1}{N}\sum_{k_2=0}^{s_2-1}{s_2\choose k_2}{s_2\choose k_2+1}(-\overrightarrow{\partial}_z)^{s_2-k_2-1}(-\overrightarrow{\partial}_z)\Laplace^{-1}(J_{O^{'\lambda\,E}_{s_2}}+J_{\tilde{O}^{'\lambda\,E}_{s_2}})(-\overrightarrow{\partial}_z)^{k_2}\right)^{-1}\\
		&\quad \sum_{k_3 = 0}^{s_3-1}{s_3\choose k_3}{s_3+1\choose k_3+2} (-1)^{s_3-1}
		(-\overrightarrow{\partial}_z)^{s_3-k_3-1}(-\overrightarrow{\partial}_z)\Laplace^{-1}\bar{J}_{{M^{'E}}_{s_3}} (-\overrightarrow{\partial}_z)^{k_3+1} \Bigg]\\
		&-\frac{N^2-1}{2}\log\Det\Bigg[I+\frac{1}{N^2}\left(I+\frac{1}{N}\sum_{k=0}^{s-2}{s\choose k}{s\choose k+2}(-\overrightarrow{\partial}_z)^{s-k-1}\Laplace^{-1}(J_{O^{'A\,E}_{s}}+J_{\tilde{O}^{'A\,E}_{s}})(-\overrightarrow{\partial}_z)^{k+1}\right)^{-1}\\
		&\quad \sum_{k_1 = 0}^{s_1-1}{s_1+1\choose k_1}{s_1\choose k_1+1} 
		(-\overrightarrow{\partial}_z)^{s_1-k_1}\Laplace^{-1}\bar{J}_{{M^{'E}}_{s_1}} (-\overrightarrow{\partial}_z)^{k_1} \\
		&\quad\left(I-\frac{1}{N}\sum_{k_2=0}^{s_2-1}{s_2\choose k_2}{s_2\choose k_2+1}(-\overrightarrow{\partial}_z)^{s_2-k_2-1}(-\overrightarrow{\partial}_z)\Laplace^{-1}(J_{O^{'\lambda\,E}_{s_2}}-J_{\tilde{O}^{'\lambda\,E}_{s_2}})(-\overrightarrow{\partial}_z)^{k_2}\right)^{-1}\\
		&\quad \sum_{k_3 = 0}^{s_3-1}{s_3+1\choose k_3}{s_3\choose k_3+1} 
		(-\overrightarrow{\partial}_z)^{s_3-k_3-1}(-\overrightarrow{\partial}_z)\Laplace^{-1}J_{{\bar M^{'E}}_{s_3}} (-\overrightarrow{\partial}_z)^{k_3}\Bigg]\,,
					\end{aligned}$
	} 
	\end{equation}

with
\begin{equation}
	\Laplace= \delta_{\mu\nu}\partial_{\mu}\partial_{\nu}=\partial_4^2+\sum_{i=1}^{3}\partial_i^2
\end{equation}
and
\begin{equation}
	-\Laplace^{-1} \rightarrow \frac{1}{4\pi^2}\frac{1}{(x-y)^2} \,.
\end{equation}

\subsection{Analytic continuation of $\Gamma_{\text{conf}}$}

The Euclidean conformal generating functional is obtained by performing the analytic continuation of Eqs.\eqref{genMg1} and \eqref{genMg2}
	\begin{equation}
		\label{EgenMg1}
						\resizebox{0.98\textwidth}{!}{%
			$\begin{aligned}
		&\Gamma^E_{\text{conf}}\left[j_{O^{A\,E}},j_{\tilde{O}^{A\,E}},j_{O^{\lambda\,E}},j_{\tilde{O}^{\lambda\,E}},\bar{j}_{{M^E}},j_{\bar{M}^E}\right] \\
		&
		=-\frac{N^2-1}{2}\log\Det \left[\mathbb{I}+\mathcal{D}_{E\,A}^{-1}j_{O^{A\,E}}+\mathcal{D}_{E\,A}^{-1}j_{\tilde{O}^{A\,E}}\right]-\frac{N^2-1}{2}\log\Det\left[\mathbb{I}+\mathcal{D}_{E\,A}^{-1}j_{O^{A\,E}}-\mathcal{D}_{E\,A}^{-1}j_{\tilde{O}^{A\,E}}\right]\\
		&\quad+\frac{N^2-1}{2}\log\Det \left[\mathbb{I}-\mathcal{D}_{E\,\lambda}^{-1} j_{O^{\lambda\,E}}-\mathcal{D}_{E\,\lambda}^{-1} j_{\tilde{O}^{\lambda\,E}}\right]+\frac{N^2-1}{2}\log\Det\left[\mathbb{I}-\mathcal{D}_{E\,\lambda}^{-1} j_{O^{\lambda\,E}}+\mathcal{D}_{E\,\lambda}^{-1} j_{\tilde{O}^{\lambda\,E}}\right]\\
		&\quad+\frac{N^2-1}{2}\log\Det\Big[\mathbb{I}+\left(\mathbb{I}-\mathcal{D}_{E\,\lambda}^{-1} j_{{O}^{\lambda\,E}}-\mathcal{D}_{E\,\lambda}^{-1} j_{\tilde{{O}}^{\lambda\,E}}\right)^{-1}\mathcal{D}_{E\,M}^{-1}\bar{j}_{M}\left(\mathbb{I}+\mathcal{D}_{E\,A}^{-1}j_{O^{A\,E}}+\mathcal{D}_{E\,A}^{-1}j_{\tilde{O}^{A\,E}}\right)^{-1}\mathcal{D}_{E\,\bar{M}}^{-1}j_{\bar{M}}\Big]\\
		&\quad+\frac{N^2-1}{2}\log\Det\Big[\mathbb{I}+\left(\mathbb{I}-\mathcal{D}_{E\,\lambda}^{-1} j_{{O}^{\lambda\,E}}+\mathcal{D}_{E\,\lambda}^{-1} j_{\tilde{{O}}^{\lambda\,E}}\right)^{-1}\mathcal{D}_{E\,M}^{-1}j_{\bar{M}}\left(\mathbb{I}+\mathcal{D}_{E\,A}^{-1}j_{O^{A\,E}}-\mathcal{D}_{E\,A}^{-1}j_{\tilde{O}^{A\,E}}\right)^{-1}\mathcal{D}_{E\,\bar{M}}^{-1}\bar{j}_{M}\Big]
							\end{aligned}$
	} 
	\end{equation}
	and
	\begin{equation}
		\label{EgenMg2}
						\resizebox{0.98\textwidth}{!}{%
			$\begin{aligned}
		&\Gamma^E_{\text{conf}}\left[j_{O^{A\,E}},j_{\tilde{O}^{A\,E}},j_{O^{\lambda\,E}},j_{\tilde{O}^{\lambda\,E}},\bar{j}_{{M^E}},j_{\bar{M}^E}\right] \\
		&= 
		-\frac{N^2-1}{2}\log\Det \left[\mathbb{I}+\mathcal{D}_{E\,A}^{-1}j_{O^{A\,E}}+\mathcal{D}_{E\,A}^{-1}j_{\tilde{O}^{A\,E}}\right]-\frac{N^2-1}{2}\log\Det\left[\mathbb{I}+\mathcal{D}_{E\,A}^{-1}j_{O^{A\,E}}-\mathcal{D}_{E\,A}^{-1}j_{\tilde{O}^{A\,E}}\right]\\
		&\quad+\frac{N^2-1}{2}\log\Det \left[\mathbb{I}-\mathcal{D}_{E\,\lambda}^{-1} j_{O^{\lambda\,E}}-\mathcal{D}_{E\,\lambda}^{-1} j_{\tilde{O}^{\lambda\,E}}\right]+\frac{N^2-1}{2}\log\Det\left[\mathbb{I}-\mathcal{D}_{E\,\lambda}^{-1} j_{O^{\lambda\,E}}+\mathcal{D}_{E\,\lambda}^{-1} j_{\tilde{O}^{\lambda\,E}}\right]\\
		&\quad-\frac{N^2-1}{2}\log\Det\Big[\mathbb{I}+\left(\mathbb{I}+\mathcal{D}_{E\,A}^{-1}j_{O^{A\,E}}+\mathcal{D}_{E\,A}^{-1}j_{\tilde{O}^{A\,E}}\right)^{-1}\mathcal{D}_{E\,\bar{M}}^{-1}j_{\bar{M}}\left(\mathbb{I}-\mathcal{D}_{E\,\lambda}^{-1} j_{{O}^{\lambda\,E}}-\mathcal{D}_{E\,\lambda}^{-1} j_{\tilde{{O}}^{\lambda\,E}}\right)^{-1}\mathcal{D}^{-1}_{{M^E}}\bar{j}_{M}\Big]\\
		&\quad-\frac{N^2-1}{2}\log\Det\Big[\mathbb{I}+\left(\mathbb{I}+\mathcal{D}_{E\,A}^{-1}j_{O^{A\,E}}-\mathcal{D}_{E\,A}^{-1}j_{\tilde{O}^{A\,E}}\right)^{-1}\mathcal{D}_{E\,\bar{M}}^{-1}\bar{j}_{M}\left(\mathbb{I}-\mathcal{D}_{E\,\lambda}^{-1} j_{{O}^{\lambda\,E}}+\mathcal{D}_{E\,\lambda}^{-1} j_{\tilde{{O}}^{\lambda\,E}}\right)^{-1}\mathcal{D}_{E\,M}^{-1} j_{\bar{M}}\Big],
							\end{aligned}$
	} 
	\end{equation}
	with the kernels: 
	\begin{itemize}
		\item \underline{gluon-gluon kernel}
		\subitem 
		\begin{align}
			\label{EkernelG}
			&	\mathcal{D}^{-1}_{E\,A} =\frac{1}{2}\frac{\Gamma(3)\Gamma(s_1+3)}{\Gamma(5)\Gamma(s_1+1)}{s_1\choose k_1}{s_2\choose k_2+2}\partial_{z}^{s_1-k_1+k_2}\Laplace^{-1}\nonumber\\ &\rightarrow \mathcal{D}^{-1}_{E\,A\,s_1k_1,s_2k_2}(x-y)=-\frac{1}{8\pi^2}\frac{\Gamma(3)\Gamma(s_1+3)}{\Gamma(5)\Gamma(s_1+1)}{s_1\choose k_1}{s_2\choose k_2+2}\partial_{z}^{s_1-k_1+k_2}\frac{1}{ (x-y)^2}
		\end{align}
		\item \underline{gluino-gluino kernel}
		\subitem
		\begin{align}
			\label{EkernelL}
			&\mathcal{D}^{-1}_{E\,\lambda} =\frac{1}{2}\frac{s_1+1}{2}{s_1\choose k_1}{s_2\choose k_2+1}\partial_{z}^{s_1-k_1+k_2-1}\partial_{z}\Laplace^{-1}\nonumber\\ &\rightarrow\mathcal{D}^{-1}_{E\,\lambda\,s_1k_1,s_2k_2}(x-y)=-\frac{1}{8\pi^2}\frac{s_1+1}{2}{s_1\choose k_1}{s_2\choose k_2+1}\partial_{z}^{s_1-k_1+k_2-1}\partial_{z}\frac{1}{ (x-y)^2}
		\end{align}
		\item \underline{gluon-gluino kernels}
		\begin{align}
			&\mathcal{D}^{-1}_{E\,M}=\frac{i}{2}{s_1+1\choose k_1+2}{s_2+1\choose k_2}(-\partial_{z})^{s_1-k_1+k_2}\partial_{z}\Laplace^{-1}\nonumber\\ &\rightarrow\mathcal{D}^{-1}_{E\,M\,s_1k_1,s_2k_2}(x-y) =-\frac{i}{8\pi^2}{s_1+1\choose k_1+2}{s_2+1\choose k_2}(-\partial_{z})^{s_1-k_1+k_2}\partial_{z}\frac{1}{ (x-y)^2}\\
			&\vspace{5cm}\nonumber\\
			&\mathcal{D}^{-1}_{E\,\bar{M}}=-\frac{i}{2}{s_1\choose k_1}{s_2\choose k_2+1}(-\partial_{z})^{s_1-k_1+k_2}\Laplace^{-1}\nonumber\\ &\rightarrow\mathcal{D}^{-1}_{E\,\bar{M}\,s_1k_1,s_2k_2}(x-y)=\frac{i}{8\pi^2}{s_1\choose k_1}{s_2\choose k_2+1}(-\partial_{z})^{s_1-k_1+k_2}\frac{1}{(x-y)^2}\,.
		\end{align}
	\end{itemize}

\section{RG-improvement of the Euclidean correlators}  \label{s0}
\subsection{Operator mixing}
\label{SectMix}

We briefly summarize the construction of the RG-improved asymptotic generating functional following \cite{BPSpaper2,MB1}.
The renormalized Euclidean correlators
\begin{equation}\label{key}
	\langle \mathcal{O}_{k_1}(x_1)\ldots \mathcal{O}_{k_n}(x_n) \rangle = G^{(n)}_{k_1 \ldots k_n}( x_1,\ldots,  x_n; \mu, g(\mu))
\end{equation}
satisfy the Callan-Symanzik equation
\begin{align}\label{CallanSymanzik}
	& \Big(\sum_{\alpha = 1}^n x_\alpha \cdot \frac{\partial}{\partial x_\alpha} + \beta(g)\frac{\partial}{\partial g} + \sum_{\alpha = 1}^n D_{\mathcal{O}_\alpha}\Big)G^{(n)}_{k_1 \ldots k_n} + \nonumber\\
	& \sum_a \Big(\gamma_{k_1a}(g) G^{(n)}_{ak_2 \ldots k_n} + \gamma_{k_2a}(g) G^{(n)}_{k_1 a k_3 \ldots k_n} \cdots +\gamma_{k_n a}(g) G^{(n)}_{k_1 \ldots a}\Big) = 0\,,
\end{align}
with solution
\begin{align}\label{csformula}
	&G^{(n)}_{k_1 \ldots k_n}(\lambda x_1,\ldots, \lambda x_n; \mu, g(\mu)) \nonumber \\
	&= \sum_{j_1 \ldots j_n} Z_{k_1 j_1} (\lambda)\ldots Z_{k_n j_n}(\lambda)\hspace{0.1cm} \lambda^{-\sum_{i=1}^nD_{\mathcal{O}_i}} G^{(n)}_{j_1 \ldots j_n }( x_1, \ldots, x_n; \mu, g(\frac{\mu}{\lambda}))\,,
\end{align}
where $D_{\mathcal{O}_i}$ is the canonical dimension of $\mathcal{O}_i(x)$ and $\gamma(g)=\gamma_0 g^2+ \cdots$ the matrix of the anomalous dimensions, with
\begin{equation}\label{eqZ}
	\Bigg(\frac{\partial}{\partial g} + \frac{\gamma(g)}{\beta(g)}\Bigg)Z(\lambda) = 0
\end{equation}
in matrix notation, and
\begin{equation} \label{ZZ}
	Z(\lambda) = P\exp \Big(\int_{g(\mu)}^{g(\frac{\mu}{\lambda})}\frac{\gamma(g')}{\beta(g')} dg'\Big)\,.
\end{equation}
Eq. \eqref{csformula} greatly simplifies if a renormalization scheme exists where $Z(\lambda)$ is diagonalizable to all orders of perturbation theory, and specifically one-loop exact, with eigenvalues $Z_{\mathcal{O}_i}(\lambda)$ \cite{MB1}
\begin{equation} \label{zz}
	Z_{\mathcal{O}_i}(\lambda) = \Bigg(\frac{g(\mu)}{g(\frac{\mu}{\lambda})}\Bigg)^{\frac{\gamma_{0\mathcal{O}_i}}{\beta_0}} \,,
\end{equation}
where $\gamma_{0\mathcal{O}_i}$ are the eigenvalues of $\gamma_0$. Indeed, in the above scheme Eq. \eqref{csformula} contains only one term 
\begin{align}\label{csformuladiag}
	G^{(n)}_{j_1 \ldots j_n}(\lambda x_1,\ldots, \lambda x_n; \mu, g(\mu))=Z_{\mathcal{O}_{j_1}}(\lambda) \ldots Z_{\mathcal{O}_{j_n}}(\lambda) \hspace{0.1cm} \lambda^{-\sum_{i=1}^nD_{\mathcal{O}_i}}G^{(n)}_{j_1 \ldots j_n }( x_1, \ldots, x_n; \mu, g(\frac{\mu}{\lambda}))\,.
\end{align}
Then, as $\lambda \rightarrow 0$, in any renormalization scheme, the renormalized correlator in the right-hand side above admits the perturbative asymptotic expansion in terms of the renormalized coupling $g(\frac{\mu}{\lambda})$ at the scale $\frac{\mu}{\lambda}$
\begin{align} \label{eq:expansion}
	G^{(n)}_{j_1 \ldots j_n }( x_1, \ldots, x_n; \mu, g(\frac{\mu}{\lambda}))\, \sim\, &{\mathcal G}^{(n,0)}_{j_1 \ldots j_n }( x_1, \ldots, x_n; \mu)+g^2(\frac{\mu}{\lambda}) \, {\mathcal G}^{(n,2)}_{j_1 \ldots j_n }( x_1, \ldots, x_n; \mu)\nonumber\\
	&+ g^4(\frac{\mu}{\lambda}) \, {\mathcal G}^{(n,4)}_{j_1 \ldots j_n }( x_1, \ldots, x_n; \mu)
	+\cdots\,.
\end{align}
Of course, the first term in the above expansion, being independent of the coupling, is the conformal contribution at zero coupling
\begin{equation}
	{\mathcal G}^{(n,0)}_{j_1 \ldots j_n }( x_1, \ldots, x_n; \mu) = G^{(n)}_{\text{conf} \, j_1 \ldots j_n }( x_1, \ldots, x_n) \, ,
\end{equation}
since the renormalized operators at zero coupling coincide with the conformal ones. \par
The higher-order corrections in Eq. \eqref{eq:expansion} arise from the nonconformal interaction due to the nonvanishing beta function, so that the conformal contribution is corrected at higher orders in the renormalized coupling as displayed, the renormalized operators being nonconformal at higher orders in any renormalization scheme.
\par
Yet, provided that the conformal contribution is nonvanishing, for fixed $x_1, \ldots, x_n$, all the higher-order terms in Eq. \eqref{eq:expansion} are suppressed with respect to the conformal one by powers of 
\begin{align} 
	g^2(\frac{\mu}{\lambda}) &\sim \dfrac{1}{\beta_0 \log(\frac{\mu^2}{\lambda^2\Lambda_{SYM}^2})}\left(1-\frac{\beta_1}{\beta_0^2}\frac{\log\log(\frac{\mu^2}{\lambda^2\Lambda_{SYM}^2})}{\log(\frac{\mu^2}{\lambda^2\Lambda_{SYM}^2})}\right)\nonumber\\
	&\sim \dfrac{1}{\beta_0 \log(\frac{1}{\lambda^2})}\left(1-\frac{\beta_1}{\beta_0^2}\frac{\log\log(\frac{1}{\lambda^2})}{\log(\frac{1}{\lambda^2})}\right)\,, \nonumber \\
\end{align}
-- i.e., asymptotically by inverse powers of $\log \lambda$ -- despite being in general nonconformal.
\par
Hence, the corresponding UV asymptotics in Eq. \eqref{csformuladiag}, with fixed $x_1, \ldots, x_n$, reads as $\lambda \rightarrow 0$
\begin{align} \label{eqrg}
	\langle \mathcal{O}_{j_1}(\lambda x_1)\ldots\mathcal{O}_{j_n}(\lambda x_n)\rangle\sim \frac{Z_{\mathcal{O}_{j_1}}(\lambda)\ldots Z_{\mathcal{O}_{j_n}}(\lambda)}{\lambda^{D_{\mathcal{O}_1}+\cdots+D_{\mathcal{O}_n}}} G^{(n)}_{\text{conf}\,j_1 \ldots j_n }( x_1, \ldots, x_n)
\end{align}
that is the RG-improved asymptotic correlator in the aforementioned renormalization scheme. \par
We refer to the above scheme as nonresonant diagonal \cite{MB1}, according to the Poincar\'e-Dulac theorem that is involved in the following differential-geometric interpretation of operator mixing \cite{MB1}.
We interpret a finite change of basis of the renormalized operators
\begin{equation}\label{linearcomb}
	\mathcal{O}'(x) = S(g) \mathcal{O}(x)
\end{equation}
in matrix notation as a real-analytic invertible gauge transformation $S(g)$ that depends on $g$. Then, the matrix
\begin{equation}
	A(g) = -\frac{\gamma(g)}{\beta(g)} = \frac{1}{g} \Big(\frac{\gamma_0}{\beta_0} + \cdots\Big)
\end{equation}
that enters the differential equation for $Z(\lambda)$
\begin{equation}
	\Big(\frac{\partial}{\partial g} - A(g)\Big) Z(\lambda) = 0
\end{equation}
defines a connection $A(g)$
\begin{eqnarray} \label{sys2}
	A(g)= \frac{1}{g} \left(A_0 + \sum^{\infty}_ {n=1} A_{2n} g^{2n} \right)\,,
\end{eqnarray}
with a regular singularity at $g = 0$ that transforms as
\begin{equation}
	A'(g) = S(g)A(g)S^{-1}(g) + \frac{\partial S(g)}{\partial g} S^{-1}(g)
\end{equation}
under the gauge transformation $S(g)$, with
\begin{equation}
	\mathcal{D} = \frac{\partial }{\partial g} - A(g)
\end{equation}
the corresponding covariant derivative.
Consequently, $Z(\lambda)$ is a Wilson line that transforms as
\begin{equation}
	Z'(\lambda) = S(g(\mu))Z(\lambda)S^{-1}(g(\frac{\mu}{\lambda}))\,.
\end{equation}
It follows from the Poincar\'e-Dulac theorem~\cite{MB1} that, if any two eigenvalues $\lambda_1, \lambda_2, \ldots$ of the matrix $\frac{\gamma_0}{\beta_0}$, in nonincreasing order $\lambda_1 \geq \lambda_2 \geq \ldots$, do not differ by a positive even integer
\begin{equation} \label{nr}
	\lambda_i - \lambda_j - 2k \neq 0
\end{equation}
for $i \leq j$ and $k$ a positive integer -- i.e., they are nonresonant -- then a gauge transformation exists that sets $A(g)$ in the canonical nonresonant form \cite{MB1}
\begin{equation} \label{1loop}
	A'(g) = \frac{\gamma_0}{\beta_0}\frac{1}{g}
\end{equation}
that is one-loop exact to all orders of perturbation theory. Hence, if in addition $\frac{\gamma_0}{\beta_0}$ is diagonalizable, Eq. \eqref{zz} follows.

\subsection{Nonresonant diagonal renormalization scheme \label{NonRes}}

To make the present paper self-contained we provide the construction order by order in perturbation theory of the nonresonant diagonal scheme \cite{MB1}.\par
The construction proceeds by induction on $k=1,2, \cdots$ by demonstrating that, once $A_0$ and the first $k-1$ matrix coefficients $A_2,\cdots,A_{2(k-1)}$ in Eq. \eqref{sys2} have been set in the canonical nonresonant form in Eq. \eqref{1loop} -- i.e., $A_0$ diagonal and $ A_2,\cdots,A_{2(k-1)}=0$ -- a real-analytic gauge transformation exists that leaves them invariant and sets the $k$-th coefficient $A_{2k}$ to $0$ as well. \par
The $0$ step of the induction consists in putting $A_0$ in diagonal form -- with the eigenvalues in nonincreasing order -- by a constant gauge transformation. \par
At the $k$-th step we choose the gauge transformation
\begin{eqnarray}
	S_k(g)=1+ g^{2k} H_{2k}\,,
\end{eqnarray}
with $H_{2k}$ a matrix to be found below. Its inverse reads
\begin{equation}
	S^{-1}_k(g)= (1+ g^{2k} H_{2k})^{-1} = 1- g^{2k} H_{2k} + \cdots\,,
\end{equation}
where the dots represent terms of order higher than $g^{2k}$.
The action of $S_k(g)$ on the connection $A(g)$ furnishes

	\begin{eqnarray} \label{ind}
		A'(g) &= & 2k g^{2k-1} H_{2k} ( 1- g^{2k} H_{2k})^{-1}+  (1+ g^{2k} H_{2k}) A(g)( 1- g^{2k} H_{2k})^{-1} \nonumber \\
		&= & 2k g^{2k-1} H_{2k} ( 1- g^{2k} H_{2k})^{-1} +  (1+ g^{2k} H_{2k})  \frac{1}{g} \left(A_0 + \sum^{\infty}_ {n=1} A_{2n} g^{2n} \right)  ( 1- g^{2k} H_{2k})^{-1} \nonumber \\
		&= & 2k g^{2k-1} H_{2k} ( 1- \cdots) +  (1+ g^{2k} H_{2k})  \frac{1}{g} \left(A_0 + \sum^{\infty}_ {n=1} A_{2n} g^{2n} \right)  ( 1- g^{2k} H_{2k}+\cdots) \nonumber \\
		&= & 2k g^{2k-1} H_{2k}  +    \frac{1}{g} \left(A_0 + \sum^k_ {n=1} A_{2n} g^{2n} \right) + g^{2k-1} (H_{2k}A_0-A_0H_{2k}) + \cdots\nonumber \\
		&= &  g^{2k-1} (2k H_{2k} + H_{2k} A_0 - A_0 H_{2k})  + A_{2(k-1)}(g) + g^{2k-1} A_{2k}+ \cdots\,,\nonumber \\
	\end{eqnarray}

where we have skipped all the terms that contribute to an order higher than $g^{2k-1}$, with
\begin{align}
	A_{2(k-1)}(g) =  \frac{1}{g} \left(A_0 + \sum^{k-1}_ {n=1} A_{2n} g^{2n} \right)
\end{align}
that is the part of $A(g)$ that is not affected by the gauge transformation $S_k(g)$, and therefore verifies the hypotheses of the induction --  i.e., that $A_2, \cdots, A_{2(k-1)}$ vanish. \par
Thus, by Eq. \eqref{ind} the $k$-th matrix coefficient $A_{2k}$ may be eliminated from the expansion of $A'(g)$ to order $g^{2k-1}$ provided that an $H_{2k}$ exists such that
\begin{align}
	A_{2k}+(2k H_{2k} + H_{2k} A_0 - A_0 H_{2k})= A_{2k}+ (2k-ad_{A_0}) H_{2k}=0\,,
\end{align}
with $ad_{A_0}Y=[A_0,Y]$.
If the inverse of $ad_{A_0}-2k$ exists, the unique solution for $H_{2k}$ is
\begin{eqnarray}
	H_{2k}=(ad_{A_0}-2k)^{-1} A_{2k}\,.
\end{eqnarray}
Hence, to complete the induction, we should demonstrate that, if the eigenvalues of $A_0$ are nonresonant, $ad_{A_0}-2k$ is invertible, i.e., its kernel is trivial. \par
Now $ad_{\Lambda}-2k$, as a linear operator that acts on matrices, is diagonal, with eigenvalues $\lambda_{i}-\lambda_{j}-2k$ and the matrices $E_{ij}$, whose only nonvanishing entries are $(E_{ij})_{ij}$, as eigenvectors. The eigenvectors $E_{ij}$, normalized so that $(E_{ij})_{ij}=1$, form an orthonormal basis for the matrices. Therefore, $E_{ij}$ belongs to the kernel of $ad_{\Lambda}-2k$ if and only if $\lambda_{i}-\lambda_{j}-2k=0$. Consequently, since  $\lambda_{i}-\lambda_{j}-2k \neq 0$ for every $i,j$ by assumption, the kernel of $ad_{\Lambda}-2k$ only contains $0$, and the construction is complete.

\subsection{Anomalous dimensions of twist-$2$ operators 
	in $\mathcal{N} = 1$  SUSY YM theory }

The eigenvalues of $\gamma_0$ for $\mathcal{O}_s = S^{(1)}_s,S^{(2)}_s,\tilde{S}^{(1)}_s,\tilde{S}^{(2)}_s,M_s,\bar{M}_s$ read
\cite{Belitsky:2004sc}
\begin{equation}
	\gamma_{0\,\mathcal{O}_s} = \frac{1}{4 \pi^2} \Big( \tilde{\gamma}^0_{\mathcal{O}_s} - \frac{3}{2}\Big)\,
\end{equation}
with
\begin{align}
	&\tilde{\gamma}_{0\,S^{(1)}_s} = \psi(s + 2)+\psi(s-1) - 2\psi(1) - \frac{2(-1)^s}{(s+1)s(s-1)} \nonumber\\
	&\tilde{\gamma}_{0\,S^{(2)}_s} = \psi(s + 3)+\psi(s) - 2\psi(1) + \frac{2(-1)^s}{(s+2)(s+1)s}
\end{align}
and ~\cite{Belitsky:2004sc,Belitsky:2003sh}
\begin{align}
	&\tilde{\gamma}_{0\,{\tilde{S}}^{(i)}_s}=\tilde{\gamma}_{0\,S^{(i)}_s}\nonumber\\
	&\tilde{\gamma}_{0\,M_s} =\tilde{\gamma}_{0\,S^{(2)}_s} 
\end{align}
Besides \cite{Belitsky:2004sc},
\begin{equation}
	\gamma_{0\,\tilde{O}^\lambda_1} = \frac{1}{4 \pi^2}  \frac{2}{3} = \frac{1 }{6 \pi^2} \,.
\end{equation}
We have verified that the first $10^4$ eigenvalues of $\frac{\gamma_0}{\beta_0}$ are nonresonant, with $\beta_0 =\frac{3}{(4\pi)^2}$. Moreover, the proof of the nonresonant condition \cite{S1} in the pure YM case applies with minor modifications to the balanced twist-$2$ operators in the present paper.

\subsection{Conformal generating functional of correlators of supermultiplet operators}

By noticing that
\begin{align}
	&S^{(1)'\,E}_sJ_{S^{(1)'\,E}_s}+S^{(2)'\,E}_sJ_{S^{(2)'\,E}_s} \nonumber\\
	&= 6O^{'A\,E}_s\left(\frac{J_{S^{(1)'\,E}_s}}{s-1}+\frac{J_{S^{(2)'\,E}_s}}{s+2}\right)+O^{'\lambda\,E}_s\left(-J_{S^{(1)'\,E}_s}+J_{S^{(2)'\,E}_s}\right)
\end{align}
and
\begin{align}
	&\tilde{S}^{(1)'\,E}_sJ_{\tilde{S}^{(1)'\,E}_s}+\tilde{S}^{(2)'\,E}_sJ_{\tilde{S}^{(2)'\,E}_s} \nonumber\\
	&=- 6{\tilde{O}}^{'A\,E}_s\left(\frac{J_{\tilde{S}^{(1)'\,E}_s}}{s-1}+\frac{J_{\tilde{S}^{(2)'\,E}_s}}{s+2}\right)+{\tilde{O}}^{'\lambda\,E}_s\left(-J_{\tilde{S}^{(1)'\,E}_s}+J_{\tilde{S}^{(2)'\,E}_s}\right)\,,
\end{align}
we get
\begin{align}
	&J_{O^{'A\,E}_s}\,= \, 6\left(\frac{J_{S^{(1)'\,E}_s}}{s-1}+\frac{J_{S^{(2)'\,E}_s}}{s+2}\right)\nonumber\\
	&J_{\tilde{O}^{'A\,E}_s}\,= \, -6\left(\frac{J_{\tilde{S}^{(1)'\,E}_s}}{s-1}+\frac{J_{\tilde{S}^{(2)'\,E}_s}}{s+2}\right)
\end{align}
and
\begin{align}
	&J_{O^{'\lambda\,E}_s}\, = \,-J_{S^{(1)'\,E}_s}+J_{S^{(2)'\,E}_s}\nonumber\\
	&J_{\tilde{O}^{'\lambda\,E}_s}\, = \,-J_{\tilde{S}^{(1)'\,E}_s}+J_{\tilde{S}^{(2)'\,E}_s}\,.
\end{align}
Substituting the above formulas into Eqs. (\ref{wexplicit1E}) and (\ref{wexplicit2E}), we obtain the generating functional of Euclidean conformal correlators of supermultiplet operators.\par We display some important special cases, the complete expression being reported in Appendix \ref{rggen}.\par
For bosonic operators

	\begin{align} \label{122}
		&{\mathcal{W}^E_{\text{conf}}}\left[0,J_{S^{(1)'\,E}},0,0,0,0,0\right] \nonumber \\
		&= -(N^2-1)\log\Det\Bigg(I+\frac{1}{N}6\sum_{k=0}^{s-2}{s\choose k}{s\choose k+2}(-\overrightarrow{\partial}_z)^{s-k-1}\Laplace^{-1}\frac{J_{S^{(1)'\,E}_s}}{s-1}(-\overrightarrow{\partial}_z)^{k+1} \Bigg)\nonumber\\
		&\quad +(N^2-1)\log\Det\Bigg(I+\frac{1}{N}\sum_{k=0}^{s-1}{s\choose k}{s\choose k+1}(-\overrightarrow{\partial}_z)^{s-k-1}(-\overrightarrow{\partial}_z)\Laplace^{-1}J_{S^{(1)'\,E}_s}(-\overrightarrow{\partial}_z)^{k} \Bigg)
	\end{align}
	and
	\begin{align} \label{123}
		&{\mathcal{W}^E_{\text{conf}}}\left[0 ,0,0,J_{S^{(2)'\,E}},0,0,0\right] \nonumber \\
		&= -(N^2-1)\log\Det\Bigg(I+\frac{1}{N}6\sum_{k=0}^{s-2}{s\choose k}{s\choose k+2}(-\overrightarrow{\partial}_z)^{s-k-1}\Laplace^{-1}\frac{J_{S^{(2)'\,E}_s}}{s+2}(-\overrightarrow{\partial}_z)^{k+1} \Bigg)\nonumber\\
		&\quad +(N^2-1)\log\Det\Bigg(I-\frac{1}{N}\sum_{k=0}^{s-1}{s\choose k}{s\choose k+1}(-\overrightarrow{\partial}_z)^{s-k-1}(-\overrightarrow{\partial}_z)\Laplace^{-1}J_{S^{(2)'\,E}_s}(-\overrightarrow{\partial}_z)^{k} \Bigg) \ .
	\end{align}
	For fermionic operators
	\begin{align} \label{124}
		&{\mathcal{W}^E_{\text{conf}}}\left[0 ,0,0,0,0,\bar{J}_{{M^{'E}}},J_{\bar M^{'E}}\right] \nonumber\\
		&= +(N^2-1)\log\Det\Bigg(I+\frac{1}{N^2}
		\sum_{k_1 = 0}^{s_1-1}{s_1\choose k_1}{s_1+1\choose k_1+2} 
		(-\overrightarrow{\partial}_z)^{s_1-k_1-1}(-\overrightarrow{\partial}_z)\Laplace^{-1}J_{ \bar{M}^{'E}_{s_1}}(-\overrightarrow{\partial}_z)^{k_1+1}   \nonumber\\
		&\quad   \sum_{k_3 = 0}^{s_3-1}{s_3+1\choose k_3}{s_3\choose k_3+1} 
		(-\overrightarrow{\partial}_z)^{s_3-k_3}\Laplace^{-1}\bar{J}_{M^{'E}_{s_3}} (-\overrightarrow{\partial}_z)^{k_3}\Bigg)
	\end{align}

	\section{RG-improved generating functionals} \label{IX}
	
	\subsection{$\mathcal{W}$}
	
	In 't Hooft large-$N$ expansion the generating functional of the Euclidean $n$-point correlators decomposes into its planar $\mathcal{W}^E_{\text{sphere}}$ and leading-order nonplanar $\mathcal{W}^E_{\text{torus}}$ contributions \cite{BPSpaper2,BPSL,QCD24}. The UV asymptotics as $\lambda \rightarrow 0$ of $\mathcal{W}^E_{\text{sphere}}$ and $\mathcal{W}^E_{\text{torus}}$ follows from the RG-improved correlators in Eq. \eqref{eqrg} and from the conformal generating functionals in Eqs. \eqref{122}, \eqref{123} and \eqref{124}, with
	\begin{equation}
		\mathcal{W}^E_{\text{asym}}[J_{\mathcal{O'}^E},\lambda]= \mathcal{W}^E_{\text{asym sphere}}[J_{\mathcal{O'}^E},\lambda]+ \mathcal{W}^E_{\text{asym torus}}[J_{\mathcal{O'}^E},\lambda] 
	\end{equation}
	and
	\begin{equation}
		\mathcal{W}^E_{\text{asym sphere}}[J_{\mathcal{O'}^E},\lambda]=-N^2 \mathcal{W}^E_{\text{asym torus}}[J_{\mathcal{O'}^E},\lambda]\,.
	\end{equation}
	For bosonic operators
	\begin{equation}
				\resizebox{0.98\textwidth}{!}{%
			$\begin{aligned}
		&{\mathcal{W}^E_{\text{asym}}}\left[0,J_{S^{(1)'\,E}},0,0,0,0,0,\lambda\right] =\\
		& -(N^2-1)\log\Det\Bigg(I+\frac{1}{N}6\sum_{k=0}^{s-2}{s\choose k}{s\choose k+2}(-\overrightarrow{\partial}_z)^{s-k-1}\frac{\Laplace^{-1}}{\lambda^{s+2}}Z_{S^{(1)'\,E}}(\lambda)\frac{J_{S^{(1)'\,E}_s}}{s-1}(-\overrightarrow{\partial}_z)^{k+1} \Bigg)\\
		&+(N^2-1)\log\Det\Bigg(I+\frac{1}{N}\sum_{k=0}^{s-1}{s\choose k}{s\choose k+1}(-\overrightarrow{\partial}_z)^{s-k-1}\frac{(-\overrightarrow{\partial}_z)\Laplace^{-1}}{\lambda^{s+2}}Z_{S^{(1)'\,E}}(\lambda)J_{S^{(1)'\,E}_s}(-\overrightarrow{\partial}_z)^{k} \Bigg)
					\end{aligned}$
	} 
	\end{equation}
	and
	\begin{equation}
				\resizebox{0.98\textwidth}{!}{%
			$\begin{aligned}
		&{\mathcal{W}^E_{\text{asym}}}\left[0 ,0,0,J_{S^{(2)'\,E}},0,0,0,\lambda\right] =\\
		& -(N^2-1)\log\Det\Bigg(I+\frac{1}{N}6\sum_{k=0}^{s-2}{s\choose k}{s\choose k+2}(-\overrightarrow{\partial}_z)^{s-k-1}\frac{\Laplace^{-1}}{\lambda^{s+2}}Z_{S^{(2)'\,E}}(\lambda)\frac{J_{S^{(2)'\,E}_s}}{s+2}(-\overrightarrow{\partial}_z)^{k+1} \Bigg)\\
		&+(N^2-1)\log\Det\Bigg(I-\frac{1}{N}\sum_{k=0}^{s-1}{s\choose k}{s\choose k+1}(-\overrightarrow{\partial}_z)^{s-k-1}\frac{(-\overrightarrow{\partial}_z)\Laplace^{-1}}{\lambda^{s+2}}Z_{S^{(2)'\,E}}(\lambda)J_{S^{(2)'\,E}_s}(-\overrightarrow{\partial}_z)^{k} \Bigg) \,.
					\end{aligned}$
	} 
	\end{equation}
	For fermionic operators
	\begin{equation}
				\resizebox{0.98\textwidth}{!}{%
			$\begin{aligned}
		&{\mathcal{W}^E_{\text{asym}}}\left[0 ,0,0,0,0,\bar{J}_{{M^{'E}}},J_{\bar M^{'E}},\lambda\right] =\\
		&+(N^2-1)\log\Det\Bigg(I+\frac{1}{N^2}
		\sum_{k_1 = 0}^{s_1-1}{s_1\choose k_1}{s_1+1\choose k_1+2} 
		(-\overrightarrow{\partial}_z)^{s_1-k_1-1}\frac{(-\overrightarrow{\partial}_z)\Laplace^{-1}}{\lambda^{s_1+2}}Z_{M}(\lambda)J_{\bar{M}^{'E}_{s_1}}(-\overrightarrow{\partial}_z)^{k_1+1}  \\
		& \sum_{k_2 = 0}^{s_2-1}{s_2+1\choose k_2}{s_2\choose k_2+1} 
		(-\overrightarrow{\partial}_z)^{s_2-k_2}\frac{\Laplace^{-1}}{\lambda^{s_2+2}}Z_{M}(\lambda)\bar{J}_{ M^{'E}_{s_2}} (-\overrightarrow{\partial}_z)^{k_2}\Bigg) \,.
					\end{aligned}$
	} 
	\end{equation}
	
	\subsection{$\Gamma$}\label{gamma}
	
	We also provide a more compact formula for the asymptotic RG-improved generating functionals expressed as Fredholm (super)determinants. \par
	For bosonic operators
	\begin{equation}
		\label{eq:EgenMgasym}
						\resizebox{0.98\textwidth}{!}{%
			$\begin{aligned}
		&\Gamma^E_{\text{asym}}\left[0,j_{S^{(1)'\,E}},0,0,0,0,0,\lambda\right] \\
		&
		=-(N^2-1)\log\Det \left[I \delta_{s_1s_2}\delta_{k_1k_2}+\frac{1}{N}6{s_1\choose k_1}{s_2\choose k_2+2}\partial_{z}^{s_1-k_1+k_2}\frac{\Laplace^{-1}}{\lambda^{s_2+2}}Z_{S^{(1)'\,E}}(\lambda)\frac{j_{S^{(1)'\,E}_{s_2k_2}}}{s_2-1}\right]\\
		&\quad+(N^2-1)\log\Det \left[I \delta_{s_1s_2}\delta_{k_1k_2}+\frac{1}{N}{s_1\choose k_1}{s_2\choose k_2+1}\partial_{z}^{s_1-k_1+k_2-1}\frac{\partial_{z}\Laplace^{-1}}{\lambda^{s_2+2}}Z_{S^{(1)'\,E}}(\lambda)j_{S^{(1)'\,E}_{s_2k_2}}\right] 
							\end{aligned}$
	} 
	\end{equation}
	and
	\begin{equation}
		\label{eq:EgenMgasym2}
						\resizebox{0.98\textwidth}{!}{%
			$\begin{aligned}
		&\Gamma^E_{\text{asym}}\left[0 ,0,0,j_{S^{(2)'\,E}},0,0,0,\lambda\right]  \\
		&
		=-(N^2-1)\log\Det \left[I \delta_{s_1s_2}\delta_{k_1k_2}+\frac{1}{N}6{s_1\choose k_1}{s_2\choose k_2+2}\partial_{z}^{s_1-k_1+k_2}\frac{\Laplace^{-1}}{\lambda^{s_2+2}}Z_{S^{(2)'\,E}}(\lambda)\frac{j_{S^{(2)'\,E}_{s_2k_2}}}{s_2-1}\right]\\
		&\quad+(N^2-1)\log\Det \left[I \delta_{s_1s_2}\delta_{k_1k_2}-\frac{1}{N}{s_1\choose k_1}{s_2\choose k_2+1}\partial_{z}^{s_1-k_1+k_2-1}\frac{\partial_{z}\Laplace^{-1}}{\lambda^{s_2+2}}Z_{S^{(2)'\,E}}(\lambda)j_{S^{(2)'\,E}_{s_2k_2}}\right] \ .
							\end{aligned}$
	} 
	\end{equation}
	For fermionic operators
	\begin{equation}
		\label{eq:EgenMgasym3}
						\resizebox{0.98\textwidth}{!}{%
			$\begin{aligned}
		&\Gamma^E_{\text{asym}}\left[0 ,0,0,0,0,\bar{j}_{{M^{'E}}},j_{\bar M^{'E}},\lambda\right] \\
		&=+(N^2-1)\log\Det\Big[I \delta_{s_1s_2}\delta_{k_1k_2}\\
		&\quad+\frac{1}{N^2}{s_1+1\choose k_1+2}{s_2+1\choose k_2}(-\partial_{z})^{s_1-k_1+k}\frac{\partial_{z}\Laplace^{-1}}{\lambda^{s+2}}Z_{M}(\lambda)j_{\bar{M}_{sk}}{s\choose k}{s_2\choose k_2+1}(-\partial_{z})^{s-k+k_2}\frac{\Laplace^{-1}}{\lambda^{s_2+2}}Z_{M}(\lambda) \bar{j}_{M_{s_2k_2}}\Big]
							\end{aligned}$
	} 
	\end{equation}		
	that follow from the corresponding conformal objects (Sec. \ref{euclgen}).
	
	\section{Conclusions}  \label{Conc}

	We are now ready to state the main results of the present paper.\par
	Nonperturbatively, the generating functional of the glueball/gluinoball one-loop correlators reads (Sec. \ref{NP})
	\begin{equation}
		\label{glueballW1loop_tot}
		\resizebox{0.98\textwidth}{!}{%
			$\begin{aligned}
				&\mathcal{W}^E_{\text{glueball/gluinoball 1-loop }}[J_{\Phi},J_{\Psi}] \\
				&=\frac{1}{2}\log\text{sDet}
				\begin{pmatrix}\ast'_2(-\Delta+M^2)+\frac{1}{N}\ast'_3\Phi_J\ast'_3& \frac{1}{N}\ast'_3\ast'_3\Psi_J\\ 
					\frac{1}{N}\ast'_3\ast'_3\Psi_J&\ast_2(-\Delta+M^2)+\frac{1}{N}\ast_3\Phi_J\ast_3 \end{pmatrix} \\
				&=+\frac{1}{2}\log\text{Det}
				\left(\ast'_2(-\Delta+M^2)+\frac{1}{N}\ast'_3\Phi_J\ast'_3\right)-\frac{1}{2}\log\text{Det}
				\left(\ast_2(-\Delta+M^2)+\frac{1}{N}\ast_3\Phi_J\ast_3\right)\\
				&\quad-\frac{1}{2}\log\text{Det}
				\left[\mathcal{I}-\frac{1}{N}\left(\ast_2(-\Delta+M^2)+\frac{1}{N}\ast_3\Phi_J\ast_3\right)^{-1}\ast'_3\ast'_3\Psi_J\left(\ast'_2(-\Delta+M^2)+\frac{1}{N}\ast'_3\Phi_J\ast'_3\right)^{-1}\frac{1}{N}\ast'_3\ast'_3\Psi_J\right] ,
			\end{aligned}$
		} 
	\end{equation} 
	where we have employed the second equality in Eq. \eqref{eq:sdet}. Setting $J_{\Psi}=0$, we get for bosonic operators up to an irrelevant constant
	\begin{align}
		\label{glueballW1loop_glue}
		\mathcal{W}^E_{\text{glueball/gluinoball 1-loop }}[J_{\Phi},0] 
		=&+\frac{1}{2}\log\text{Det}
		\left(\mathcal{I}+(\ast'_2(-\Delta+M^2))^{-1}\frac{1}{N}\ast'_3\Phi_J\ast'_3\right)\nonumber\\
		&-\frac{1}{2}\log\text{Det}
		\left(\mathcal{I}+(\ast_2(-\Delta+M^2))^{-1}\frac{1}{N}\ast_3\Phi_J\ast_3\right)\,.
	\end{align}
	Similarly, setting $J_{\Phi}=0$, we obtain for fermionic operators up to an irrelevant constant
	\begin{equation}
		\label{glueballW1loop_gluino}
		\resizebox{0.98\textwidth}{!}{%
			$\begin{aligned}
				\mathcal{W}^E_{\text{glueball/gluinoball 1-loop }}[0,J_{\Psi}] =-\frac{1}{2}\log\text{Det}
				\left[\mathcal{I}-\left(\ast_2(-\Delta+M^2)\right)^{-1}\frac{1}{N}\ast'_3\ast'_3\Psi_J\left(\ast'_2(-\Delta+M^2)\right)^{-1}\frac{1}{N}\ast'_3\ast'_3\Psi_J\right] .
			\end{aligned}$
		} 
	\end{equation}
	The corresponding RG-improved objects that follow from Eqs. \eqref{eq:EgenMgasym}, \eqref{eq:EgenMgasym2}, \eqref{eq:EgenMgasym3} and the identification in Eq. \eqref{74} read for bosonic operators
	\begin{align}
		\label{EgenMgasymtorus}
		&\Gamma^E_{\text{asym torus}}\left[0,j_{S^{(1)'\,E}},0,0,0,0,0,\lambda\right] \nonumber\\
		&
		=+\log\Det \left[I \delta_{s_1s_2}\delta_{k_1k_2}+\frac{1}{N}6{s_1\choose k_1}{s_2\choose k_2+2}\partial_{z}^{s_1-k_1+k_2}\frac{\Laplace^{-1}}{\lambda^{s_2+2}}Z_{S^{(1)'\,E}}(\lambda)\frac{j_{S^{(1)'\,E}_{s_2k_2}}}{s_2-1}\right]\nonumber\\
		&\quad-\log\Det \left[I \delta_{s_1s_2}\delta_{k_1k_2}+\frac{1}{N}{s_1\choose k_1}{s_2\choose k_2+1}\partial_{z}^{s_1-k_1+k_2-1}\frac{\partial_{z}\Laplace^{-1}}{\lambda^{s_2+2}}Z_{S^{(1)'\,E}}(\lambda)j_{S^{(1)'\,E}_{s_2k_2}}\right] 
	\end{align}
	and
	\begin{align}
		\label{EgenMgasymtorus2}
		&\Gamma^E_{\text{asym torus}}\left[0 ,0,0,j_{S^{(2)'\,E}},0,0,0,\lambda\right]  \nonumber\\
		&
		=+\log\Det \left[I \delta_{s_1s_2}\delta_{k_1k_2}+\frac{1}{N}6{s_1\choose k_1}{s_2\choose k_2+2}\partial_{z}^{s_1-k_1+k_2}\frac{\Laplace^{-1}}{\lambda^{s_2+2}}Z_{S^{(2)'\,E}}(\lambda)\frac{j_{S^{(2)'\,E}_{s_2k_2}}}{s_2-1}\right]\nonumber\\
		&\quad-\log\Det \left[I \delta_{s_1s_2}\delta_{k_1k_2}-\frac{1}{N}{s_1\choose k_1}{s_2\choose k_2+1}\partial_{z}^{s_1-k_1+k_2-1}\frac{\partial_{z}\Laplace^{-1}}{\lambda^{s_2+2}}Z_{S^{(2)'\,E}}(\lambda)j_{S^{(2)'\,E}_{s_2k_2}}\right]  \,,
	\end{align}
	and for fermionic operators
	\begin{equation}
		\label{EgenMgasymtorus3}
				\resizebox{0.98\textwidth}{!}{%
			$\begin{aligned}
		&\Gamma^E_{\text{asym torus}}\left[0 ,0,0,0,0,\bar{j}_{{M^{'E}}},j_{\bar M^{'E}},\lambda\right]  \\
		&
		=
		-\log\Det\Big[I \delta_{s_1s_2}\delta_{k_1k_2}\\
		&+{s_1+1\choose k_1+2}{s_2+1\choose k_2}(-\partial_{z})^{s_1-k_1+k}\frac{\partial_{z}\Laplace^{-1}}{\lambda^{s+2}}Z_{M}(\lambda)\frac{1}{N}j_{\bar{M}_{sk}}{s\choose k}{s_2\choose k_2+1}(-\partial_{z})^{s-k+k_2}\frac{\Laplace^{-1}}{\lambda^{s_2+2}}Z_{M}(\lambda) \frac{1}{N}\bar{j}_{M_{s_2k_2}}\Big] \, .		
		\end{aligned}$
	} 
	\end{equation}		

They are are UV asymptotic as $\lambda \rightarrow 0$ to the above corresponding nonperturbative objects according to the AF, where in the following, to keep the notation simple, the rescaling of the coordinates by the factor of $\lambda$ in the nonperturbative generating functional is understood: Respectively,
\begin{align}
	\Gamma^E_{\text{asym torus}} \left[\;0,\ldots, {j}_{\VarPrimeSup{S}{(i)}{E}},\;0,\ldots; \lambda \right]  \sim \mathcal{W}^E_{\text{glueball/gluinoball 1-loop }}[J_{\Phi^{(i)}},0] 
\end{align}
and
\begin{align}
	\Gamma^E_{\text{asym torus}}\left[0,0,0,0,0,\bar{j}_{M^E},j_{\bar{M}^E},\lambda\right]
	 \sim \mathcal{W}^E_{\text{glueball/gluinoball 1-loop }}[0,J_{\Psi}] \,.
\end{align}
for a suitable choice of the glueball, $\Phi^{(i)}, i=1,2$, and gluinoball, $\Psi$, interpolating fields (Sec. \ref{NP}).\par
Hence, the matching of the $\log \Det$ structure of the above nonperturbative and UV-asymptotic RG-improved generating functionals of correlators of balanced twist-$2$ operators in the large-$N$ expansion to the leading-nonplanar order sets strong qualitative and quantitative UV constraints on the yet-to-come nonperturbative solution of large-$N$ $\mathcal{N} = 1$ SUSY YM theory and it may be an essential guide for the search of such a solution.

\section*{Acknowledgments}

We would like to thank Loris Del Grosso for participating in the early stages of this project.

\newpage
\pagebreak
\appendix

\section{Notation and Wick rotation \label{appN}}

We follow the conventions in \cite{Braun:2003rp}. We choose the mostly minus Minkowskian metric
\begin{equation}
	(g_{\mu\nu}) = \text{diag}(1,-1,-1,-1)
\end{equation}
and define the light-cone coordinates
\begin{equation}
	x^{\pm} = \frac{x^0\pm x^3}{\sqrt{2}}=x_{\mp}\,.
\end{equation}
The corresponding Minkowskian (squared) distance is
\begin{equation}
	\label{mod2}
	\rvert x \rvert^2 = 2 x^+ x^- -x_{\perp}^2\,,
\end{equation}
with
\begin{equation}
	x^2_\perp=(x^1)^2+(x^2)^2\,.
\end{equation}
We denote the derivative with respect to $x^+$ by
\begin{equation}
	\partial_+ = \frac{\partial}{\partial x^+} = \partial_{x^+} =\frac{\partial}{\partial x_-} = \partial_{x_-}
\end{equation}
and define the light-like vectors $n^\mu$ and $\bar{n}^\mu$
\begin{equation}
	n_\mu n^\mu = \bar{n}_\mu \bar{n}^\mu = 0 \qquad n_\mu \bar{n}^\mu = 1
\end{equation}
that respectively read $(n^{\mu}) = \frac{1}{\sqrt{2}}(1,0,0,1)$ and $(\bar{n}^{\mu}) = \frac{1}{\sqrt{2}}(1,0,0,-1)$. Correspondingly, the Minkowskian metric decomposes into transverse and longitudinal components with respect to the light-like vectors
\begin{equation}
	g_{\mu\nu} =g^\perp_{\mu\nu}+n_\mu \bar{n}_\nu +n_\nu \bar{n}_\mu \,.
\end{equation} 
The Euclidean metric reads
\begin{equation}
	(\delta_{\mu\nu}) =  \text{diag}(1,1,1,1)\,
\end{equation}
with Euclidean (squared) distance
\begin{equation}
	x^2 = 2x^z x^{\bar{z}}+x_\perp^2\,
\end{equation}
where
\begin{equation}
	x^z= \frac{x^4+ix^3}{\sqrt{2}}=\frac{x_4+ix_3}{\sqrt{2}}=x_{\bar z}
\end{equation}
and
\begin{equation}
	x^{\bar z}=  \frac{x^4-ix^3}{\sqrt{2}}=\frac{x_4-ix_3}{\sqrt{2}}=x_z\,.
\end{equation}
We define the Wick rotation by
\begin{align}
	\label{wick1}
	&x^0=x_0 \rightarrow  - i x^4=-i x_4 
\end{align} 
and
\begin{align}
	\label{wick2}
	&p_0=p^0 \rightarrow  i p_4= i p^4\,.
\end{align} 
Eq. (\ref{wick1}) ensures that $\exp(iS_M)\rightarrow \exp(-S_E)$, where $S_M$ and $S_E$ are respectively the Minkowskian and Euclidean actions, with $S_E$ positive definite. By defining $ p \cdot x = p_{\mu} x^{\mu}$ and $\langle p x \rangle= p_{\mu} x^{\mu}$ respectively in Minkowskian and Euclidean space-time, Eq. (\ref{wick2}) ensures that by the Wick rotation $ p \cdot x  \rightarrow \langle p x \rangle$, so that the pairings $p \cdot x$ and $\langle p x \rangle$ are actually independent of the Minkowskian and Euclidean metric. 
Therefore, by a slight abuse of notation, we also write $ p \cdot x$ in Euclidean space-time, instead of $\langle p x \rangle$. Besides, $|x|^2 \rightarrow - x^2$ and  $|p|^2 \rightarrow - p^2$. As a consequence, the Wick rotation of the scalar propagator of mass $m$ in Minkowskian space-time
\begin{equation}
	\langle \phi(x) \phi(y)\rangle = \int \frac{d^4p}{(2\pi)^4}\, e^{i  p \cdot (x-y)}\, \frac{i}{|p|^2-m^2+i \epsilon}
\end{equation}
reads in Euclidean space-time
\begin{align}
	\langle \phi^E(x) \phi^E(y)\rangle &= \int \frac{d^4p}{(2\pi)^4}\, e^{i p \cdot(x-y)}\, \frac{i^2}{- p^2-m^2}\nonumber\\
	&= \int \frac{d^4p}{(2\pi)^4}\, e^{i p \cdot(x-y)}\, \frac{1}{p^2+m^2}
\end{align}
as it should be. The Wick rotation of the light-cone coordinates
\begin{align}
	&x^+=x_-\rightarrow -ix^z= -i x_{\bar z} \qquad 
\end{align}
and
\begin{align}
	&x^-=x_+\rightarrow -i x^{\bar{z}}= -i x_{z}\qquad
\end{align}
implies the Wick rotation of the derivative with respect to $x^+$
\begin{align}
	\label{wickderivative}
	\partial_+\rightarrow i \partial_{z} = i\frac{\partial}{\partial x^z}\,.
\end{align}

\section{Jacobi and Gegenbauer polynomials \label{appB}}

We work out some formulas for the Jacobi and Gegenbauer polynomials employed in the present paper. 
For $x$ real the Jacobi polynomials $P^{(\alpha,\beta)}_l(x)$ admit the representation \cite{szego1959orthogonal}
\begin{align}
	\label{jaco1}
	P^{(\alpha,\beta)}_l(x)
	 = \sum_{k = 0}^{l}{l+\alpha\choose k}{l+\beta\choose k+\beta}\left(\frac{x-1}{2}\right)^k\left(\frac{x+1}{2}\right)^{l-k}\,,
\end{align}
with $\alpha, \beta$ real and $l$ a natural number. Besides, they satisfy the symmetry property
\begin{align}
	\label{jacox}
	P^{(\alpha,\beta)}_l(-x) = (-1)^{l}P^{(\beta,\alpha)}_l(x)\,.
\end{align} 
The Gegenbauer polynomials $C^{\alpha'}_l(x)$ are a special case of the Jacobi polynomials
\begin{equation} \label{GPol}
	C^{\alpha'}_l(x) = \frac{\Gamma(l+2\alpha')\Gamma(\alpha'+\frac{1}{2})}{\Gamma(2\alpha')\Gamma(l+\alpha'+\frac{1}{2})}P_l^{(\alpha'-\frac{1}{2},\alpha'-\frac{1}{2})}(x)\,,
\end{equation}
with the symmetry property
\begin{align}
	\label{symmgegen}
	C^{\alpha'}_l(-x) = (-1)^l C^{\alpha'}_l(x) \,.
\end{align}
We set
\begin{equation}
	x = \frac{b-a}{a+b}\,,
\end{equation}
so that
\begin{equation}
	\left(\frac{x-1}{2}\right)^k\left(\frac{x+1}{2}\right)^{l-k} = (-1)^{l-k} \frac{a^{l-k} b^k}{(a+b)^l}\,.
\end{equation}
Eq. (\ref{jaco1}) becomes
\begin{equation}
	\label{jaco2}
	P^{(\alpha,\beta)}_l(x) = \sum_{k = 0}^{l}{l+\alpha\choose k}{l+\beta\choose k+\beta}(-1)^{l-k} \frac{a^{l-k} b^k}{(a+b)^l}\,.
\end{equation}
Moreover, putting $l = J-\alpha'+\frac{1}{2}$ in Eq. \eqref{GPol} and $\alpha=\beta=\alpha'-\frac{1}{2}$ in Eq. \eqref{jaco1}, we obtain
\begin{align}
	C^{\alpha'}_{J-\alpha'+\frac{1}{2}}\left(x\right) =\frac{\Gamma(J+\frac{1}{2}+\alpha')\Gamma(\alpha'+\frac{1}{2})}{\Gamma(2\alpha')\Gamma(J+1)}\sum_{k=0}^{J-\alpha'+\frac{1}{2}} {J\choose k}{J\choose k+\alpha'-\frac{1}{2}}
	(-1)^{J-\alpha'+\frac{1}{2}-k} \frac{a^{J-\alpha'+\frac{1}{2}-k} b^k}{(a+b)^{J-\alpha'+\frac{1}{2}}}\,.
\end{align}
Specializing the above equation to $J = s$ and $\alpha' = \frac{5}{2}$, we get
\begin{align}
	\label{physicalgegen}
	C^{\frac{5}{2}}_{s-2}\left(x\right) =  \frac{\Gamma(s+3)\Gamma(3)}{\Gamma(5)\Gamma(s+1)}
	\sum_{k=0}^{s-2} {s\choose k}{s\choose k+2}(-1)^{s-k} \frac{a^{s-k-2} b^k}{(a+b)^{s-2}}\,.
\end{align}
Moreover, for $J = s$ and $\alpha' = \frac{3}{2}$, we obtain 
\begin{align}
	\label{physicalgegen2}
	C^{\frac{3}{2}}_{s-1}\left(x\right) =\frac{\Gamma(s+2)\Gamma(1)}{\Gamma(3)\Gamma(s+1)}
	\sum_{k=0}^{s-1} {s\choose k}{s\choose k+1}(-1)^{s-k-1} \frac{a^{s-k-1} b^k}{(a+b)^{s-1}}\,.
\end{align}
We restrict $\alpha, \beta$ to the natural numbers and, correspondingly, $\alpha'$ to the positive half-integers and $J$ to the natural numbers. 
By employing the identity
\begin{align}
	{l+\alpha\choose k}{l+\beta\choose k+\beta} 
	=\frac{(l+\beta)!(l+\alpha)!}{l!(l+\alpha+\beta)!}{l\choose k}{l+\beta+\alpha\choose k+\beta}\,,
\end{align}
it follows from Eq. \eqref{jaco2} that
\begin{align}
	\label{jaco3}
	P^{(\alpha,\beta)}_l(x) = \frac{(l+\beta)!(l+\alpha)!}{l!(l+\alpha+\beta)!} 
 \sum_{k = 0}^{l} {l\choose k}{l+\beta+\alpha\choose k+\beta}(-1)^{l-k} \frac{a^{l-k} b^k}{(a+b)^l}\,.
\end{align}
Correspondingly, Eq. \eqref{GPol} reads
\begin{align}
	C^{\alpha'}_l\left(x\right) =& \frac{\Gamma(l+2\alpha')\Gamma(\alpha'+\frac{1}{2})}{\Gamma(2\alpha')\Gamma(l+\alpha'+\frac{1}{2})}  \frac{(l+\alpha'-\frac{1}{2})!(l+\alpha'-\frac{1}{2})!}{l!(l+2\alpha'-1)!}\nonumber\\
	&\sum_{k = 0}^{l} {l\choose k}{l+2\alpha'-1\choose k+\alpha'-\frac{1}{2}}(-1)^{l-k} \frac{a^{l-k} b^k}{(a+b)^l}
\end{align}
that reduces to
\begin{align} \label{bin}
	C^{\alpha'}_l\left(x\right) 
	= \frac{\Gamma(\alpha'+\frac{1}{2})\Gamma(l+\alpha'+\frac{1}{2})}{\Gamma(2\alpha')\Gamma(l+1)}
	\sum_{k = 0}^{l} {l\choose k}{l+2\alpha'-1\choose k+\alpha'-\frac{1}{2}}(-1)^{l-k} \frac{a^{l-k} b^k}{(a+b)^l}\,.
\end{align}

\section{Conformal properties of the standard basis} \label{B}

The gauge-invariant collinear twist-$2$ operators in the light-cone gauge that enter the balanced superfields read in the standard basis~\cite{BPS1,Belitsky:2004sc, Belitsky:2003sh}
\begin{equation}\label{1000}
	\resizebox{0.67\textwidth}{!}{%
		$\begin{aligned}
			O^A_s &= \frac{1}{2} \partial_+ \bar{A}^a(i\overrightarrow{\partial}_++ i\overleftarrow{\partial}_+)^{s-2}C^{\frac{5}{2}}_{s-2}\Bigg(\frac{\overrightarrow{\partial}_+- \overleftarrow{\partial}_+}{\overrightarrow{\partial_+}+\overleftarrow{\partial}_+}\Bigg)\partial_+ A^a \\
			\tilde{O}^A_s &= \frac{1}{2} \partial_+ \bar{A}^a(i\overrightarrow{\partial}_++ i\overleftarrow{\partial}_+)^{s-2}C^{\frac{5}{2}}_{s-2}\Bigg(\frac{\overrightarrow{\partial}_+- \overleftarrow{\partial}_+}{\overrightarrow{\partial_+}+\overleftarrow{\partial}_+}\Bigg)\partial_+ A^a\\
			O^\lambda_s &=  \frac{1}{2} \bar{\lambda}^a(i\overrightarrow{\partial}_++ i\overleftarrow{\partial}_+)^{s-1}C^{\frac{3}{2}}_{s-1}\Bigg(\frac{\overrightarrow{\partial}_+- \overleftarrow{\partial}_+}{\overrightarrow{\partial_+}+\overleftarrow{\partial}_+}\Bigg) \lambda^a\\
			\tilde{O}^\lambda_s &=  \frac{1}{2} \bar{\lambda}^a(i\overrightarrow{\partial}_++ i\overleftarrow{\partial}_+)^{s-1}C^{\frac{3}{2}}_{s-1}\Bigg(\frac{\overrightarrow{\partial}_+- \overleftarrow{\partial}_+}{\overrightarrow{\partial_+}+\overleftarrow{\partial}_+}\Bigg) \lambda^a\\
			M_s &= \frac{1}{2}\hspace{0.08cm} \partial_+A^a (i\overrightarrow{\partial}_++ i\overleftarrow{\partial}_+)^{s-1}P^{(2,1)}_{s-1}\Bigg(\frac{\overrightarrow{\partial}_+- \overleftarrow{\partial}_+}{\overrightarrow{\partial_+}+\overleftarrow{\partial}_+}\Bigg) \lambda^a \\
			\bar{M}_s &=  \hspace{0.08cm}\frac{1}{2}\bar{\lambda}^a(i\overrightarrow{\partial}_++ i\overleftarrow{\partial}_+)^{s-1}P^{(1,2)}_{s-1}\Bigg(\frac{\overrightarrow{\partial}_+- \overleftarrow{\partial}_+}{\overrightarrow{\partial_+}+\overleftarrow{\partial}_+}\Bigg) \partial_+\bar{A}^a\,,
		\end{aligned}$
	} 
\end{equation}
where $C^{\alpha'}_l(x)$ are Gegenbauer polynomials (appendix \ref{appB}) with the symmetry properties
\begin{align}
	C_{l}^{\alpha'}(-x)=(-1)^{l}C_{l}^{\alpha'}(x)\,.
\end{align}
They are the restriction to the components with maximal-spin projection $s$ along the $p_+$ direction of linear combinations of twist-$2$ operators of the kind
\begin{align} \label{1}
	&O^{A\,\mathcal{T}=2}_{s} \quad=\quad \Tr\, F^\mu_{(\rho_1}\overleftarrow{D}_{\rho_2}\ldots \overrightarrow{D}_{\rho_{s-1}}F_{\rho_s)\mu}-\,\text{traces}\qquad\qquad \nonumber\\
	&\tilde{O}^{A\,\mathcal{T}=2}_{s} \quad=\quad \Tr\, \tilde{F}^\mu_{(\rho_1}\overleftarrow{D}_{\rho_2}\ldots \overrightarrow{D}_{\rho_{s-1}}F_{\rho_s)\mu}-\,\text{traces}\qquad\qquad \nonumber\\
	&O^{\lambda\,\mathcal{T}=2}_{s} \quad=\quad \Tr\, \bar{\chi}\gamma_{(\rho_1}\overleftarrow{D}_{\rho_2}\ldots \overrightarrow{D}_{\rho_{s-1})}\chi-\,\text{traces}\qquad\qquad \nonumber\\
	&\tilde{O}^{\lambda\,\mathcal{T}=2}_{s} \quad=\quad \Tr\, \bar{\chi}\gamma_{(\rho_1}\gamma_5\overleftarrow{D}_{\rho_2}\ldots \overrightarrow{D}_{\rho_{s-1})}\chi-\,\text{traces}\qquad\qquad \nonumber\\
	&M^{\mathcal{T}=2}_{s} \quad=\quad \Tr\, F_{(\rho_1}^{\nu}\overleftarrow{D}_{\rho_2}\ldots \overrightarrow{D}_{\rho_{s-1})}\sigma_{\nu}\lambda-\,\text{traces}\nonumber\\
	&\bar{M}^{\mathcal{T}=2}_{s} \quad=\quad \Tr\,\bar{\lambda}\,\bar{\sigma}_{\nu} \overleftarrow{D}_{(\rho_{s-1}}\ldots \overrightarrow{D}_{\rho_{2}} F_{\rho_{1})}^{\nu}-\,\text{traces}\,,
\end{align}
with all the possible combinations of right and left derivatives \cite{makeenko,Belitsky:2007jp}, where the parentheses stand for symmetrization of all the indices in between and the subtraction of the traces ensures that the contraction of any two indices vanishes. \par
Suitable linear combinations of the above twist-$2$ operators are conserved  \cite{makeenko,Belitsky:2007jp} to the leading order of perturbation theory and automatically transform  \cite{makeenko,Belitsky:2007jp} as primary operators with respect to the conformal group \cite{Beisert:2004fv}.
By projecting on the maximal-spin component along the $p_+$ direction they restrict to 
\begin{equation} \label{OO}
	\resizebox{0.67\textwidth}{!}{%
		$\begin{aligned}
			O^A_s &= \frac{1}{2}  f_{11}^a(i\overrightarrow{\partial}_++ i\overleftarrow{\partial}_+)^{s-2}C^{\frac{5}{2}}_{s-2}\Bigg(\frac{\overrightarrow{\partial}_+- \overleftarrow{\partial}_+}{\overrightarrow{\partial_+}+\overleftarrow{\partial}_+}\Bigg)f_{\dot{1}\dot{1}}^a \\
			\tilde{O}^A_s &= \frac{1}{2}  f_{11}^a(i\overrightarrow{\partial}_++ i\overleftarrow{\partial}_+)^{s-2}C^{\frac{5}{2}}_{s-2}\Bigg(\frac{\overrightarrow{\partial}_+- \overleftarrow{\partial}_+}{\overrightarrow{\partial_+}+\overleftarrow{\partial}_+}\Bigg)f_{\dot{1}\dot{1}}^a\\
			O^\lambda_s &=  \frac{1}{2} \bar{\lambda}^a(i\overrightarrow{\partial}_++ i\overleftarrow{\partial}_+)^{s-1}C^{\frac{3}{2}}_{s-1}\Bigg(\frac{\overrightarrow{\partial}_+- \overleftarrow{\partial}_+}{\overrightarrow{\partial_+}+\overleftarrow{\partial}_+}\Bigg) \lambda^a\\
			\tilde{O}^\lambda_s &=  \frac{1}{2} \bar{\lambda}^a(i\overrightarrow{\partial}_++ i\overleftarrow{\partial}_+)^{s-1}C^{\frac{3}{2}}_{s-1}\Bigg(\frac{\overrightarrow{\partial}_+- \overleftarrow{\partial}_+}{\overrightarrow{\partial_+}+\overleftarrow{\partial}_+}\Bigg) \lambda^a \\
			M_s &=- \frac{1}{2}\hspace{0.08cm}  f_{\dot{1}\dot{1}}^a (i\overrightarrow{\partial}_++ i\overleftarrow{\partial}_+)^{s-1}P^{(2,1)}_{s-1}\Bigg(\frac{\overrightarrow{\partial}_+- \overleftarrow{\partial}_+}{\overrightarrow{\partial_+}+\overleftarrow{\partial}_+}\Bigg) \lambda^a \\
			\bar{M}_s &=  -\hspace{0.08cm}\frac{1}{2}\bar{\lambda}^a(i\overrightarrow{\partial}_++ i\overleftarrow{\partial}_+)^{s-1}P^{(1,2)}_{s-1}\Bigg(\frac{\overrightarrow{\partial}_+- \overleftarrow{\partial}_+}{\overrightarrow{\partial_+}+\overleftarrow{\partial}_+}\Bigg) f_{11}^a
		\end{aligned}$
	} 
\end{equation}
that are primaries with respect to the collinear conformal subgroup $SL(2,R)$ \cite{Belitsky:1998gc}
and reduce in the light-cone gauge to the operators in eq. \eqref{1000} with $f_{11}=- \partial_+ \bar A$. By allowing operator mixing with derivatives of twist-2 operators of lower spin they may be extended to primary conformal operators to the next-to-leading order in the conformal renormalization scheme \cite{Braun:2003rp} that differs from the $\overline{MS}$ scheme by a finite renormalization.\par

	\section{Minkowskian conformal correlators \label{npoint}}

	We derive from the generating functional in Eqs. (\ref{wgen1}) and (\ref{wgen2}) the $n$-point conformal correlators in several sectors.  
	
	\subsection{$O^A$ and $\tilde{O}^A$ correlators}
	
	We get
	\begin{align}
		&\langle O^A_{s_1}(x_1)\ldots O^A_{s_n}(x_n)\tilde{O}^A_{s_{n+1}}(x_{n+1})\ldots \tilde{O}^A_{s_{n+2m}}(x_{n+2m})\rangle =\nonumber\\
		& \frac{\delta}{\delta J_{O^A_{s_1}}(x_1)}\cdots\frac{\delta}{\delta J_{O^A_{s_n}}(x_n)}\frac{\delta}{\delta J_{\tilde{O}^A_{s_{n+1}}}(x_{n+1})}\cdots\frac{\delta}{\delta J_{\tilde{O}^A_{s_{n+2m}}}(x_{n+2m})}\mathcal{W}_{\text{conf}}\left[J_{O^A},J_{\tilde{O}^A},0,0,0,0\right]\nonumber\\
		&= -\frac{1}{2} \frac{\delta}{\delta J_{O^A_{s_1}}(x_1)}\cdots\frac{\delta}{\delta J_{O^A_{s_n}}(x_n)}\frac{\delta}{\delta J_{\tilde{O}^A_{s_{n+1}}}(x_{n+1})}\cdots\frac{\delta}{\delta J_{\tilde{O}^A_{s_{n+2m}}}(x_{n+2m})}\nonumber\\
		&\quad\Bigg[\log\Det \Big(\mathcal{I} +\frac{1}{2}i\square^{-1}J_{O^A_{s}}\otimes\mathcal{Y}_{s-2}^{\frac{5}{2}}+\frac{1}{2}i\square^{-1}J_{\tilde{O}^A_{s}}\otimes\mathcal{H}_{s-2}^{\frac{5}{2}} \Big) \nonumber\\
		&\quad\quad+\log\Det \Big( \mathcal{I}+\frac{1}{2}i\square^{-1}J_{O^A_{s}}\otimes\mathcal{Y}_{s-2}^{\frac{5}{2}}-\frac{1}{2}i\square^{-1}J_{\tilde{O}^A_{s}}\otimes\mathcal{H}_{s-2}^{\frac{5}{2}} \Big)    \Bigg]
	\end{align}
	that reproduces the known result \cite{BPS1,BPSpaper2}
	\begin{align}
		&\langle O^A_{s_1}(x_1)\ldots O^A_{s_n}(x_n)\tilde{O}^A_{s_{n+1}}(x_{n+1})\ldots \tilde{O}^A_{s_{n+2m}}(x_{n+2m})\rangle \nonumber\\
		& =\frac{1}{(4\pi^2)^{n+2m}}\frac{N^2-1}{2^{n+2m}}2^{\sum_{l=1}^{n+2m} s_l}i^{\sum_{l=1}^{n+2m} s_l}\frac{\Gamma(3)\Gamma(s_1+3)}{\Gamma(5)\Gamma(s_1+1)} \ldots\frac{\Gamma(3)\Gamma(s_{n+2m}+3)}{\Gamma(5)\Gamma(s_{n+2m}+1)}\nonumber\\
		&\quad\sum_{k_1=0}^{s_1-2}\ldots \sum_{k_{n+2m} = 0}^{s_{n+2m}-2}{s_1\choose k_1}{s_1\choose k_1+2}\ldots{s_{n+2m}\choose k_{n+2m}}{s_{n+2m}\choose k_{n+2m}+2}\nonumber\\
		&\quad\frac{(-1)^{n+2m}}{n+{2m}}\sum_{\sigma\in P_{n+2m}}(s_{\sigma(1)}-k_{\sigma(1)}+k_{\sigma(2)})!\ldots(s_{\sigma(n+{2m})}-k_{\sigma(n+{2m})}+k_{\sigma(1)})!\nonumber\\
		&\quad\frac{(x_{\sigma(1)}-x_{\sigma(2)})_+^{s_{\sigma(1)}-k_{\sigma(1)}+k_{\sigma(2)}}}{\left(\rvert x_{\sigma(1)}-x_{\sigma(2)}\rvert^2\right)^{s_{\sigma(1)}-k_{\sigma(1)}+k_{\sigma(2)}+1}}\ldots\frac{(x_{\sigma(n+2m)}-x_{\sigma(1)})_+^{s_{\sigma(n+{2m})}-k_{\sigma(n+{2m})}+k_{\sigma(1)}}}{\left(\rvert x_{\sigma(n+{2m})}-x_{\sigma(1)}\rvert^2\right)^{s_{\sigma(n+{2m})}-k_{\sigma(n+{2m})}+k_{\sigma(1)}+1}}\,,
	\end{align}
	where $P_n$ is the set of permutations of $\{1,\ldots,n\}$ and the $-i\epsilon$ prescription in the propagators is understood.
	
	\subsection{$O^{\lambda}$ and $\tilde{O}^\lambda$ correlators}
	
	Similarly, from 
	\begin{align}
		&\langle O^{\lambda}_{s_1}(x_1)\ldots O^{\lambda}_{s_n}(x_n)\rangle = \frac{\delta}{\delta J_{O^{\lambda}_{s_1}}(x_1)}\cdots\frac{\delta}{\delta J_{O^{\lambda}_{s_n}}(x_n)}\mathcal{W}_{\text{conf}}\left[0,0,J_{O^{\lambda}},0,0,0\right]\nonumber\\
		&= \frac{\delta}{\delta J_{O^{\lambda}_{s_1}}(x_1)}\cdots\frac{\delta}{\delta J_{O^{\lambda}_{s_n}}(x_n)}\log\Det \Big(\mathcal{I}+\frac{1}{2}\partial_+\square^{-1}J_{O^\lambda_{s}}\otimes\mathcal{Y}_{s-1}^{\frac{3}{2}}\Big)  \nonumber\\
		&=(N^2-1)\frac{\delta}{\delta J_{O^{\lambda}_{s_1}}(x_1)}\cdots\frac{\delta}{\delta J_{O^{\lambda}_{s_n}}(x_n)}\sum_{l=1}^{\infty}\frac{(-1)^{l+1}}{l}\frac{1}{2^l}\sum_{s'_1}\ldots\sum_{s'_l}\nonumber\\
		&\quad\int d^4y_1\ldots d^4y_l \partial_{y_1^+}\Box^{-1}(y_1-y_2)\mathcal{Y}_{s'_2-1}^{\frac{3}{2}}\otimes J_{O^{\lambda}_{s'_2}}(y_2)\ldots \partial_{y_l^+}\Box^{-1}(y_l-y_1)\mathcal{Y}_{s'_1-1}^{\frac{3}{2}}\otimes J_{O^{\lambda}_{s_1'}}(y_1)
	\end{align}
	we obtain
	\begin{align}
		&\langle O^{\lambda}_{s_1}(x_1)\ldots O^{\lambda}_{s_n}(x_n)\rangle \nonumber\\
		&=(N^2-1)\sum_{\sigma \in P_n}\frac{(-1)^{n+1}}{n}\frac{1}{2^n} \partial_{x_{\sigma(1)}^+}\Box^{-1}(x_{\sigma(1)}-x_{\sigma(2)})\mathcal{Y}_{s_{\sigma(2)}-1}^{\frac{3}{2}}(\overleftarrow{\partial}_{x_{\sigma(2)}^+},\overrightarrow{\partial}_{x_{\sigma(2)}^+}) \nonumber\\
		&\quad\ldots \partial_{x_{\sigma(n)}^+}\Box^{-1}(x_{\sigma(n)}-x_{\sigma(1)})\mathcal{Y}_{s_{\sigma(1)}-1}^{\frac{3}{2}}(\overleftarrow{\partial}_{x_{\sigma(1)}^+},\overrightarrow{\partial}_{x_{\sigma(1)}^+}) \,.
	\end{align}
	Employing the definition in Eq. (\ref{defY2}), we get
	\begin{align}
		&\langle O^{\lambda}_{s_1}(x_1)\ldots O^{\lambda}_{s_n}(x_n)\rangle \nonumber\\
		&=\frac{N^2-1}{2^n}\frac{(-1)^{n+1}}{n}i^{\sum_{l=1}^n s_l-n}\frac{(s_1+1)}{2}\ldots \frac{(s_n+1)}{2}\nonumber\\
		&\quad\sum_{k_1 = 0}^{s_1-1}{s_1\choose k_1}{s_1\choose k_1+1}(-1)^{s_1-k_1-1}\ldots \sum_{k_n = 0}^{s_n-1}{s_n\choose k_n}{s_n\choose k_n+1}(-1)^{s_n-k_n-1}\nonumber\\
		&\quad \sum_{\sigma \in P_n}\partial_{x_{\sigma(1)}^+}\Box^{-1}(x_{\sigma(1)}-x_{\sigma(2)})\overleftarrow{\partial}_{x^+_{\sigma(2)}}^{s_{\sigma(2)}-k_{\sigma(2)}-1}\overrightarrow{\partial}_{x^+_{\sigma(2)}}^{k_{\sigma(2)}}\nonumber\\
		&\quad\ldots \partial_{x_{\sigma(n)}^+}\Box^{-1}(x_{\sigma(n)}-x_{\sigma(1)})\overleftarrow{\partial}_{x^+_{\sigma(1)}}^{s_{\sigma(1)}-k_{\sigma(1)}-1}\overrightarrow{\partial}_{x^+_{\sigma(1)}}^{k_{\sigma(1)}}\,.
	\end{align}
	It follows from Eq. (\ref{propagator})
	\begin{align}
		&\langle O^{\lambda}_{s_1}(x_1)\ldots O^{\lambda}_{s_n}(x_n)\rangle\nonumber\\ &=\frac{N^2-1}{2^n}\frac{(-i)^n}{(4\pi^2)^n}\frac{(-1)^{n+1}}{n}i^{\sum_{l=1}^n s_l-n}\frac{(s_1+1)}{2}\ldots \frac{(s_n+1)}{2}\nonumber\\
		&\quad\sum_{k_1 = 0}^{s_1-1}{s_1\choose k_1}{s_1\choose k_1+1}(-1)^{s_1-k_1-1}\ldots \sum_{k_n = 0}^{s_n-1}{s_n\choose k_n}{s_n\choose k_n+1}(-1)^{s_n-k_n-1}\nonumber\\
		&\quad \sum_{\sigma \in P_n}\partial_{x_{\sigma(1)}^+}\frac{1}{\rvert x_{\sigma(1)}-x_{\sigma(2)} \rvert^2}\overleftarrow{\partial}_{x^+_{\sigma(2)}}^{s_{\sigma(2)}-k_{\sigma(2)}-1}\overrightarrow{\partial}_{x^+_{\sigma(2)}}^{k_{\sigma(2)}}\nonumber\\
		&\quad\ldots \partial_{x_{\sigma(n)}^+}\frac{1}{\rvert x_{\sigma(n)}-x_{\sigma(1)} \rvert^2}\overleftarrow{\partial}_{x^+_{\sigma(1)}}^{s_{\sigma(1)}-k_{\sigma(1)}-1}\overrightarrow{\partial}_{x^+_{\sigma(1)}}^{k_{\sigma(1)}}\,.
	\end{align}
	We now employ \cite{BPS1}
	\begin{align}
		\nonumber
		\label{doubleder}
		&\partial_{x^+}^{a}\partial_{y^+}^{b}\frac{1}{\rvert x-y\rvert^2} =\partial_{x^+}^{a}\partial_{y^+}^{b}\frac{1}{2 (x-y)_+(x-y)_--(x-y)^2_\perp} \\
		&=(-1)^{a} (a+b)!\,2^{a+b} \frac{(x-y)_+^{a+b}}{(\rvert x-y\rvert^2)^{a+b+1}}\,,
	\end{align}
	so that
	\begin{align}
		&\langle O^{\lambda}_{s_1}(x_1)\ldots O^{\lambda}_{s_n}(x_n)\rangle\nonumber\\ &=\frac{N^2-1}{2^n}\frac{(-i)^n}{(4\pi^2)^n}\frac{(-1)^{n+1}}{n}i^{\sum_{l=1}^n s_l-n}\frac{(s_1+1)}{2}\ldots \frac{(s_n+1)}{2}\nonumber\\
		&\quad\sum_{k_1 = 0}^{s_1-1}{s_1\choose k_1}{s_1\choose k_1+1}(-1)^{s_1-k_1-1}\ldots \sum_{k_n = 0}^{s_n-1}{s_n\choose k_n}{s_n\choose k_n+1}(-1)^{s_n-k_n-1}\nonumber\\
		& \quad\sum_{\sigma \in P_n}(-1)^{k_{\sigma(1)}+1}(s_{\sigma(2)}-k_{\sigma(2)}+k_{\sigma(1)})!2^{s_{\sigma(2)}-k_{\sigma(2)}+k_{\sigma(1)}}\frac{(x_{\sigma(1)}-x_{\sigma(2)})_+^{s_{\sigma(2)}-k_{\sigma(2)}+k_{\sigma(1)}}}{(\rvert x_{\sigma(1)}-x_{\sigma(2)}\rvert^2)^{s_{\sigma(2)}-k_{\sigma(2)}+k_{\sigma(1)}+1}}\nonumber\\
		&\quad\ldots(-1)^{k_{\sigma(n)}+1}(s_{\sigma(1)}-k_{\sigma(1)}+k_{\sigma(n)})!2^{s_{\sigma(1)}-k_{\sigma(1)}+k_{\sigma(n)}}\frac{(x_{\sigma(n)}-x_{\sigma(1)})_+^{s_{\sigma(1)}-k_{\sigma(1)}+k_{\sigma(n)}}}{(\rvert x_{\sigma(n)}-x_{\sigma(1)}\rvert^2)^{s_{\sigma(1)}-k_{\sigma(1)}+k_{\sigma(n)}+1}}\,.
	\end{align}
	Relabeling the permutations, $\sigma(n)\rightarrow\sigma(2)$,  $\sigma(n-1)\rightarrow\sigma(3)$ and so on, while keeping $\sigma(1)$ fixed, we obtain
	\begin{align}
		&\langle O^{\lambda}_{s_1}(x_1)\ldots O^{\lambda}_{s_n}(x_n)\rangle\nonumber\\ &=\frac{N^2-1}{2^n}\frac{(-i)^n}{(4\pi^2)^n}\frac{(-1)^{n+1}}{n}i^{\sum_{l=1}^n s_l-n}\frac{(s_1+1)}{2}\ldots \frac{(s_n+1)}{2}\nonumber\\
		&\quad\sum_{k_1 = 0}^{s_1-1}{s_1\choose k_1}{s_1\choose k_1+1}(-1)^{s_1-k_1-1}\ldots \sum_{k_n = 0}^{s_n-1}{s_n\choose k_n}{s_n\choose k_n+1}(-1)^{s_n-k_n-1}\nonumber\\
		&\quad \sum_{\sigma \in P_n}(-1)^{k_{\sigma(1)}+1}(s_{\sigma(n)}-k_{\sigma(n)}+k_{\sigma(1)})!2^{s_{\sigma(n)}-k_{\sigma(n)}+k_{\sigma(1)}}\frac{(x_{\sigma(1)}-x_{\sigma(n)})_+^{s_{\sigma(n)}-k_{\sigma(n)}+k_{\sigma(1)}}}{(\rvert x_{\sigma(1)}-x_{\sigma(n)}\rvert^2)^{s_{\sigma(n)}-k_{\sigma(n)}+k_{\sigma(1)}+1}}\nonumber\\
		&\quad\ldots(-1)^{k_{\sigma(2)}+1}(s_{\sigma(1)}-k_{\sigma(1)}+k_{\sigma(2)})!2^{s_{\sigma(1)}-k_{\sigma(1)}+k_{\sigma(2)}}\frac{(x_{\sigma(2)}-x_{\sigma(1)})_+^{s_{\sigma(1)}-k_{\sigma(1)}+k_{\sigma(2)}}}{(\rvert x_{\sigma(2)}-x_{\sigma(1)}\rvert^2)^{s_{\sigma(1)}-k_{\sigma(1)}+k_{\sigma(2)}+1}}
	\end{align}
	that simplifies to
	\begin{align}
		&\langle O^{\lambda}_{s_1}(x_1)\ldots O^{\lambda}_{s_n}(x_n)\rangle \nonumber\\
		&=-\frac{N^2-1}{2^n}\frac{1}{(4\pi^2)^n}i^{\sum_{l=1}^n s_l}2^{\sum_{l=1}^n s_l}\frac{(s_1+1)}{2}\ldots \frac{(s_n+1)}{2}\nonumber\\
		&\quad\sum_{k_1 = 0}^{s_1-1}{s_1\choose k_1}{s_1\choose k_1+1}\ldots \sum_{k_n = 0}^{s_n-1}{s_n\choose k_n}{s_n\choose k_n+1}\nonumber\\
		&\quad \frac{1}{n}\sum_{\sigma \in P_n}(s_{\sigma(1)}-k_{\sigma(1)}+k_{\sigma(2)})!\ldots(s_{\sigma(n)}-k_{\sigma(n)}+k_{\sigma(1)})!\nonumber\\
		&\quad\frac{(x_{\sigma(1)}-x_{\sigma(2)})_+^{s_{\sigma(1)}-k_{\sigma(1)}+k_{\sigma(2)}}}{(\rvert x_{\sigma(1)}-x_{\sigma(2)}\rvert^2)^{s_{\sigma(1)}-k_{\sigma(1)}+k_{\sigma(2)}+1}}\cdots\frac{(x_{\sigma(n)}-x_{\sigma(1)})_+^{s_{\sigma(n)}-k_{\sigma(n)}+k_{\sigma(1)}}}{(\rvert x_{\sigma(n)}-x_{\sigma(1)}\rvert^2)^{s_{\sigma(n)}-k_{\sigma(n)}+k_{\sigma(1)}+1}}\,.
	\end{align}
	The $2$-point correlators follow
	\begin{align}
		\label{two-pts_lambda}
		&\langle O^{\lambda}_{s_1}(x_1) O^{\lambda}_{s_2}(x_2)\rangle\nonumber\\ &=-\frac{N^2-1}{4}\frac{1}{(4\pi^2)^2}i^{s_1+s_2}2^{s_1+s_2}\frac{(s_1+1)(s_2+1)}{4}\nonumber\\
		&\quad \sum_{k_1 = 0}^{s_1-1}\sum_{k_2 = 0}^{s_2-1}{s_1\choose k_1}{s_1\choose k_1+1} {s_2\choose k_2}{s_2\choose k_2+1}(-1)^{s_2-k_2+k_1}(s_{1}-k_{1}+k_{2})!(s_{2}-k_{2}+k_{1})!\nonumber\\
		&\quad\frac{(x_{1}-x_{2})_+^{s_{1}+s_2}}{(\rvert x_{1}-x_{2}\rvert^2)^{s_{1}+s_2+2}}
	\end{align}
	that can be rewritten as
	\begin{align}
		\label{2pointlambda}
		&\langle O^{\lambda}_{s_1}(x_1) O^{\lambda}_{s_2}(x_2)\rangle\nonumber\\ &=-\frac{N^2-1}{4}\frac{1}{(4\pi^2)^2}i^{s_1+s_2}2^{s_1+s_2}\frac{(s_1+1)(s_2+1)}{4}\nonumber\\
		&\quad \sum_{k_1 = 0}^{s_1-1}\sum_{k_2 = 0}^{s_2-1}{s_1\choose k_1}{s_1\choose k_1+1} {s_2\choose k_2}{s_2\choose k_2+1}(-1)^{k_1+k_2+1}(s_{1}+s_2-k_{1}-k_{2}-1)!(k_{2}+k_{1}+1)!\nonumber\\
		&\quad\frac{(x_{1}-x_{2})_+^{s_{1}+s_2}}{(\rvert x_{1}-x_{2}\rvert^2)^{s_{1}+s_2+2}}
	\end{align}
	by substituting
	\begin{equation}
		k'_2 = s_2-k_2-1
	\end{equation}
	into Eq. \eqref{two-pts_lambda} and dropping the primed index. By employing (Appendix \ref{appGGresum})
	\begin{align}
		& \delta_{s_1s_2}\frac{s_1}{s_1+1}=\sum_{k_1 = 0}^{s_1-1}\sum_{k_2 = 0}^{s_2-1}{s_1\choose k_1}{s_1\choose k_1+1}{s_2\choose k_2}{s_2\choose k_2+1}(-1)^{k_1+k_2}\frac{1}{{s_1+s_2\choose k_1+k_2+1}}
	\end{align}
	the $2$-point correlators become
	\begin{align}
		&\langle O^{\lambda}_{s_1}(x_1) O^{\lambda}_{s_2}(x_2)\rangle =\delta_{s_1s_2}\frac{N^2-1}{4}\frac{1}{(4\pi^2)^2}(-1)^{s_1}2^{2s_1}(2s_1)!\frac{s_1(s_1+1)}{4}\frac{(x_{1}-x_{2})_+^{2s_1}}{(\rvert x_{1}-x_{2}\rvert^2)^{2s_1+2}}\,.
	\end{align}
	Similarly, we obtain the correlators of $O^\lambda$ and $\tilde{O}^\lambda$ from their definition
	\begin{align}
		&\langle O^\lambda_{s_1}(x_1)\ldots O^\lambda_{s_n}(x_n)\tilde{O}^\lambda_{s_{n+1}}(x_{n+1})\ldots \tilde{O}^\lambda_{s_{n+2m}}(x_{n+2m})\rangle =\nonumber\\
		& \frac{\delta}{\delta J_{O^\lambda_{s_1}}(x_1)}\cdots\frac{\delta}{\delta J_{O^\lambda_{s_n}}(x_n)}\frac{\delta}{\delta J_{\tilde{O}^\lambda_{s_{n+1}}}(x_{n+1})}\cdots\frac{\delta}{\delta J_{\tilde{O}^\lambda_{s_{n+2m}}}(x_{n+2m})}\mathcal{W}_{\text{conf}}\left[0,0,J_{O^\lambda},J_{\tilde{O}^\lambda},0,0\right]\nonumber\\
		&= \frac{1}{2} \frac{\delta}{\delta J_{O^\lambda_{s_1}}(x_1)}\cdots\frac{\delta}{\delta J_{O^\lambda_{s_n}}(x_n)}\frac{\delta}{\delta J_{\tilde{O}^\lambda_{s_{n+1}}}(x_{n+1})}\cdots\frac{\delta}{\delta J_{\tilde{O}^\lambda_{s_{n+2m}}}(x_{n+2m})}\nonumber\\
		&\quad\Bigg[\log\Det \Big(\mathcal{I}+\frac{1}{2}\partial_+\square^{-1}J_{O^\lambda_{s}}\otimes\mathcal{Y}_{s-1}^{\frac{3}{2}}+\frac{1}{2}\partial_+\square^{-1}J_{\tilde{O}^\lambda_{s}}\otimes\mathcal{H}_{s-1}^{\frac{3}{2}}  \Big) \nonumber\\
		&\quad\quad+\log\Det \Big(\mathcal{I}+\frac{1}{2}\partial_+\square^{-1}J_{O^\lambda_{s}}\otimes\mathcal{Y}_{s-1}^{\frac{3}{2}}-\frac{1}{2}\partial_+\square^{-1}J_{\tilde{O}^\lambda_{s}}\otimes\mathcal{H}_{s-1}^{\frac{3}{2}}   \Big)    \Bigg]
	\end{align}
	that yields
	\begin{align}
		&\langle O^\lambda_{s_1}(x_1)\ldots O^\lambda_{s_n}(x_n)\tilde{O}^\lambda_{s_{n+1}}(x_{n+1})\ldots \tilde{O}^\lambda_{s_{n+2m}}(x_{n+2m})\rangle \nonumber\\
		&=-\frac{N^2-1}{2^{n+2m}}\frac{1}{(4\pi^2)^{n+2m}}i^{\sum_{l=1}^{n+2m} s_l}2^{\sum_{l=1}^{n+2m} s_l}\frac{(s_1+1)}{2}\ldots \frac{(s_{n+2m}+1)}{2}\nonumber\\
		&\quad\sum_{k_1 = 0}^{s_1-1}{s_1\choose k_1}{s_1\choose k_1+1}\ldots \sum_{k_{n+2m} = 0}^{s_{n+2m}-1}{s_{n+2m}\choose k_{n+2m}}{s_{n+2m}\choose k_{n+2m}+1}\nonumber\\
		&\quad \frac{1}{{n+2m}}\sum_{\sigma \in P_{n+2m}}(s_{\sigma(1)}-k_{\sigma(1)}+k_{\sigma(2)})!\ldots(s_{\sigma({n+2m})}-k_{\sigma({n+2m})}+k_{\sigma(1)})!\nonumber\\
		&\quad\frac{(x_{\sigma(1)}-x_{\sigma(2)})_+^{s_{\sigma(1)}-k_{\sigma(1)}+k_{\sigma(2)}}}{(\rvert x_{\sigma(1)}-x_{\sigma(2)}\rvert^2)^{s_{\sigma(1)}-k_{\sigma(1)}+k_{\sigma(2)}+1}}\cdots\frac{(x_{\sigma({n+2m})}-x_{\sigma(1)})_+^{s_{\sigma({n+2m})}-k_{\sigma({n+2m})}+k_{\sigma(1)}}}{(\rvert x_{\sigma({n+2m})}-x_{\sigma(1)}\rvert^2)^{s_{\sigma({n+2m})}-k_{\sigma({n+2m})}+k_{\sigma(1)}+1}}\,.
	\end{align}

	\subsection{$M$ and $\bar{M}$ correlators}

	The correlators of gluon-gluino operators involve the same number of $M_s$ and $\bar{M}_s$, otherwise they vanish, as follows from the generating functional. Therefore,  the nonvanishing correlators read
	\begin{align}
		&\langle M_{s_1}(x_1)\bar{M}_{s'_1}(y_1) M_{s_2}(x_2)\bar{M}_{s'_2}(y_2)\ldots M_{s_n}(x_n) \bar{M}_{s'_n}(y_n)\rangle \nonumber\\
		&= \frac{\delta}{\delta \bar{J}_{M_{s_1}}(x_1)}\left(-\frac{\delta}{\delta J_{\bar M_{s'_1}}(y_1)}\right)\cdots\frac{\delta}{\delta \bar{J}_{M_{s_n}}(x_n)}\left(-\frac{\delta}{\delta J_{\bar M_{s'_n}}(y_n)}\right)\mathcal{W}_{\text{conf}}\left[0,0,J_{\bar M},\bar{J}_{M}\right]\nonumber\\
		&= \frac{\delta}{\delta \bar{J}_{M_{s_1}}(x_1)}\left(-\frac{\delta}{\delta J_{\bar M_{s'_1}}(y_1)}\right)\cdots\frac{\delta}{\delta \bar{J}_{M_{s_n}}(x_n)}\left(-\frac{\delta}{\delta J_{\bar M_{s'_n}}(y_n)}\right)\nonumber\\
		&\quad\frac{1}{2}\Bigg[\log\Det \Big(\mathcal{I}+\frac{1}{4} i\partial_+i\square^{-1}\bar{J}_{M_{s_1}}\otimes\mathcal{G}_{s_1-1}^{(1,2)}(-1)^{s_1-1}i\square^{-1} J_{\bar M_{s_3}}\otimes\mathcal{G}_{s_2-1}^{(2,1)}(-1)^{s_2-1} \Big)\nonumber\\
		&\quad+\log\Det \Big(\mathcal{I}  + \frac{1}{4}i\partial_+i\square^{-1}J_{\bar M_{s_1}}\otimes\mathcal{G}_{s_1-1}^{(1,2)} i\square^{-1}\bar{J}_{M_{s_3}}\otimes\mathcal{G}_{s_3-1}^{(2,1)}\Big)\Bigg]\,.\nonumber\\
	\end{align}
	The two determinants above are equal (appendix \ref{appM})
	\begin{align}
		&\log\Det \Big(\mathcal{I}+\frac{1}{4} i\partial_+i\square^{-1}\bar{J}_{M_{s_1}}\otimes\mathcal{G}_{s_1-1}^{(1,2)}(-1)^{s_1-1}i\square^{-1} J_{\bar M_{s_3}}\otimes\mathcal{G}_{s_3-1}^{(2,1)}(-1)^{s_3-1} \Big)\nonumber\\
		&=\log\Det \Big(\mathcal{I}  + \frac{1}{4}i\partial_+i\square^{-1}J_{\bar M_{s_1}}\otimes\mathcal{G}_{s_1-1}^{(1,2)} i\square^{-1}\bar{J}_{M_{s_3}}\otimes\mathcal{G}_{s_3-1}^{(2,1)}\Big)\label{2det}\,,
	\end{align}
	so that
	\begin{align}
		&\langle M_{s_1}(x_1)\bar{M}_{s'_1}(y_1) M_{s_2}(x_2)\bar{M}_{s'_2}(y_2)\ldots M_{s_n}(x_n) \bar{M}_{s'_n}(y_n)\rangle \nonumber\\
		&= \frac{\delta}{\delta \bar{J}_{M_{s_1}}(x_1)}\left(-\frac{\delta}{\delta J_{\bar M_{s'_1}}(y_1)}\right)\cdots\frac{\delta}{\delta \bar{J}_{M_{s_n}}(x_n)}\left(-\frac{\delta}{\delta J_{\bar M_{s'_n}}(y_n)}\right)\mathcal{W}_{\text{conf}}\left[0,0,J_{\bar M},\bar{J}_{M}\right]\nonumber\\
		&= \frac{\delta}{\delta \bar{J}_{M_{s_1}}(x_1)}\left(-\frac{\delta}{\delta J_{\bar M_{s'_1}}(y_1)}\right)\cdots\frac{\delta}{\delta \bar{J}_{M_{s_n}}(x_n)}\left(-\frac{\delta}{\delta J_{\bar M_{s'_n}}(y_n)}\right)\nonumber\\
		&\quad\log\Det \Big(\mathcal{I}  + \frac{1}{4}i\partial_+i\square^{-1}J_{\bar M_{s_1}}\otimes\mathcal{G}_{s_1-1}^{(1,2)} i\square^{-1}\bar{J}_{M_{s_2}}\otimes\mathcal{G}_{s_2-1}^{(2,1)}\Big)\nonumber\\
		&=\frac{\delta}{\delta \bar{J}_{M_{s_1}}(x_1)}\left(-\frac{\delta}{\delta J_{\bar M_{s'_1}}(y_1)}\right)\cdots\frac{\delta}{\delta \bar{J}_{M_{s_n}}(x_n)}\left(-\frac{\delta}{\delta J_{\bar M_{s'_n}}(y_n)}\right)\nonumber\\
		&\quad(N^2-1)\sum_{l=1}^{\infty}\frac{(-1)^{l+1}}{l}\frac{1}{2^{2l}}\int d^4u_1\ldots d^4u_ld^4v_1\ldots d^4v_l \sum_{s_1}\ldots \sum_{s_l}\sum_{s'_1}\ldots\sum_{s'_l} \nonumber\\
		&\quad i\partial_{u_1^+}i\Box^{-1}(u_1-v_1)
		\mathcal{G}_{s'_1-1}^{(1,2)}(\overleftarrow{\partial}_{v_1^+},\overrightarrow{\partial}_{v_1^+})\otimes J_{\bar M_{s'_1}}(v_1)\nonumber\\
		&\quad i\Box^{-1}(v_1-u_2) \mathcal{G}_{s_2-1}^{(2,1)}(\overleftarrow{\partial}_{u_2^+},\overrightarrow{\partial}_{u_2^+})\otimes \bar{J}_{M_{s_2}}(u_2)\nonumber\\
		&\quad\ldots i\partial_{u_l^+}i\Box^{-1}(u_l-v_l)\mathcal{G}_{s'_l-1}^{(1,2)}(\overleftarrow{\partial}_{v_l^+},\overrightarrow{\partial}_{v_l^+})\otimes J_{\bar M_{s'_l}}(v_l)\nonumber\\
		&\quad i\Box^{-1}(v_l-u_1) \mathcal{G}_{s_1-1}^{(2,1)}(\overleftarrow{\partial}_{u_1^+},\overrightarrow{\partial}_{u_1^+})\otimes \bar{J}_{M_{s_1}}(u_1)\,.
	\end{align}
	Performing the functional derivatives, we obtain
	\begin{align}
		&\langle M_{s_1}(x_1)\bar{M}_{s'_1}(y_1) M_{s_2}(x_2)\bar{M}_{s'_2}(y_2)\ldots M_{s_n}(x_n) \bar{M}_{s'_n}(y_n)\rangle \nonumber\\
		&= (N^2-1)\sum_{\sigma\in P_n}\sum_{\rho\in P_n}\frac{(-1)^{2n+1}}{n}\frac{1}{2^{2n}}\text{sgn}(\sigma)\text{sgn}(\rho)\nonumber\\
		& \quad i\partial_{x_{\sigma(1)}^+}i\Box^{-1}(x_{\sigma(1)}-y_{\rho(1)})
		\mathcal{G}_{s'_{\rho(1)}-1}^{(1,2)}(\overleftarrow{\partial}_{y_{\rho(1)}^+},\overrightarrow{\partial}_{y_{\rho(1)}^+})\nonumber\\
		&\quad i\Box^{-1}(y_{\rho(1)}-x_{\sigma(2)}) \mathcal{G}_{s_{\sigma(2)}-1}^{(2,1)}(\overleftarrow{\partial}_{x_{\sigma(2)}^+},\overrightarrow{\partial}_{x_{\sigma(2)}^+})\nonumber\\
		&\quad\ldots i\partial_{x_{\sigma(n)}^+}i\Box^{-1}(x_{\sigma(n)}-y_{\rho(n)})\mathcal{G}_{s'_{\rho(n)}-1}^{(1,2)}(\overleftarrow{\partial}_{y_{\rho(n)}^+},\overrightarrow{\partial}_{y_{\rho(n)}^+})\nonumber\\
		&\quad i\Box^{-1}(y_{\rho(n)}-x_{\sigma(1)}) \mathcal{G}_{s_{\sigma(1)}-1}^{(2,1)}(\overleftarrow{\partial}_{x_{\sigma(1)}^+},\overrightarrow{\partial}_{x_{\sigma(1)}^+})\,,
	\end{align}
	where $\text{sgn}(\sigma)$ denotes the sign of the permutation $\sigma$. Substituting Eq.\eqref{propagator} in the above equation, we get
	\begin{align}
		&\langle M_{s_1}(x_1)\bar{M}_{s'_1}(y_1) M_{s_2}(x_2)\bar{M}_{s'_2}(y_2)\ldots M_{s_n}(x_n) \bar{M}_{s'_n}(y_n)\rangle \nonumber\\
		&= (N^2-1)\frac{i^n}{(4\pi^2)^{2n}}\sum_{\sigma\in P_n}\sum_{\rho\in P_n}\frac{(-1)^{2n+1}}{n}\frac{1}{2^{2n}}\text{sgn}(\sigma)\text{sgn}(\rho)\nonumber\\
		&\quad\partial_{x_{\sigma(1)}^+}\frac{1}{\rvert x_{\sigma(1)}-y_{\rho(1)}\rvert^2}	\mathcal{G}_{s'_{\rho(1)}-1}^{(1,2)}(\overleftarrow{\partial}_{y_{\rho(1)}^+},\overrightarrow{\partial}_{y_{\rho(1)}^+})\frac{1}{\rvert y_{\rho(1)}-x_{\sigma(2)}\rvert^2}\mathcal{G}_{s_{\sigma(1)}-1}^{(2,1)}(\overleftarrow{\partial}_{x_{\sigma(2)}^+},\overrightarrow{\partial}_{x_{\sigma(2)}^+})\nonumber\\
		&\quad\ldots \partial_{x_{\sigma(n)}^+}\frac{1}{\rvert x_{\sigma(n)}-y_{\rho(n)}\rvert^2}\mathcal{G}_{s'_{\rho(n)}-1}^{(1,2)}(\overleftarrow{\partial}_{y_{\rho(n)}^+},\overrightarrow{\partial}_{y_{\rho(n)}^+})\frac{1}{\rvert y_{\rho(n)}-x_{\sigma(1)}\rvert^2} \mathcal{G}_{s_{\sigma(1)}-1}^{(2,1)}(\overleftarrow{\partial}_{x_{\sigma(1)}^+},\overrightarrow{\partial}_{x_{\sigma(1)}^+})\,.
	\end{align}
	Employing Eqs. \eqref{gdef1} and \eqref{gdef2}
	\begin{align}
		&\langle M_{s_1}(x_1)\bar{M}_{s'_1}(y_1) M_{s_2}(x_2)\bar{M}_{s'_2}(y_2)\ldots M_{s_n}(x_n) \bar{M}_{s'_n}(y_n)\rangle  \nonumber\\
		&=(N^2-1)\frac{i^n}{(4\pi^2)^{2n}}i^{s'_1-1} \sum_{k'_1 = 0}^{s'_1-1}{s'_1\choose k'_1}{s'_1+1\choose k'_1+2} (-1)^{s'_1-k'_1-1}\ldots i^{s'_n-1} \sum_{k'_n = 0}^{s'_n-1}{s'_n\choose k'_n}{s'_n+1\choose k'_n+2} (-1)^{s'_n-k'_n-1}\nonumber\\
		&\quad i^{s_1-1}\sum_{k_1 = 0}^{s_1-1}{s_1+1\choose k_1}{s_1\choose k_1+1} (-1)^{s_1-k_1-1}\ldots i^{s_n-1}\sum_{k_n = 0}^{s_n-1}{s_n+1\choose k_n}{s_n\choose k_n+1} (-1)^{s_n-k_n-1}\nonumber\\
		&\quad\sum_{\sigma\in P_n}\sum_{\rho\in P_n}\frac{(-1)^{2n+1}}{n}\frac{1}{2^{2n}}\text{sgn}(\sigma)\text{sgn}(\rho)\nonumber\\
		&\quad\partial_{x_{\sigma(1)}^+}\frac{1}{\rvert x_{\sigma(1)}-y_{\rho(1)}\rvert^2}\overleftarrow{\partial}_{y^+_{\rho(1)}}^{s'_{\rho(1)}-k'_{\rho(1)}-1}\overrightarrow{\partial}_{y^+_{\rho(1)}}^{k'_{\rho(1)}+1}\frac{1}{\rvert y_{\rho(1)}-x_{\sigma(2)}\rvert^2}\overleftarrow{\partial}_{x^+_{\sigma(2)}}^{s_{\sigma(2)-}k_{\sigma(2)}}\overrightarrow{\partial}_{x^+_{\sigma(2)}}^{k_{\sigma(2)}} \nonumber\\
		&\quad\ldots\partial_{x_{\sigma(n)}^+}\frac{1}{\rvert x_{\sigma(n)}-y_{\rho(n)}\rvert^2}\overleftarrow{\partial}_{y^+_{\rho(n)}}^{s'_{\rho(n)}-k'_{\rho(n)}-1}\overrightarrow{\partial}_{y^+_{\rho(n)}}^{k'_{\rho(n)}+1}\frac{1}{\rvert y_{\rho(n)}-x_{\sigma(1)}\rvert^2}\overleftarrow{\partial}_{x^+_{\sigma(1)}}^{s_{\sigma(1)-}k_{\sigma(1)}}\overrightarrow{\partial}_{x^+_{\sigma(1)}}^{k_{\sigma(1)}}
	\end{align}
	and Eq. (\ref{doubleder}), we obtain
	\begin{align}
		&\langle M_{s_1}(x_1)\bar{M}_{s'_1}(y_1) M_{s_2}(x_2)\bar{M}_{s'_2}(y_2)\ldots M_{s_n}(x_n) \bar{M}_{s'_n}(y_n)\rangle \nonumber\\
		&= (N^2-1)\frac{i^n}{(4\pi^2)^{2n}}
		i^{s'_1-1} \sum_{k'_1 = 0}^{s'_1-1}{s'_1\choose k'_1}{s'_1+1\choose k'_1+2} (-1)^{s'_1-k'_1-1}\ldots i^{s'_n-1} \sum_{k'_n = 0}^{s'_n-1}{s'_n\choose k'_n}{s'_n+1\choose k'_n+2} (-1)^{s'_n-k'_n-1}\nonumber\\
		&\quad i^{s_1-1}\sum_{k_1 = 0}^{s_1-1}{s_1+1\choose k_1}{s_1\choose k_1+1} (-1)^{s_1-k_1-1}\ldots i^{s_n-1}\sum_{k_n = 0}^{s_n-1}{s_n+1\choose k_n}{s_n\choose k_n+1} (-1)^{s_n-k_n-1}\nonumber\\
		&\quad\sum_{\sigma\in P_n}\sum_{\rho\in P_n}\frac{(-1)^{2n+1}}{n}\frac{1}{2^{2n}}\text{sgn}(\sigma)\text{sgn}(\rho)\nonumber\\
		&\quad(-1)^{k_{\sigma(1)}+1}2^{s'_{\rho(1)}-k'_{\rho(1)}+k_{\sigma(1)}}(s'_{\rho(1)}-k'_{\rho(1)}+k_{\sigma(1)})! \frac{(x_{\sigma(1)}-y_{\rho(1)})_+^{s'_{\rho(1)}-k'_{\rho(1)}+k_{\sigma(1)}}}{(\rvert x_{\sigma(1)}-y_{\rho(1)}\rvert^2)^{s'_{\rho(1)}-k'_{\rho(1)}+k_{\sigma(1)}+1}}\nonumber\\
		&\quad(-1)^{k'_{\rho(1)}+1}2^{s_{\sigma(2)}-k_{\sigma(2)}+k'_{\rho(1)}+1}(s_{\sigma(2)}-k_{\sigma(2)}+k'_{\rho(1)}+1)! \frac{(y_{\rho(1)}-x_{\sigma(2)})_+^{s_{\sigma(2)}-k_{\sigma(2)}+k'_{\rho(1)}+1}}{(\rvert y_{\rho(1)}-x_{\sigma(2)}\rvert^2)^{s_{\sigma(2)}-k_{\sigma(2)}+k'_{\rho(1)}+2}}\nonumber\\
		&\quad\ldots\nonumber\\
		&\quad(-1)^{k_{\sigma(n)+1}}2^{s'_{\rho(n)}-k'_{\rho(n)}+k_{\sigma(n)}}(s'_{\rho(n)}-k'_{\rho(n)}+k_{\sigma(n)})! \frac{(x_{\sigma(n)}-y_{\rho(n)})_+^{s'_{\rho(n)}-k'_{\rho(n)}+k_{\sigma(n)}}}{(\rvert x_{\sigma(n)}-y_{\rho(n)}\rvert^2)^{s'_{\rho(n)}-k'_{\rho(n)}+k_{\sigma(n)}+1}}\nonumber\\
		&\quad(-1)^{k'_{\rho(n)}+1}2^{s_{\sigma(1)}-k_{\sigma(1)}+k'_{\rho(n)}+1}(s_{\sigma(1)}-k_{\sigma(1)}+k'_{\rho(n)}+1)! \frac{(y_{\rho(n)}-x_{\sigma(1)})_+^{s_{\sigma(1)}-k_{\sigma(1)}+k'_{\rho(n)}+1}}{(\rvert y_{\rho(n)}-x_{\sigma(1)}\rvert^2)^{s_{\sigma(1)}-k_{\sigma(1)}+k'_{\rho(n)}+2}}
	\end{align}
	that simplifies to
	\begin{align}
		&\langle M_{s_1}(x_1)\bar{M}_{s'_1}(y_1) M_{s_2}(x_2)\bar{M}_{s'_2}(y_2)\ldots M_{s_n}(x_n) \bar{M}_{s'_n}(y_n)\rangle \nonumber\\
		&= (N^2-1)\frac{1}{(4\pi^2)^{2n}}2^{\sum_{l=1}^n s_l+\sum_{l=1}^n s'_{l}-n}i^{\sum_{l=1}^n s_l+\sum_{l=1}^n s'_{l}-n}(-1)^{\sum_{l=1}^n s_l+\sum_{l=1}^n s'_{l}}\nonumber\\
		&\quad\sum_{k'_1 = 0}^{s'_1-1}{s'_1\choose k'_1}{s'_1+1\choose k'_1+2} \ldots  \sum_{k'_n = 0}^{s'_n-1}{s'_n\choose k'_n}{s'_n+1\choose k'_n+2} \nonumber\\
		&\quad\sum_{k_1 = 0}^{s_1-1}{s_1+1\choose k_1}{s_1\choose k_1+1} \ldots \sum_{k_n = 0}^{s_n-1}{s_n+1\choose k_n}{s_n\choose k_n+1} \nonumber\\
		&\quad\sum_{\sigma\in P_n}\sum_{\rho\in P_n}\frac{(-1)^{2n+1}}{n}\text{sgn}(\sigma)\text{sgn}(\rho)\nonumber\\
		&\quad(s'_{\rho(1)}-k'_{\rho(1)}+k_{\sigma(1)})! \frac{(x_{\sigma(1)}-y_{\rho(1)})_+^{s'_{\rho(1)}-k'_{\rho(1)}+k_{\sigma(1)}}}{(\rvert x_{\sigma(1)}-y_{\rho(1)}\rvert^2)^{s'_{\rho(1)}-k'_{\rho(1)}+k_{\sigma(1)}+1}}\nonumber\\
		&\quad(s_{\sigma(2)}-k_{\sigma(2)}+k'_{\rho(1)}+1)! \frac{(y_{\rho(1)}-x_{\sigma(2)})_+^{s_{\sigma(2)}-k_{\sigma(2)}+k'_{\rho(1)}+1}}{(\rvert y_{\rho(1)}-x_{\sigma(2)}\rvert^2)^{s_{\sigma(2)}-k_{\sigma(2)}+k'_{\rho(1)}+2}}\nonumber\\
		&\quad\ldots\nonumber\\
		&\quad(s'_{\rho(n)}-k'_{\rho(n)}+k_{\sigma(n)})! \frac{(x_{\sigma(n)}-y_{\rho(n)})_+^{s'_{\rho(n)}-k'_{\rho(n)}+k_{\sigma(n)}}}{(\rvert x_{\sigma(n)}-y_{\rho(n)}\rvert^2)^{s'_{\rho(n)}-k'_{\rho(n)}+k_{\sigma(n)}+1}}\nonumber\\
		&\quad(s_{\sigma(1)}-k_{\sigma(1)}+k'_{\rho(n)}+1)! \frac{(y_{\rho(n)}-x_{\sigma(1)})_+^{s_{\sigma(1)}-k_{\sigma(1)}+k'_{\rho(n)}+1}}{(\rvert y_{\rho(n)}-x_{\sigma(1)}\rvert^2)^{s_{\sigma(1)}-k_{\sigma(1)}+k'_{\rho(n)}+2}}\,.
	\end{align}
	By substituting
	\begin{equation}
		k_{\sigma(i)}\rightarrow s_{\sigma(i)}-1-k_{\sigma(i)}\qquad\qquad k'_{\rho(i)}\rightarrow s'_{\rho(i)}-1-k'_{\rho(i)}\,,
	\end{equation}
	we get
	\begin{align}
		&\langle M_{s_1}(x_1)\bar{M}_{s'_1}(y_1) M_{s_2}(x_2)\bar{M}_{s'_2}(y_2)\ldots M_{s_n}(x_n) \bar{M}_{s'_n}(y_n)\rangle \nonumber\\
		&= (N^2-1)\frac{1}{(4\pi^2)^{2n}}2^{\sum_{l=1}^n s_l+\sum_{l=1}^n s'_{l}-n}i^{\sum_{l=1}^n s_l+\sum_{l=1}^n s'_{l}-n}(-1)^{\sum_{l=1}^n s_l+\sum_{l=1}^n s'_{l}}\nonumber\\
		&\quad\sum_{k'_1 = 0}^{s'_1-1}{s'_1+1\choose k'_1}{s'_1\choose k'_1+1} \ldots \sum_{k'_n = 0}^{s'_n-1}{s'_n+1\choose k'_n}{s'_n\choose k'_n+1} \nonumber\\
		&\quad\sum_{k_1 = 0}^{s_1-1}{s_1\choose k_1}{s_1+1\choose k_1+2} \ldots  \sum_{k_n = 0}^{s_n-1}{s_n\choose k_n}{s_n+1\choose k_n+2} \nonumber\\
		&\quad\sum_{\sigma\in P_n}\sum_{\rho\in P_n}\frac{(-1)^{2n+1}}{n}\text{sgn}(\sigma)\text{sgn}(\rho)\nonumber\\
		&\quad(s_{\sigma(1)}-k_{\sigma(1)}+k'_{\rho(1)})! \frac{(x_{\sigma(1)}-y_{\rho(1)})_+^{s_{\sigma(1)}-k_{\sigma(1)}+k'_{\rho(1)}}}{(\rvert x_{\sigma(1)}-y_{\rho(1)}\rvert^2)^{s_{\sigma(1)}-k_{\sigma(1)}+k'_{\rho(1)}+1}}\nonumber\\
		&\quad(s'_{\rho(1)}-k'_{\rho(1)}+k_{\sigma(2)}+1)! \frac{(y_{\rho(1)}-x_{\sigma(2)})_+^{s'_{\rho(1)}-k'_{\rho(1)}+k_{\sigma(2)}+1}}{(\rvert y_{\rho(1)}-x_{\sigma(2)}\rvert^2)^{s'_{\rho(1)}-k'_{\rho(1)}+k_{\sigma(2)}+2}}\nonumber\\
		&\quad\ldots\nonumber\\
		&\quad(s_{\sigma(n)}-k_{\sigma(n)}+k'_{\rho(n)})! \frac{(x_{\sigma(n)}-y_{\rho(n)})_+^{s_{\sigma(n)}-k_{\sigma(n)}+k'_{\rho(n)}}}{(\rvert x_{\sigma(n)}-y_{\rho(n)}\rvert^2)^{s_{\sigma(n)}-k_{\sigma(n)}+k'_{\rho(n)}+1}}\nonumber\\
		&\quad(s'_{\rho(n)}-k'_{\rho(n)}+k_{\sigma(1)}+1)! \frac{(y_{\rho(n)}-x_{\sigma(1)})_+^{s'_{\rho(n)}-k'_{\rho(n)}+k_{\sigma(1)}+1}}{(\rvert y_{\rho(n)}-x_{\sigma(1)}\rvert^2)^{s'_{\rho(n)}-k'_{\rho(n)}+k_{\sigma(1)}+2}}\,.
	\end{align}
	The $2$-point correlators read
	\begin{align}
		\langle M_{s_1}(x)\bar{M}_{s_2}(y)\rangle
		&= i(N^2-1)\frac{1}{(4\pi^2)^{2}}2^{s_1+s_2-1}i^{s_1+s_2}(-1)^{s_1+s_2}\nonumber\\
		&\quad\sum_{k_1 = 0}^{s_1-1}\sum_{k_2 = 0}^{s_2-1}{s_1\choose k_1}{s_1+1\choose k_1+2}{s_2+1\choose k_2}{s_2\choose k_2+1}(s_{1}-k_{1}+k_2)!(s_2-k_2+k_{1}+1)! \nonumber\\
		& \quad\frac{(x-y)_+^{s_{1}-k_{1}+k_2}}{(\rvert x-y\rvert^2)^{s_{1}-k_{1}+k_2+1}}\frac{(y-x)_+^{s_2-k_2+k_{1}+1}}{(\rvert y-x\rvert^2)^{s_2-k_2+k_{1}+2}}
	\end{align}
	that become
	\begin{align}
		\label{twopointmix}
		\langle M_{s_1}(x)\bar{M}_{s_2}(y)\rangle
		&= -i(N^2-1)\frac{1}{(4\pi^2)^{2}}2^{s_1+s_2-1}i^{s_1+s_2}\nonumber\\
		&\quad\sum_{k_1 = 0}^{s_1-1}\sum_{k_2 = 0}^{s_2-1}{s_1\choose k_1}{s_1+1\choose k_1+2}{s_2+1\choose k_2}{s_2\choose k_2+1}(s_{1}-k_{1}+k_2)!(s_2-k_2+k_{1}+1)!\nonumber\\
		&\quad (-1)^{s_1-k_2+k_{1}}\frac{(x-y)_+^{s_{1}+s_2+1}}{(\rvert x-y\rvert^2)^{s_{1}+s_2+3}}\,.
	\end{align}
	Finally, employing (Appendix \ref{appppintegral})
	\begin{align}
		&-\delta_{s_1s_2}\frac{s_1}{s_1+2}=\sum_{k_1 = 0}^{s_1-1}\sum_{k_2 = 0}^{s_2-1}{s_1\choose k_1}{s_1+1\choose k_1+2}{s_2+1\choose k_2}{s_2\choose k_2+1}(-1)^{s_1-k_2+k_1}\frac{1}{{s_1+s_2+1\choose s_1-k_1+k_2}}
	\end{align}
	we obtain
	\begin{align}
		\label{twopointmix2}
		&\langle M_{s_1}(x)\bar{M}_{s_2}(y)\rangle= \delta_{s_1s_2}i(N^2-1)\frac{1}{(4\pi^2)^{2}}2^{2s_1-1}(-1)^{s_1}\frac{s_1}{s_1+2}(2s_1+1)!\frac{(x-y)^{2s_1+1}}{(\rvert x-y\rvert^2)^{2s_1+3}}\,.
	\end{align}

	\subsection{Vanishing correlators}
	
	By inspecting Eqs. \eqref{wgen1} and \eqref{wgen2}, the mixed conformal correlators of $O^A$, $\tilde{O}^A$ and $O^{\lambda}$, $\tilde{O}^\lambda$ vanish for $n,m\geq 1$ and $l,k \geq 1$
	\begin{align}
		&\langle O^A_{s_1}(x_1)\ldots O^A_{s_n}(x_n)\tilde{O}^A_{s'_1}(y_1)\ldots \tilde{O}^A_{s'_m}(y_m)  O^{\lambda}_{s''_1}(z_1)\ldots O^{\lambda}_{s''_l}(z_l)\tilde{O}^{\lambda}_{s'''_1}(t_1)\ldots \tilde{O}^{\lambda}_{s'''_k}(t_k)\rangle = 0\,.
	\end{align}
	They also vanish for $l,k=0$ and $m$ odd or $n,m =0$ and $k$ odd.

	\section{Euclidean conformal correlators} \label{npointE}

	We work out a few examples of the Euclidean $n$-point correlators. 
	
	\subsection{$O^{A\,E}$ correlators}
	
	The $O^{A\,E}$ correlators read
	\begin{align}
		& \langle O^{A\,E}_{s_1}(x_1)\ldots O^{A\,E}_{s_n}(x_n)\rangle \nonumber\\
		&=\frac{1}{(4\pi^2)^n}\frac{N^2-1}{2^n}2^{\sum_{l=1}^{n}s_l} (-1)^{\sum_{l=1}^{n}s_l}\frac{2(s_1+1)(s_1+2)}{4!}\ldots \frac{2(s_n+1)(s_n+2)}{4!}\nonumber\\
		&\quad\sum_{k_1=0}^{s_1-2} {s_1\choose k_1}{s_1\choose k_1+2}\ldots \sum_{k_n=0}^{s_n-2} {s_n\choose k_n}{s_n\choose k_n+2}\nonumber\\
		&\quad\frac{1}{n}\sum_{\sigma\in P_n}(s_{\sigma(1)}-k_{\sigma(1)}+k_{\sigma(2)})!\ldots(s_{\sigma(n)}-k_{\sigma(n)}+k_{\sigma(1)})!\nonumber\\
		&\quad\frac{(x_{\sigma(1)}-x_{\sigma(2)})_z^{s_{\sigma(1)}-k_{\sigma(1)}+k_{\sigma(2)}}}{(( x_{\sigma(1)}-x_{\sigma(2)})^2)^{s_{\sigma(1)}-k_{\sigma(1)}+k_{\sigma(2)}+1}}\cdots\frac{(x_{\sigma(n)}-x_{\sigma(1)})_z^{s_{\sigma(n)}-k_{\sigma(n)}+k_{\sigma(1)}}}{(( x_{\sigma(n)}-x_{\sigma(1)})^2)^{s_{\sigma(n)}-k_{\sigma(n)}+k_{\sigma(1)}+1}}\,.
	\end{align}
	
	\subsection{$O^{\lambda\,E}$ correlators}
	
	The $O^{\lambda\,E}$ correlators read
	\begin{align}
		&\langle O^{\lambda\,E}_{s_1}(x_1)\ldots O^{\lambda\,E}_{s_n}(x_n)\rangle \nonumber\\
		&=-\frac{N^2-1}{2^n}\frac{1}{(4\pi^2)^n}(-1)^{\sum_{l=1}^n s_l}2^{\sum_{l=1}^n s_l}\frac{(s_1+1)}{2}\ldots \frac{(s_n+1)}{2}\nonumber\\
		&\quad\sum_{k_1 = 0}^{s_1-1}{s_1\choose k_1}{s_1\choose k_1+1}\ldots \sum_{k_n = 0}^{s_n-1}{s_n\choose k_n}{s_n\choose k_n+1}\nonumber\\
		&\quad \frac{(-1)^n}{n}\sum_{\sigma \in P_n}(s_{\sigma(1)}-k_{\sigma(1)}+k_{\sigma(2)})!\ldots(s_{\sigma(n)}-k_{\sigma(n)}+k_{\sigma(1)})!\nonumber\\
		&\quad\frac{(x_{\sigma(1)}-x_{\sigma(2)})_z^{s_{\sigma(1)}-k_{\sigma(1)}+k_{\sigma(2)}}}{(( x_{\sigma(1)}-x_{\sigma(2)})^2)^{s_{\sigma(1)}-k_{\sigma(1)}+k_{\sigma(2)}+1}}\cdots\frac{(x_{\sigma(n)}-x_{\sigma(1)})_z^{s_{\sigma(n)}-k_{\sigma(n)}+k_{\sigma(1)}}}{(( x_{\sigma(n)}-x_{\sigma(1)})^2)^{s_{\sigma(n)}-k_{\sigma(n)}+k_{\sigma(1)}+1}}\,.
	\end{align}
	
	\subsection{$M^E$ and ${\bar M}^E$ correlators}
	
	The $2n$-point correlators of $M^E$ and ${\bar M}^E$ read
	\begin{align}
		&\langle M^E_{s_1}(x_1)\bar{M}^E_{s'_1}(y_1) M^E_{s_2}(x_2)\bar{M}^E_{s'_2}(y_2)\ldots M^E_{s_n}(x_n) \bar{M}^E_{s'_n}(y_n)\rangle \nonumber\\
		&= (N^2-1)\frac{1}{(4\pi^2)^{2n}}2^{\sum_{l=1}^n s_l+\sum_{l=1}^n s'_{l}-n}\nonumber\\
		&\quad\sum_{k'_1 = 0}^{s'_1-1}{s'_1+1\choose k'_1}{s'_1\choose k'_1+1} \ldots \sum_{k'_n = 0}^{s'_n-1}{s'_n+1\choose k'_n}{s'_n\choose k'_n+1} \nonumber\\
		&\quad\sum_{k_1 = 0}^{s_1-1}{s_1\choose k_1}{s_1+1\choose k_1+2} \ldots  \sum_{k_n = 0}^{s_n-1}{s_n\choose k_n}{s_n+1\choose k_n+2} \nonumber\\
		&\quad\sum_{\sigma\in P_n}\sum_{\rho\in P_n}\frac{(-1)^{2n+1}}{n}\text{sgn}(\sigma)\text{sgn}(\rho)\nonumber\\
		&\quad(s_{\sigma(1)}-k_{\sigma(1)}+k'_{\rho(1)})! \frac{(x_{\sigma(1)}-y_{\rho(1)})_z^{s_{\sigma(1)}-k_{\sigma(1)}+k'_{\rho(1)}}}{(( x_{\sigma(1)}-y_{\rho(1)})^2)^{s_{\sigma(1)}-k_{\sigma(1)}+k'_{\rho(1)}+1}}\nonumber\\
		&\quad(s'_{\rho(1)}-k'_{\rho(1)}+k_{\sigma(2)}+1)! \frac{(y_{\rho(1)}-x_{\sigma(2)})_z^{s'_{\rho(1)}-k'_{\rho(1)}+k_{\sigma(2)}+1}}{(( y_{\rho(1)}-x_{\sigma(2)})^2)^{s'_{\rho(1)}-k'_{\rho(1)}+k_{\sigma(2)}+2}}\nonumber\\
		&\quad\ldots\nonumber\\
		&\quad(s_{\sigma(n)}-k_{\sigma(n)}+k'_{\rho(n)})! \frac{(x_{\sigma(n)}-y_{\rho(n)})_z^{s_{\sigma(n)}-k_{\sigma(n)}+k'_{\rho(n)}}}{(( x_{\sigma(n)}-y_{\rho(n)})^2)^{s_{\sigma(n)}-k_{\sigma(n)}+k'_{\rho(n)}+1}}\nonumber\\
		&\quad(s'_{\rho(n)}-k'_{\rho(n)}+k_{\sigma(n)}+1)! \frac{(y_{\rho(n)}-x_{\sigma(1)})_z^{s'_{\rho(n)}-k'_{\rho(n)}+k_{\sigma(n)}+1}}{(( y_{\rho(n)}-x_{\sigma(1)})^2)^{s'_{\rho(n)}-k'_{\rho(n)}+k_{\sigma(n)}+2}}\,.
	\end{align}

\section{Normalization of $2$-point correlators}
We compute the normalization of the $2$-point correlators by a technique \cite{Kazakov:2012ar} involving the orthogonality of Gegenbauer and Jacobi polynomias that makes manifest the vanishing of correlators of operators with different spins. 

\subsection{$2$-point correlators of gluino-gluino operators \label{appGGresum}}

The $2$-point point correlators of gluino-gluino operators read
\begin{align}
	\langle O^{\lambda}_{s_1}(x) O^{\lambda}_{s_2}(y)\rangle
	=&-\frac{1}{4}\frac{N^2-1}{(4\pi^2)^2} \mathcal{Y}_{s_1-1}^{\frac{3}{2}}(\partial_{x_1^+},\partial_{x_2^+})\mathcal{Y}_{s_2-1}^{\frac{3}{2}}(\partial_{y_1^+},\partial_{y_2^+})\nonumber\\
	&\partial_{x_1^+}\partial_{x_2^+}\frac{1}{\rvert x_1-y_2\rvert^2}\frac{1}{\rvert x_2-y_1\rvert^2}\Big\rvert_{x_1=x_2=x}^{y_1=y_2=y}\,.
\end{align}
Initially, we restrict to $(x-y)_{\perp} = 0$, so that $\rvert x-y\rvert^2 = 2(x-y)_+(x-y)_-$. For $x_- > y_-$ we get
\begin{equation}
	\label{C2}
	\frac{\Gamma(k)}{(x-y)_-^k} = \int_{0}^{\infty}d\tau\, \tau^{k-1} e^{-\tau(x-y)_-}\,.
\end{equation}
By the above equation we convert derivatives into multiplications
\begin{align}
	\label{C3}
	\nonumber
	&\partial_{x_1^+} \rightarrow -\tau_1, \qquad \partial_{x_2^+} \rightarrow -\tau_2\\
	&\partial_{y_1^+}  \rightarrow \tau_2, \qquad \partial_{y_2^+}  \rightarrow \tau_1\,.
\end{align}
By the symmetry properties (Appendix \ref{appB})
\begin{align}
	&\mathcal{Y}_{s-1}^{\frac{3}{2}}(-\tau_1,-\tau_2) = (-1)^{s_1-1} (i\tau_1+i\tau_2)^{s_1-1}C^{\frac{3}{2}}_{s_1-1}\Bigg(\frac{\tau_1-\tau_2}{\tau_1+\tau_2}\Bigg)\nonumber\\
	&\mathcal{Y}_{s-1}^{\frac{3}{2}}(\tau_2,\tau_1) = (-1)^{s_2-1} (i\tau_1+i\tau_2)^{s_2-1}C^{\frac{3}{2}}_{s_2-1}\Bigg(\frac{\tau_1-\tau_2}{\tau_1+\tau_2}\Bigg)\,,
\end{align}
we obtain
\begin{align}
	\langle O^{\lambda}_{s_1}(x) O^{\lambda}_{s_2}(y)\rangle
	&= -\frac{1}{4}\frac{N^2-1}{(4\pi^2)^2}i^{s_1+s_2-2}\frac{1}{4(x-y)^2_+} (-1)^{s_1+s_2}\nonumber\\
	&\quad\int_{0}^{\infty}d\tau_1d\tau_2\, (\tau_1+\tau_2)^{s_1+s_2-2}\tau_1 \tau_2\, e^{-(\tau_1+\tau_2)(x-y)_-}\nonumber\\
	&\quad C_{s_1-1}^{\frac{3}{2}}\Bigg(\frac{\tau_1-\tau_2}{\tau_1+\tau_2}\Bigg) C_{s_2-1}^{\frac{3}{2}}\Bigg(\frac{\tau_1-\tau_2}{\tau_1+\tau_2}\Bigg)\,.
\end{align}
Changing variables
\begin{equation}
	\tau_1 = \tau \alpha,\qquad \tau_2 = \tau (1-\alpha)\,,
\end{equation}
we get
\begin{align}
	\langle O^{\lambda}_{s_1}(x) O^{\lambda}_{s_2}(y)\rangle
	&= -\frac{1}{4}\frac{N^2-1}{(4\pi^2)^2}i^{s_1+s_2-2}\frac{1}{4(x-y)^2_+} (-1)^{s_1+s_2}\nonumber\\
	&\quad\int_{0}^{\infty}d\tau \int_{0}^{1}d\alpha\, \tau^{s_1+s_2+1}\alpha(1-\alpha) e^{-\tau(x-y)_-}\nonumber\\ &\quad C_{s_1-1}^{\frac{3}{2}}(1-2\alpha)C_{s_2-1}^{\frac{3}{2}}(1-2\alpha)\,.
\end{align}
The $\tau$ integral yields
\begin{equation}
	\int_{0}^{\infty}d\tau\, \tau^{s_1+s_2+1} e^{-\tau(x-y)_-} = \frac{\Gamma(s_1+s_2+2)}{(x-y)_-^{s_1+s_2+2}}\,.
\end{equation}
The $\alpha$ integral
\begin{equation}
	\int_{0}^{1}d\alpha\, \alpha(1-\alpha) C_{s_1-1}^{\frac{3}{2}}(1-2\alpha)C_{s_2-1}^{\frac{3}{2}}(1-2\alpha)
\end{equation}
can be written as
\begin{equation}
	\int_{-1}^{1}\frac{du}{2}\, \left(\frac{1-u^2}{4}\right) C_{s_1-1}^{\frac{3}{2}}(u)C_{s_2-1}^{\frac{3}{2}}(u)\,
\end{equation}
where $u = 1-2\alpha$. The orthonormality property of Gegenbauer polynomials reads
\begin{align}
\int_{-1}^{1}dz\,(1-z^2)^{\alpha-\frac{1}{2}} C_{n_1}^{\alpha}(z)C_{n_2}^{\alpha}(z) 
	= \delta_{n_1,n_2}\frac{\pi 2^{1-2\alpha} \Gamma(n_1+2\alpha)}{n_1!(n_1+\alpha)\Gamma(\alpha)^2}\,
\end{align}
so that for $\alpha = \frac{3}{2}$ we obtain
\begin{align}
	\int_{-1}^{1}\frac{du}{2}\, \frac{\left(1-u^2\right)}{4} C_{s_1-1}^{\frac{3}{2}}(u)C_{s_2-1}^{\frac{3}{2}}(u)
	=\delta_{s_1,s_2} \frac{1}{4}\frac{s_1 (s_1+1)}{2 s_1+1}\,.
\end{align}
Collecting all the above factors
\begin{align}
	\langle O^{\lambda}_{s_1}(x) O^{\lambda}_{s_2}(y)\rangle
	&= -\frac{1}{4}\frac{N^2-1}{(4\pi^2)^2}i^{s_1+s_2-2} (-1)^{s_1+s_2}\nonumber \\
	&\quad\frac{1}{4}s_1 (s_1+1)\delta_{s_1,s_2}\frac{\Gamma(s_1+s_2+2)}{2s_1+1} \nonumber\\
	&\quad\frac{1}{4(x-y)^2_+(x-y)_-^{s_1+s_2+2}}
\end{align}
and employing the identity
\begin{equation}
	\frac{1}{4(x-y)^2_+(x-y)_-^{s_1+s_2+2}} = 2^{s_1+s_2}\frac{(x-y)_+^{s_1+s_2}}{(\rvert x-y\rvert^2)^{s_1+s_2+2}}
\end{equation}
yields the desired result
\begin{align}
	\label{2pointlambdare}
	\langle O^{\lambda}_{s_1}(x) O^{\lambda}_{s_2}(y)\rangle
	&= -\delta_{s_1s_2}\frac{1}{4}\frac{N^2-1}{(4\pi^2)^2}i^{2s_1-2} (-1)^{2s_1}2^{2s_1}\nonumber\\
	&\quad\frac{s_1 (s_1+1)}{4}\frac{\Gamma(2s_1+2)}{2s_1+1} \frac{(x-y)_+^{2s_1}}{(\rvert x-y\rvert^2)^{2s_1+2}}\,.
\end{align}
Incidentally, equating the above equation to Eq.~(\ref{2pointlambda}) we deduce the identity
\begin{align}
	\label{thlambda}
	\delta_{s_1 s_2} \frac{s_1}{s_1+1}
	=\sum_{k_1 = 0}^{s_1-1}\sum_{k_2 = 0}^{s_2-1}{s_1\choose k_1}{s_1\choose k_1+1}{s_2\choose k_2}{s_2\choose k_2+1}(-1)^{k_1+k_2}\frac{1}{{s_1+s_2\choose k_1+k_2+1}}\,.
\end{align}

\subsection{$2$-point correlators of gluon-gluino operators \label{appppintegral}}

The $2$-point point correlators of gluon-gluino operators read
\begin{align}
	\langle M_{s_1}(x)\bar{M}_{s_2}(y) \rangle 
	&= (N^2-1)\frac{i}{(4\pi^2)^{2}}\frac{1}{2^2}\partial_{y^+_1}\frac{1}{\rvert y_1-x_2\rvert^2}\mathcal{G}_{s_{2}-1}^{(1,2)}(\partial_{y_1^+},\partial_{y_2^+})\nonumber\\
	&\quad\frac{1}{\rvert y_2-x_1\rvert^2} \mathcal{G}_{s_{1}-1}^{(2,1)}(\partial_{x_1^+},\partial_{x_2^+})\Big\rvert_{x_1=x_2=x}^{y_1=y_2=y}\,,
\end{align}
where
\begin{align}
	&\mathcal{G}_{s-1}^{(1,2)}(\partial_{y_1^+},\partial_{y_2^+}) =  (i\partial_{y_1^+}+i\partial_{y_2^+})^{s-1}P^{(1,2)}_{s-1}\Bigg(\frac{\partial_{y_1^+}-\partial_{y_2^+}}{\partial_{y_1^+}+\partial_{y_2^+}}\Bigg){\partial}_{y_2^+}\nonumber\\
	&\mathcal{G}_{s-1}^{(2,1)}(\partial_{x_1^+},\partial_{x_2^+}) =  \partial_{x_1^+}(i\partial_{x_1^+}+i\partial_{x_2^+})^{s-1}P^{(2,1)}_{s-1}\Bigg(\frac{\partial_{x_1^+}-\partial_{x_2^+}}{\partial_{x_1^+}+\partial_{x_2^+}}\Bigg)\,.
\end{align}
Employing Eqs. \eqref{C2}, \eqref{C3} and  the symmetry properties (appendix \ref{appB})
\begin{align}
	&\mathcal{G}_{s-1}^{(1,2)}(-\tau_1,-\tau_2)= (-1)^{s_1} \tau_2(i\tau_1+i\tau_2)^{s_1-1}P^{(1,2)}_{s_1-1}\Bigg(\frac{\tau_1-\tau_2}{\tau_1+\tau_2}\Bigg)\nonumber\\
	&\mathcal{G}_{s-1}^{(2,1)}(\tau_2,\tau_1)= (-1)^{s_2-1} (i\tau_1+i\tau_2)^{s_2-1}P^{(1,2)}_{s_2-1}\Bigg(\frac{\tau_1-\tau_2}{\tau_1+\tau_2}\Bigg)\tau_2\,,
\end{align}
we obtain
\begin{align}
	\langle M_{s_1}(x)\bar{M}_{s_2}(y) \rangle 
	&=  (N^2-1)\frac{i}{(4\pi^2)^{2}}\frac{1}{2^2}i^{s_1+s_2-2}\frac{1}{4(y-x)^2_+} (-1)^{s_1+s_2}\nonumber\\
	&\quad\int_{0}^{\infty}d\tau_1d\tau_2\, (\tau_1+\tau_2)^{s_1+s_2-2} \tau_1\tau_2^2\, e^{-(\tau_1+\tau_2)(y-x)_-}\nonumber\\ &\quad P^{(1,2)}_{s_1-1}\Bigg(\frac{\tau_1-\tau_2}{\tau_1+\tau_2}\Bigg)P^{(1,2)}_{s_2-1}\Bigg(\frac{\tau_1-\tau_2}{\tau_1+\tau_2}\Bigg)\,.
\end{align}
Changing variables
\begin{equation}
	\tau_1 = \tau \alpha,\qquad \tau_2 = \tau (1-\alpha)\,,
\end{equation}
we get
\begin{align}
	\langle M_{s_1}(x)\bar{M}_{s_2}(y) \rangle 
	&=  (N^2-1)\frac{i}{(4\pi^2)^{2}}\frac{1}{2^2}i^{s_1+s_2-2}\frac{1}{4(x-y)^2_+} (-1)^{s_1+s_2}\nonumber\\
	&\quad\int_{0}^{\infty}d\tau \int_{0}^{1}d\alpha\, \tau^{s_1+s_2+2} \alpha(1-\alpha)^2\, e^{-\tau(x-y)_-}\nonumber\\ &\quad P^{(1,2)}_{s_1-1}(1-2\alpha)P^{(1,2)}_{s_2-1}(1-2\alpha)\,.
\end{align}
Integrating on $\tau$ yields
\begin{equation}
	\int_{0}^{\infty}d\tau\, \tau^{s_1+s_2+2} e^{-\tau(y-x)_-} = \frac{\Gamma(s_1+s_2+3)}{(y-x)_-^{s_1+s_2+3}}\,.
\end{equation}
Then, we rewrite
\begin{equation}
	\int_{0}^{1}d\alpha\, \alpha(1-\alpha)^2\, P^{(1,2)}_{s_1-1}(1-2\alpha)P^{(1,2)}_{s_2-1}(1-2\alpha)
\end{equation}
as
\begin{equation}
	\int_{-1}^{1}\frac{du}{2}\, \left(\frac{1-u}{2}\right)\left(\frac{1+u}{2}\right)^2\,  P^{(1,2)}_{s_1-1}(u)P^{(1,2)}_{s_2-1}(u)\,,
\end{equation}
with $u = 1-2\alpha$. 
The orthonormality property of Jacobi polynomials reads
\begin{align}
	\int_{-1}^1 (1-u)^{\alpha} (1+u)^{\beta} P_m^{(\alpha,\beta)} (u)P_n^{(\alpha,\beta)} (u)\,du =\delta_{nm}\frac{2^{\alpha+\beta+1}}{2n+\alpha+\beta+1} \frac{\Gamma(n+\alpha+1)\Gamma(n+\beta+1)}{\Gamma(n+\alpha+\beta+1)n!} \,,
\end{align}
so that for $\alpha = 1$ and $\beta=2$ we obtain
\begin{align}
	\int_{-1}^{1}\frac{du}{2}\, \left(\frac{1-u}{2}\right)\left(\frac{1+u}{2}\right)^2\,  P^{(1,2)}_{s_1-1}(u)P^{(1,2)}_{s_2-1}(u)
	 = \delta_{s_1s_2}\frac{1}{2s_1+2} \frac{\Gamma(s_1+1)\Gamma(s_1+2)}{\Gamma(s_1+3)(s_1-1)!} \,.
\end{align}
Collecting the above results
\begin{align}
	\nonumber
	\langle M_{s_1}(x)\bar{M}_{s_2}(y) \rangle 
	&= \delta_{s_1s_2} (N^2-1)\frac{i}{(4\pi^2)^{2}}\frac{1}{2^2}i^{2s_1-2} (-1)^{2s_1} \nonumber\\
	&\quad\frac{1}{2s_1+2} \frac{\Gamma(s_1+1)\Gamma(s_1+2)}{\Gamma(s_1+3)(s_1-1)!} \nonumber\\
	&\quad \frac{\Gamma(2s_1+3)}{4(y-x)^2_+(y-x)_-^{2s_1+3}}
\end{align}
and employing the identity
\begin{align}
	\frac{1}{4(y-x)^2_+(y-x)_-^{s_1+s_2+3}}
	 = 2^{s_1+s_2+1}(-1)^{s_1+s_2+1}\frac{(x-y)_+^{s_1+s_2+1}}{(\rvert x-y\rvert^2)^{s_1+s_2+3}}\,,
\end{align}
we get
\begin{align}
	\langle M_{s_1}(x)\bar{M}_{s_2}(y) \rangle 
	&= \delta_{s_1s_2} (N^2-1)\frac{i}{(4\pi^2)^{2}}i^{s_1+s_2} \Gamma(2s_1+3)2^{2s_1-1}\nonumber\\
	&\quad \frac{1}{2s_1+2}\frac{s_1}{s_1+2} \frac{(x-y)_+^{2s_1+1}}{(\rvert x-y\rvert^2)^{2s_1+3}}\,.
\end{align}
Incidentally, equating the above equation to Eq.~(\ref{twopointmix}) we deduce the identity
\begin{align}
	-\delta_{s_1s_2}\frac{s_1}{s_1+2}
	=\sum_{k_1 = 0}^{s_1-1}\sum_{k_2 = 0}^{s_2-1}{s_1\choose k_1}{s_1+1\choose k_1+2}{s_2+1\choose k_2}{s_2\choose k_2+1}(-1)^{s_1-k_2+k_{1}}\frac{1}{{s_1+s_2+1\choose s_1-k_1+k_2}}\,.
\end{align}

	\subsection{Conformal generating functional of correlators of supermultiplet operators}\label{confgen}
	
	We write the complete generating functional of Euclidean conformal correlators of supermultiplet operators
	\begin{equation}
		\label{wexplicit1ES}
		\resizebox{0.99\textwidth}{!}{%
			$\begin{aligned}
				&{\mathcal{W}^E_{\text{conf}}}\left[J_{\tilde{O}^{\lambda\,E}_1} ,J_{S^{(1)'\,E}},J_{\tilde{S}^{(1)'\,E}},J_{S^{(2)'\,E}},J_{\tilde{S}^{(2)'\,E}},\bar{J}_{{M^{'E}}},J_{\bar M^{'E}}\right] =\\
				& -\frac{N^2-1}{2}\log\Det\Bigg(I+\frac{1}{N}6\sum_{k=0}^{s-2}{s\choose k}{s\choose k+2}(-\overrightarrow{\partial}_z)^{s-k-1}\Laplace^{-1}(\frac{J_{S^{(1)'\,E}_s}}{s-1}+\frac{J_{S^{(2)'\,E}_s}}{s+2}-\frac{J_{\tilde{S}^{(1)'\,E}_s}}{s-1}-\frac{J_{\tilde{S}^{(2)'\,E}_s}}{s+2})(-\overrightarrow{\partial}_z)^{k+1} \Bigg)\\
				&-\frac{N^2-1}{2}\log\Det\Bigg(I+\frac{1}{N}6\sum_{k=0}^{s-2}{s\choose k}{s\choose k+2}(-\overrightarrow{\partial}_z)^{s-k-1}\Laplace^{-1}(\frac{J_{S^{(1)'\,E}_s}}{s-1}+\frac{J_{S^{(2)'\,E}_s}}{s+2}+\frac{J_{\tilde{S}^{(1)'\,E}_s}}{s-1}+\frac{J_{\tilde{S}^{(2)'\,E}_s}}{s+2})(-\overrightarrow{\partial}_z)^{k+1} \Bigg)\\
				&+\frac{N^2-1}{2}\log\Det\Bigg(I-\frac{1}{N}\Laplace^{-1}J_{\tilde{O}^{\lambda\,E}_1}\\
				&-\frac{1}{N}\sum_{k=0}^{s-1}{s\choose k}{s\choose k+1}(-\overrightarrow{\partial}_z)^{s-k}\Laplace^{-1}(-J_{S^{(1)'\,E}_s}+J_{S^{(2)'\,E}_s}-J_{\tilde{S}^{(1)'\,E}_s}+J_{\tilde{S}^{(2)'\,E}_s})(-\overrightarrow{\partial}_z)^{k} \Bigg)\\
				&+\frac{N^2-1}{2}\log\Det\Bigg(I+\frac{1}{N}\Laplace^{-1}J_{\tilde{O}^{\lambda\,E}_1}\\
				&-\frac{1}{N}\sum_{k=0}^{s-1}{s\choose k}{s\choose k+1}(-\overrightarrow{\partial}_z)^{s-k}\Laplace^{-1}(-J_{S^{(1)'\,E}_s}+J_{S^{(2)'\,E}_s}+J_{\tilde{S}^{(1)'\,E}_s}-J_{\tilde{S}^{(2)'\,E}_s})(-\overrightarrow{\partial}_z)^{k} \Bigg)\\
				&+\frac{N^2-1}{2}\log\Det\Bigg[I\\
				&+\frac{1}{N^2}\Bigg(I+\frac{1}{N}\Laplace^{-1}J_{\tilde{O}^{\lambda\,E}_1}-\frac{1}{N}\sum_{k=0}^{s-1}{s\choose k}{s\choose k+1}(-\overrightarrow{\partial}_z)^{s-k}\Laplace^{-1}(-J_{S^{(1)'\,E}_s}+J_{S^{(2)'\,E}_s}+J_{\tilde{S}^{(1)'\,E}_s}-J_{\tilde{S}^{(2)'\,E}_s})(-\overrightarrow{\partial}_z)^{k} \Bigg)^{-1}\\
				&\sum_{k_1 = 0}^{s_1-1}{s_1\choose k_1}{s_1+1\choose k_1+2} (-1)^{s_1-1}
				(-\overrightarrow{\partial}_z)^{s_1-k_1}\Laplace^{-1}\bar{J}_{M^{'E}_{s_1}}(-\overrightarrow{\partial}_z)^{k_1+1}  \\
				&\Bigg(I+\frac{1}{N}6\sum_{k_2=0}^{s_2-2}{s_2\choose k_2}{s_2\choose k_2+2}(-\overrightarrow{\partial}_z)^{s_2-k_2-1}\Laplace^{-1}(\frac{J_{S^{(1)'\,E}_{s_2}}}{s_2-1}+\frac{J_{S^{(2)'\,E}_{s_2}}}{s_2+2}-\frac{J_{\tilde{S}^{(1)'\,E}_{s_2}}}{s_2-1}-\frac{J_{\tilde{S}^{(2)'\,E}_{s_2}}}{s_2+2})(-\overrightarrow{\partial}_z)^{k_2+1} \Bigg)^{-1}\\
				& \sum_{k_3 = 0}^{s_3-1}{s_3+1\choose k_3}{s_3\choose k_3+1} (-1)^{s_3-1}
				(-\overrightarrow{\partial}_z)^{s_3-k_3}\Laplace^{-1}J_{\bar M^{'E}_{s_3}} (-\overrightarrow{\partial}_z)^{k_3}\Bigg]\\
				&+\frac{N^2-1}{2}\log\Det\Bigg[I\\
				&+\frac{1}{N^2}\left(I-\frac{1}{N}\Laplace^{-1}J_{\tilde{O}^{\lambda\,E}_1}-\frac{1}{N}\sum_{k=0}^{s-1}{s\choose k}{s\choose k+1}(-\overrightarrow{\partial}_z)^{s-k}\Laplace^{-1}(-J_{S^{(1)'\,E}_s}+J_{S^{(2)'\,E}_s}-J_{\tilde{S}^{(1)'\,E}_s}+J_{\tilde{S}^{(2)'\,E}_s})(-\overrightarrow{\partial}_z)^{k}\right)^{-1}\\
				& \sum_{k_1 = 0}^{s_1-1}{s_1\choose k_1}{s_1+1\choose k_1+2} 
				(-\overrightarrow{\partial}_z)^{s_1-k_1}\Laplace^{-1}J_{\bar M^{'E}_{s_1}}(-\overrightarrow{\partial}_z)^{k_1+1}   \\
				&\left(I+\frac{1}{N}\frac{6}{s_2-1}\sum_{k=0}^{s_2-2}{s_2\choose k_2}{s_2\choose k_2+2}(-\overrightarrow{\partial}_z)^{s_2-k_2-1}\Laplace^{-1}(\frac{J_{S^{(1)'\,E}_{s_2}}}{s_2-1}+\frac{J_{S^{(2)'\,E}_{s_2}}}{s_2+2}+\frac{J_{\tilde{S}^{(1)'\,E}_{s_2}}}{s_2-1}+\frac{J_{\tilde{S}^{(2)'\,E}_{s_2}}}{s_2+2})(-\overrightarrow{\partial}_z)^{k_2+1}\right)^{-1}\\
				&  \sum_{k_3 = 0}^{s_3-1}{s_3+1\choose k_3}{s_3\choose k_3+1} 
				(-\overrightarrow{\partial}_z)^{s_3-k_3}\Laplace^{-1}\bar{J}_{M^{'E}_{s_3}} (-\overrightarrow{\partial}_z)^{k_3}\Bigg]
			\end{aligned}$
		} 
	\end{equation}
	and
	\begin{equation}
		\label{wexplicit2ES}
		\resizebox{0.99\textwidth}{!}{%
			$\begin{aligned}
				&{\mathcal{W}^E_{\text{conf}}}\left[J_{\tilde{O}^{\lambda\,E}_1} ,J_{S^{(1)'\,E}},J_{\tilde{S}^{(1)'\,E}},J_{S^{(2)'\,E}},J_{\tilde{S}^{(2)'\,E}},\bar{J}_{{M^{'E}}},J_{\bar M^{'E}}\right] =\\
				& -\frac{N^2-1}{2}\log\Det\Bigg(I+\frac{1}{N}6\sum_{k=0}^{s-2}{s\choose k}{s\choose k+2}(-\overrightarrow{\partial}_z)^{s-k-1}\Laplace^{-1}(\frac{J_{S^{(1)'\,E}_s}}{s-1}+\frac{J_{S^{(2)'\,E}_s}}{s+2}-\frac{J_{\tilde{S}^{(1)'\,E}_s}}{s-1}-\frac{J_{\tilde{S}^{(2)'\,E}_s}}{s+2})(-\overrightarrow{\partial}_z)^{k+1} \Bigg)\\
				&-\frac{N^2-1}{2}\log\Det\Bigg(I+\frac{1}{N}6\sum_{k=0}^{s-2}{s\choose k}{s\choose k+2}(-\overrightarrow{\partial}_z)^{s-k-1}\Laplace^{-1}(\frac{J_{S^{(1)'\,E}_s}}{s-1}+\frac{J_{S^{(2)'\,E}_s}}{s+2}+\frac{J_{\tilde{S}^{(1)'\,E}_s}}{s-1}-\frac{J_{\tilde{S}^{(2)'\,E}_s}}{s+2})(-\overrightarrow{\partial}_z)^{k+1} \Bigg)\\
				&+\frac{N^2-1}{2}\log\Det\Bigg(I-\frac{1}{N}\Laplace^{-1}J_{\tilde{O}^{\lambda\,E}_1}\\
				&-\frac{1}{N}\sum_{k=0}^{s-1}{s\choose k}{s\choose k+1}(-\overrightarrow{\partial}_z)^{s-k}\Laplace^{-1}(-J_{S^{(1)'\,E}_s}+J_{S^{(2)'\,E}_s}-J_{\tilde{S}^{(1)'\,E}_s}+J_{\tilde{S}^{(2)'\,E}_s})(-\overrightarrow{\partial}_z)^{k} \Bigg)\\
				&+\frac{N^2-1}{2}\log\Det\Bigg(I+\frac{1}{N}\Laplace^{-1}J_{\tilde{O}^{\lambda\,E}_1}\\
				&-\frac{1}{N}\sum_{k=0}^{s-1}{s\choose k}{s\choose k+1}(-\overrightarrow{\partial}_z)^{s-k}\Laplace^{-1}(-J_{S^{(1)'\,E}_s}+J_{S^{(2)'\,E}_s}+J_{\tilde{S}^{(1)'\,E}_s}-J_{\tilde{S}^{(2)'\,E}_s})(-\overrightarrow{\partial}_z)^{k} \Bigg)\\
				&-\frac{N^2-1}{2}\log\Det\Bigg[I\\
				&+\frac{1}{N^2}\Bigg(I+\frac{1}{N}6\sum_{k=0}^{s-2}{s\choose k}{s\choose k+2}(-\overrightarrow{\partial}_z)^{s-k-1}\Laplace^{-1}(\frac{J_{S^{(1)'\,E}_s}}{s-1}+\frac{J_{S^{(2)'\,E}_s}}{s+2}-\frac{J_{\tilde{S}^{(1)'\,E}_s}}{s-1}-\frac{J_{\tilde{S}^{(2)'\,E}_s}}{s+2})(-\overrightarrow{\partial}_z)^{k+1}  \Bigg)^{-1}\\
				&\sum_{k_1 = 0}^{s_1-1}{s_1\choose k_1}{s_1+1\choose k_1+2} (-1)^{s_1-1}
				(-\overrightarrow{\partial}_z)^{s_1-k_1}\Laplace^{-1}J_{\bar M^{'E}_{s_1}}(-\overrightarrow{\partial}_z)^{k_1+1}  \\
				&\Bigg(I+\frac{1}{N}\Laplace^{-1}J_{\tilde{O}^{\lambda\,E}_1}-\frac{1}{N}\sum_{k_2=0}^{{s_2}-1}{{s_2}\choose k_2}{{s_2}\choose k_2+1}(-\overrightarrow{\partial}_z)^{{s_2}-k_2}\Laplace^{-1}(-J_{S^{(1)'\,E}_{s_2}}+J_{S^{(2)'\,E}_{s_2}}+J_{\tilde{S}^{(1)'\,E}_{s_2}}-J_{\tilde{S}^{(2)'\,E}_{s_2}})(-\overrightarrow{\partial}_z)^{k_2} \Bigg)^{-1}\\
				& \sum_{k_3 = 0}^{s_3-1}{s_3+1\choose k_3}{s_3\choose k_3+1} (-1)^{s_3-1}
				(-\overrightarrow{\partial}_z)^{s_3-k_3}\Laplace^{-1}\bar{J}_{M^{'E}_{s_3}} (-\overrightarrow{\partial}_z)^{k_3}\Bigg]\\
				&-\frac{N^2-1}{2}\log\Det\Bigg[I\\
				&+\frac{1}{N^2}\left(I+\frac{1}{N}6\sum_{k=0}^{s-2}{s\choose k}{s\choose k+2}(-\overrightarrow{\partial}_z)^{s-k-1}\Laplace^{-1}(\frac{J_{S^{(1)'\,E}_s}}{s-1}+\frac{J_{S^{(2)'\,E}_s}}{s+2}+\frac{J_{\tilde{S}^{(1)'\,E}_s}}{s-1}+\frac{J_{\tilde{S}^{(2)'\,E}_s}}{s+2})(-\overrightarrow{\partial}_z)^{k+1}\right)^{-1}\\
				& \sum_{k_1 = 0}^{s_1-1}{s_1\choose k_1}{s_1+1\choose k_1+2} 
				(-\overrightarrow{\partial}_z)^{s_1-k_1}\Laplace^{-1}\bar{J}_{M^{'E}_{s_1}}(-\overrightarrow{\partial}_z)^{k_1+1}   \\
				&\left(I-\frac{1}{N}\Laplace^{-1}J_{\tilde{O}^{\lambda\,E}_1}-\frac{1}{N}\sum_{{k_2}=0}^{{s_2}-1}{{s_2}\choose {k_2}}{{s_2}\choose {k_2}+1}(-\overrightarrow{\partial}_z)^{{s_2}-{k_2}}\Laplace^{-1}(-J_{S^{(1)'\,E}_{s_2}}+J_{S^{(2)'\,E}_{s_2}}-J_{\tilde{S}^{(1)'\,E}_{s_2}}+J_{\tilde{S}^{(2)'\,E}_{s_2}})(-\overrightarrow{\partial}_z)^{{k_2}}\right)^{-1}\\
				&  \sum_{k_3 = 0}^{s_3-1}{s_3+1\choose k_3}{s_3\choose k_3+1} 
				(-\overrightarrow{\partial}_z)^{s_3-k_3}\Laplace^{-1}J_{\bar M^{'E}_{s_3}} (-\overrightarrow{\partial}_z)^{k_3}\Bigg]\,.
			\end{aligned}$
		} 
	\end{equation}

\section{Connected generating functional in the momentum representation \label{momgen}}

The conformal generating functional in the momentum representation is defined by the functional integral
\begin{align}
	\label{Zmom}
		\mathcal{Z}_{\text{conf}}[J_{\mathcal{O}}] 
	&= \frac{1}{Z}\int \mathcal{D} A \mathcal{D} \bar{A} \mathcal{D} \lambda \mathcal{D} \bar{\lambda} e^{\int -i\bar{A}^a \square A^a+\bar{\lambda}^a\Box \partial_+^{-1} \lambda^a \, d^4x}\exp\Bigg(\int \frac{d^4p}{(2\pi)^4}\sum_i\, J_{\mathcal{O}_i}(-p)\mathcal{O}_i(p)\Bigg)\,.
\end{align}
Hence, we get for the correlators in the momentum representation \cite{BPSpaper2}
\begin{align}
	\langle \mathcal{O}_{s_1}(p_1)\ldots\mathcal{O}_{s_n}(p_n)\rangle 
	= (2\pi)^4\frac{\delta }{\delta J_{\mathcal{O}_{s_1}}(-p_1)}\cdots (2\pi)^4\frac{\delta }{\delta J_{\mathcal{O}_{s_n}}(-p_n)}\mathcal{W}[J_{\mathcal{O}}]\,.
\end{align} 
The generating functional is obtained by means of the kernels in the momentum representation
\begin{equation} \label{AB}
	i\square^{-1} \rightarrow \frac{-i}{\rvert q\rvert^2+i\epsilon} 
\end{equation}
and
\begin{equation} \label{BA}
	(i\partial_+)^{k} i\square^{-1} \rightarrow (-q_+)^{k} \frac{-i}{\rvert q\rvert^2+i\epsilon} \,,
\end{equation}
where we employ the measure in momentum space \cite{BPSpaper2}
\begin{equation}
	\int\frac{d^4q}{(2\pi)^4}\,.
\end{equation}
We also get for the identity in space-time
\begin{align}
	&I \to(2\pi)^4\delta^{(4)}(q_1-q_2) \,,
\end{align}
for the sources
\begin{align}
	&J_{\mathcal{O}}\to J_{\mathcal{O}}(q_1-q_2)
\end{align}
and the differential operators
	\begin{align}
		&\mathcal{Y}^{\frac{5}{2}}_{s-2}(\overrightarrow{\partial}_+,\overleftarrow{\partial}_+)\to\mathcal{Y}_{s-2}^{\frac{5}{2}}(q_{2+},q_{1+}) = q_{1+} (q_{2+}-q_{1+})^{s-2}C^{\frac{5}{2}}_{s-2}\left(\frac{q_{2+}+q_{1+}}{q_{2+}-q_{1+}}\right)q_{2+}\nonumber\\
		&\mathcal{Y}^{\frac{3}{2}}_{s-1}(\overrightarrow{\partial}_+,\overleftarrow{\partial}_+)\to\mathcal{Y}_{s-1}^{\frac{3}{2}}(q_{2+},q_{1+})  =(q_{2+}-q_{1+})^{s-2}C^{\frac{3}{2}}_{s-1}\left(\frac{q_{2+}+q_{1+}}{q_{2+}-q_{1+}}\right)\nonumber\\
		&\mathcal{H}^{\frac{5}{2}}_{s-2}(\overrightarrow{\partial}_+,\overleftarrow{\partial}_+)\to\mathcal{H}_{s-2}^{\frac{5}{2}}(q_{2+},q_{1+}) = q_{1+} (q_{2+}-q_{1+})^{s-2}C^{\frac{5}{2}}_{s-2}\left(\frac{q_{2+}+q_{1+}}{q_{2+}-q_{1+}}\right)q_{2+}\nonumber\\
		&\mathcal{H}^{\frac{3}{2}}_{s-1}(\overrightarrow{\partial}_+,\overleftarrow{\partial}_+)\to\mathcal{H}_{s-1}^{\frac{3}{2}}(q_{2+},q_{1+})  =(q_{2+}-q_{1+})^{s-2}C^{\frac{3}{2}}_{s-1}\left(\frac{q_{2+}+q_{1+}}{q_{2+}-q_{1+}}\right)\nonumber\\
		&	\mathcal{G}_{s-1}^{(1,2)} (\overrightarrow{\partial}_+,\overleftarrow{\partial}_+)\to	\mathcal{G}_{s-1}^{(1,2)} (q_{2+},q_{1+}) =i(q_{2+}-q_{1+})^{s-1}P^{(1,2)}_{s-1}\left(\frac{q_{2+}+q_{1+}}{q_{2+}-q_{1+}}\right) q_{2+}\nonumber\\
		&	\mathcal{G}_{s-1}^{(2,1)} (\overrightarrow{\partial}_+,\overleftarrow{\partial}_+)\to	\mathcal{G}_{s-1}^{(2,1)} (q_{2+},q_{1+}) =-iq_{1+}(q_{2+}-q_{1+})^{s-1}P^{(2,1)}_{s-1}\left(\frac{q_{2+}+q_{1+}}{q_{2+}-q_{1+}}\right) \,.
	\end{align}
	The explicit expression in the momentum representation \cite{BPSpaper2} follows from Eq. \eqref{jaco2}.
	We report the generating functional restricted to several sectors
	\begin{align}
		&\mathcal{W}_{\text{conf}}\left[J_{O^A},0,0,0,0,0\right]\nonumber\\
		&=-(N^2-1)\log\Det \Big((2\pi)^4\delta^{(4)}(q_1-q_2) -\frac{1}{2}\frac{i}{\rvert q_1\rvert^2+i\epsilon} J_{O^A_{s}}(q_1-q_2)\otimes\mathcal{Y}_{s-2}^{\frac{5}{2}}(q_{2+},q_{1+})  \Big) \,,
	\end{align}
	where by a slight abuse of notation we have displayed as argument of the functional determinant the corresponding kernel in the momentum representation \cite{BPSpaper2}.\par
	We write the generating functional of the correlators of fermionic operators, $M_s$ and $\bar{M}_s$, as a single object by means of an equality of determinants (appendix \ref{appM})
	\begin{align}
		&\mathcal{W}_{\text{conf}}\left[0,0,0,0,\bar{J}_{M},J_{\bar M}\right] =\log\Det \Big(\mathcal{I}  + \frac{1}{4}i\partial_+i\square^{-1}J_{\bar M_{s_1}}\otimes\mathcal{G}_{s_1-1}^{(1,2)} i\square^{-1}\bar{J}_{M_{s_2}}\otimes\mathcal{G}_{s_2-1}^{(2,1)}\Big)\,.
	\end{align}
	Then, we get
	\begin{align}
		&\mathcal{W}_{\text{conf}}\left[0,0,0,0,\bar{J}_{M},J_{\bar M}\right] =(N^2-1)\log\Det \Big((2\pi)^4\delta^{(4)}(q_1-q_2) \nonumber\\
		& - \frac{1}{4}\int\frac{d^4q}{(2\pi)^4} \frac{iq_{1+}}{\rvert q_1\rvert^2+i\epsilon} J_{\bar M_{s_1}}(q_1-q)\otimes\mathcal{G}_{s_1-1}^{(1,2)}(q_{+},q_{1+}) \frac{i}{\rvert q\rvert^2+i\epsilon} \bar{J}_{M_{s_2}}(q-q_2)\otimes\mathcal{G}_{s_2-1}^{(2,1)}(q_{2+},q_{+})\Big)\,.
	\end{align}
	Besides,
	\begin{align}
		&\mathcal{W}_{\text{conf}}\left[J_{S^{(2)}}\right] =\nonumber\\ &=-(N^2-1)\log\Det\bigg((2\pi)^4\delta^{(4)}(q_1-q_2)-\frac{1}{2}\frac{i}{\rvert q_1\rvert^2+i\epsilon}\frac{6}{s+2}J_{S^{(2)}_s}(q_1-q_2)\otimes\mathcal{Y}_{s-2}^{\frac{5}{2}}(q_{1\,+},q_{2\,+})\bigg) \nonumber\\
		&\quad+(N^2-1)\log\Det\bigg((2\pi)^4\delta^{(4)}(q_1-q_2)-\frac{1}{2}\frac{iq_{1\,+}}{\rvert q_1\rvert^2+i\epsilon}J_{S^{(2)}_s}(q_1-q_2)\otimes\mathcal{Y}_{s-1}^{\frac{3}{2}}(q_{1\,+},q_{2\,+})\bigg)\,.
	\end{align}
	Explicitly,
	\begin{equation}
		\resizebox{0.99\textwidth}{!}{%
			$\begin{aligned}
				\mathcal{W}_{\text{conf}}\left[J_{S^{(2)}}\right] =
				& -(N^2-1)\log\Det\bigg((2\pi)^4\delta^{(4)}(q_1-q_2)-\frac{i}{\rvert q_1\rvert^2+i\epsilon}\frac{s+1}{4}J_{S^{(2)}_s}(q_1-q_2)\sum_{k=0}^{s-2} {s\choose k}{s\choose k+2} q_{1\,+}^{s-k-1} q_{2\,+}^{k+1}\bigg) \\
				&+(N^2-1)\log\Det\bigg((2\pi)^4\delta^{(4)}(q_1-q_2)-\frac{iq_{1\,+}}{\rvert q_1\rvert^2+i\epsilon}\frac{s+1}{4}J_{S^{(2)}_s}(q_1-q_2) \sum_{k = 0}^{s-1}{s\choose k}{s\choose k+1}q_{1\,+}^{s-k-1}q_{2\,+}^{k} \bigg)\,.
			\end{aligned}$
		} 
	\end{equation}
	Correspondingly, $\Gamma_{\text{conf}}\left[J_{S^{(2)}}\right]$ reads
	\begin{align}
		&\Gamma_{conf}\left[J_{S^{(2)}}\right] \nonumber\\
		&= -(N^2-1)\log \Det \left(\delta_{s_1k_1,s_2k_2}(2\pi)^4\delta^{(4)}(q_1-q_2)+\mathcal{D}^{-1}_{A\,s_1k_1,s_2k_2}(q_1)\frac{6}{s_2+2}J_{S^{(2)}_{s_2k_2}}(q_1-q_2)\right)\nonumber\\
		& \quad+(N^2-1)\log \Det \left(\delta_{s_1k_1,s_2k_2}(2\pi)^4\delta^{(4)}(q_1-q_2)+\mathcal{D}^{-1}_{\lambda\,s_1k_1,s_2k_2}(q_1)J_{S^{(2)}_{s_2k_2}}(q_1-q_2)\right)\,,
	\end{align}
	with
	\begin{align}
		\mathcal{D}^{-1}_{A\,s_1k_1,s_2k_2}(p) &=\frac{1}{2}\frac{\Gamma(3)\Gamma(s_1+3)}{\Gamma(5)\Gamma(s_1+1)}{s_1\choose k_1}{s_2\choose k_2+2}p_{+}^{s_1-k_1+k_2}\frac{-i}{\rvert p\rvert^2+i\epsilon} \nonumber\\
		\mathcal{D}^{-1}_{\lambda\,s_1k_1,s_2k_2}(p) &=\frac{1}{2}\frac{s_1+1}{2}{s_1\choose k_1}{s_2\choose k_2+1}p_{+}^{s_1-k_1+k_2-1}\frac{-ip_+}{\rvert p\rvert^2+i\epsilon} \,.
	\end{align} 
	
	\section{RG-improved generating functional of correlators of supermultiplet operators}\label{rggen}
	
	From the above construction of the conformal generating functional and RG-improved correlators, it follows the complete generating functional of the Euclidean asymptotic correlators $\mathcal{W}^E_{\text{asym}}[J_{\mathcal{O'}^E},\lambda]$
	\begin{equation}
		\label{wexplicit1ES}
		\resizebox{1.00\textwidth}{!}{
			$\begin{aligned}
				&\mathcal{W}^E_{\text{asym torus}}\left[J_{\tilde{O}^{\lambda\,E}_1},J_{S^{(1)'\,E}},J_{\tilde{S}^{(1)'\,E}},J_{S^{(2)'\,E}},J_{\tilde{S}^{(2)'\,E}},\bar{J}_{{M^{'E}}},J_{\bar M^{'E}},\lambda\right] =\\
				&+ \frac{1}{2}\log\Det\Bigg(I\\
				&+\frac{1}{N}6\sum_{k=0}^{s-2}{s\choose k}{s\choose k+2}(-\overrightarrow{\partial}_z)^{s-k-1}\frac{\Laplace^{-1}}{\lambda^{s+2}}(Z_{S^{(1)'\,E}}(\lambda)\frac{J_{S^{(1)'\,E}_s}}{s-1}+Z_{S^{(2)'\,E}}(\lambda)\frac{J_{S^{(2)'\,E}_s}}{s+2}-Z_{\tilde{S}^{(1)'\,E}}(\lambda)\frac{J_{\tilde{S}^{(1)'\,E}_s}}{s-1}-Z_{\tilde{S}^{(2)'\,E}}(\lambda)\frac{J_{\tilde{S}^{(2)'\,E}_s}}{s+2})(-\overrightarrow{\partial}_z)^{k+1} \Bigg)\\
				& +\frac{1}{2}\log\Det\Bigg(I\\
				&+\frac{1}{N}6\sum_{k=0}^{s-2}{s\choose k}{s\choose k+2}(-\overrightarrow{\partial}_z)^{s-k-1}\frac{\Laplace^{-1}}{\lambda^{s+2}}(Z_{S^{(1)'\,E}}(\lambda)\frac{J_{S^{(1)'\,E}_s}}{s-1}+Z_{S^{(2)'\,E}}(\lambda)\frac{J_{S^{(2)'\,E}_s}}{s+2}+Z_{\tilde{S}^{(1)'\,E}}(\lambda)\frac{J_{\tilde{S}^{(1)'\,E}_s}}{s-1}+Z_{\tilde{S}^{(2)'\,E}}(\lambda)\frac{J_{\tilde{S}^{(2)'\,E}_s}}{s+2})(-\overrightarrow{\partial}_z)^{k+1} \Bigg)\\
				&-\frac{1}{2}\log\Det\Bigg(I-\frac{1}{N}\frac{Z_{\tilde{O}^\lambda_1}(\lambda)}{\lambda^3}\Laplace^{-1}J_{\tilde{O}^{\lambda\,E}_1}\\
				&-\frac{1}{N}\sum_{k=0}^{s-1}{s\choose k}{s\choose k+1}(-\overrightarrow{\partial}_z)^{s-k}\frac{\Laplace^{-1}}{\lambda^{s+2}}(-Z_{S^{(1)'\,E}}(\lambda)J_{S^{(1)'\,E}_s}+Z_{S^{(2)'\,E}}(\lambda)J_{S^{(2)'\,E}_s}-Z_{\tilde{S}^{(1)'\,E}}(\lambda)J_{\tilde{S}^{(1)'\,E}_s}+Z_{\tilde{S}^{(2)'\,E}}(\lambda)J_{\tilde{S}^{(2)'\,E}_s})(-\overrightarrow{\partial}_z)^{k} \Bigg)\\
				&-\frac{1}{2}\log\Det\Bigg(I+\frac{1}{N}\frac{Z_{\tilde{O}^\lambda_1}(\lambda)}{\lambda^3}\Laplace^{-1}J_{\tilde{O}^{\lambda\,E}_1}\\
				&-\frac{1}{N}\sum_{k=0}^{s-1}{s\choose k}{s\choose k+1}(-\overrightarrow{\partial}_z)^{s-k}\frac{\Laplace^{-1}}{\lambda^{s+2}}(-Z_{S^{(1)'\,E}}(\lambda)J_{S^{(1)'\,E}_s}+Z_{S^{(2)'\,E}}(\lambda)J_{S^{(2)'\,E}_s}+Z_{\tilde{S}^{(1)'\,E}}(\lambda)J_{\tilde{S}^{(1)'\,E}_s}-Z_{\tilde{S}^{(2)'\,E}}(\lambda)J_{\tilde{S}^{(2)'\,E}_s})(-\overrightarrow{\partial}_z)^{k} \Bigg)\\
				&-\frac{1}{2}\log\Det\Bigg[I+\frac{1}{N^2}\Bigg(I+\frac{1}{N}\frac{Z_{\tilde{O}^\lambda_1}(\lambda)}{\lambda^3}\Laplace^{-1}J_{\tilde{O}^{\lambda\,E}_1}\\
				&-\frac{1}{N}\sum_{k=0}^{s-1}{s\choose k}{s\choose k+1}(-\overrightarrow{\partial}_z)^{s-k}\frac{\Laplace^{-1}}{\lambda^{s+2}}(-Z_{S^{(1)'\,E}}(\lambda)J_{S^{(1)'\,E}_s}+Z_{S^{(2)'\,E}}(\lambda)J_{S^{(2)'\,E}_s}+Z_{\tilde{S}^{(1)'\,E}}(\lambda)J_{\tilde{S}^{(1)'\,E}_s}-Z_{\tilde{S}^{(2)'\,E}}(\lambda)J_{\tilde{S}^{(2)'\,E}_s})(-\overrightarrow{\partial}_z)^{k} \Bigg)^{-1}\\
				&\quad\sum_{k_1 = 0}^{s_1-1}{s_1\choose k_1}{s_1+1\choose k_1+2} (-1)^{s_1-1}
				(-\overrightarrow{\partial}_z)^{s_1-k_1}\frac{\Laplace^{-1}}{\lambda^{s_1+2}}Z_{M}(\lambda)\bar{J}_{M^{'E}_{s_1}}(-\overrightarrow{\partial}_z)^{k_1+1}  \\
				&\quad\Bigg(I+\frac{1}{N}6\sum_{k_2=0}^{s_2-2}{s_2\choose k_2}{s_2\choose k_2+2}(-\overrightarrow{\partial}_z)^{s_2-k_2-1}\frac{\Laplace^{-1}}{\lambda^{s_2+2}}(Z_{S^{(1)'\,E}}(\lambda)\frac{J_{S^{(1)'\,E}_{s_2}}}{s_2-1}+Z_{S^{(2)'\,E}}(\lambda)\frac{J_{S^{(2)'\,E}_{s_2}}}{s_2+2}-Z_{\tilde{S}^{(1)'\,E}}(\lambda)\frac{J_{\tilde{S}^{(1)'\,E}_{s_2}}}{s_2-1}-Z_{\tilde{S}^{(2)'\,E}}(\lambda)\frac{J_{\tilde{S}^{(2)'\,E}_{s_2}}}{s_2+2})(-\overrightarrow{\partial}_z)^{k_2+1} \Bigg)^{-1}\\
				&\quad \sum_{k_3 = 0}^{s_3-1}{s_3+1\choose k_3}{s_3\choose k_3+1} (-1)^{s_3-1}
				(-\overrightarrow{\partial}_z)^{s_3-k_3}\frac{\Laplace^{-1}}{\lambda^{s_3+2}}Z_{M}(\lambda)J_{\bar M^{'E}_{s_3}} (-\overrightarrow{\partial}_z)^{k_3}\Bigg]\\
				&-\frac{1}{2}\log\Det\Bigg[I+\frac{1}{N^2}\Bigg(I-\frac{1}{N}\frac{Z_{\tilde{O}^\lambda_1}(\lambda)}{\lambda^3}\Laplace^{-1}J_{\tilde{O}^{\lambda\,E}_1}\\
				&-\frac{1}{N}\sum_{k=0}^{s-1}{s\choose k}{s\choose k+1}(-\overrightarrow{\partial}_z)^{s-k}\frac{\Laplace^{-1}}{\lambda^{s+2}}(-Z_{S^{(1)'\,E}}(\lambda)J_{S^{(1)'\,E}_s}+Z_{S^{(2)'\,E}}(\lambda)J_{S^{(2)'\,E}_s}-Z_{\tilde{S}^{(1)'\,E}}(\lambda)J_{\tilde{S}^{(1)'\,E}_s}+Z_{\tilde{S}^{(2)'\,E}}(\lambda)J_{\tilde{S}^{(2)'\,E}_s})(-\overrightarrow{\partial}_z)^{k}\Bigg)^{-1}\\
				&\quad \sum_{k_1 = 0}^{s_1-1}{s_1\choose k_1}{s_1+1\choose k_1+2} 
				(-\overrightarrow{\partial}_z)^{s_1-k_1}\frac{\Laplace^{-1}}{\lambda^{s_1+2}}Z_{M}(\lambda)J_{\bar M^{'E}_{s_1}}(-\overrightarrow{\partial}_z)^{k_1+1}   \\
				&\quad\left(I+\frac{1}{N}6\sum_{k_2=0}^{s_2-2}{s_2\choose k_2}{s_2\choose k_2+2}(-\overrightarrow{\partial}_z)^{s_2-k_2-1}\frac{\Laplace^{-1}}{\lambda^{s_2+2}}(Z_{S^{(1)'\,E}}(\lambda)\frac{J_{S^{(1)'\,E}_{s_2}}}{s_2-1}+Z_{S^{(2)'\,E}}(\lambda)\frac{J_{S^{(2)'\,E}_{s_2}}}{s_2+2}+Z_{\tilde{S}^{(1)'\,E}}(\lambda)\frac{J_{\tilde{S}^{(1)'\,E}_{s_2}}}{s_2-1}+Z_{\tilde{S}^{(2)'\,E}}(\lambda)\frac{J_{\tilde{S}^{(2)'\,E}_{s_2}}}{s_2+2})(-\overrightarrow{\partial}_z)^{k_2+1} \right)^{-1}\\
				&\quad  \sum_{k_3 = 0}^{s_3-1}{s_3+1\choose k_3}{s_3\choose k_3+1} 
				(-\overrightarrow{\partial}_z)^{s_3-k_3}\frac{\Laplace^{-1}}{\lambda^{s_3+2}}Z_{M}(\lambda)\bar{J}_{M^{'E}_{s_3}} (-\overrightarrow{\partial}_z)^{k_3}\Bigg]
			\end{aligned}$
		}
	\end{equation}
	and
	\begin{equation}
		\label{wexplicit2ES}
		\resizebox{1.00\textwidth}{!}{
			$\begin{aligned}
				&\mathcal{W}^E_{\text{asym torus}}\left[J_{\tilde{O}^{\lambda\,E}_1},J_{S^{(1)'\,E}},J_{\tilde{S}^{(1)'\,E}},J_{S^{(2)'\,E}},J_{\tilde{S}^{(2)'\,E}},\bar{J}_{{M^{'E}}},J_{\bar M^{'E}},\lambda\right] =\\
				&+ \frac{1}{2}\log\Det\Bigg(I\\
				&+\frac{1}{N}6\sum_{k=0}^{s-2}{s\choose k}{s\choose k+2}(-\overrightarrow{\partial}_z)^{s-k-1}\frac{\Laplace^{-1}}{\lambda^{s+2}}(Z_{S^{(1)'\,E}}(\lambda)\frac{J_{S^{(1)'\,E}_s}}{s-1}+Z_{S^{(2)'\,E}}(\lambda)\frac{J_{S^{(2)'\,E}_s}}{s+2}-Z_{\tilde{S}^{(1)'\,E}}(\lambda)\frac{J_{\tilde{S}^{(1)'\,E}_s}}{s-1}-Z_{\tilde{S}^{(2)'\,E}}(\lambda)\frac{J_{\tilde{S}^{(2)'\,E}_s}}{s+2})(-\overrightarrow{\partial}_z)^{k+1} \Bigg)\\
				& +\frac{1}{2}\log\Det\Bigg(I\\
				&+\frac{1}{N}6\sum_{k=0}^{s-2}{s\choose k}{s\choose k+2}(-\overrightarrow{\partial}_z)^{s-k-1}\frac{\Laplace^{-1}}{\lambda^{s+2}}(Z_{S^{(1)'\,E}}(\lambda)\frac{J_{S^{(1)'\,E}_s}}{s-1}+Z_{S^{(2)'\,E}}(\lambda)\frac{J_{S^{(2)'\,E}_s}}{s+2}+Z_{\tilde{S}^{(1)'\,E}}(\lambda)\frac{J_{\tilde{S}^{(1)'\,E}_s}}{s-1}+Z_{\tilde{S}^{(2)'\,E}}(\lambda)\frac{J_{\tilde{S}^{(2)'\,E}_s}}{s+2})(-\overrightarrow{\partial}_z)^{k+1} \Bigg)\\
				&-\frac{1}{2}\log\Det\Bigg(I-\frac{1}{N}\frac{Z_{\tilde{O}^\lambda_1}(\lambda)}{\lambda^3}\Laplace^{-1}J_{\tilde{O}^{\lambda\,E}_1}\\
				&-\frac{1}{N}\sum_{k=0}^{s-1}{s\choose k}{s\choose k+1}(-\overrightarrow{\partial}_z)^{s-k}\frac{\Laplace^{-1}}{\lambda^{s+2}}(-Z_{S^{(1)'\,E}}(\lambda)J_{S^{(1)'\,E}_s}+Z_{S^{(2)'\,E}}(\lambda)J_{S^{(2)'\,E}_s}-Z_{\tilde{S}^{(1)'\,E}}(\lambda)J_{\tilde{S}^{(1)'\,E}_s}+Z_{\tilde{S}^{(2)'\,E}}(\lambda)J_{\tilde{S}^{(2)'\,E}_s})(-\overrightarrow{\partial}_z)^{k} \Bigg)\\
				&-\frac{1}{2}\log\Det\Bigg(I+\frac{1}{N}\frac{Z_{\tilde{O}^\lambda_1}(\lambda)}{\lambda^3}\Laplace^{-1}J_{\tilde{O}^{\lambda\,E}_1}\\
				&-\frac{1}{N}\sum_{k=0}^{s-1}{s\choose k}{s\choose k+1}(-\overrightarrow{\partial}_z)^{s-k}\frac{\Laplace^{-1}}{\lambda^{s+2}}(-Z_{S^{(1)'\,E}}(\lambda)J_{S^{(1)'\,E}_s}+Z_{S^{(2)'\,E}}(\lambda)J_{S^{(2)'\,E}_s}+Z_{\tilde{S}^{(1)'\,E}}(\lambda)J_{\tilde{S}^{(1)'\,E}_s}-Z_{\tilde{S}^{(2)'\,E}}(\lambda)J_{\tilde{S}^{(2)'\,E}_s})(-\overrightarrow{\partial}_z)^{k} \Bigg)\\
				&+\frac{1}{2}\log\Det\Bigg[I\\
				&+\frac{1}{N^2}\Bigg(I+\frac{1}{N}6\sum_{k=0}^{s-2}{s\choose k}{s\choose k+2}(-\overrightarrow{\partial}_z)^{s-k-1}\frac{\Laplace^{-1}}{\lambda^{s+2}}(Z_{S^{(1)'\,E}}(\lambda)\frac{J_{S^{(1)'\,E}_s}}{s-1}+Z_{S^{(2)'\,E}}(\lambda)\frac{J_{S^{(2)'\,E}_s}}{s+2}-Z_{\tilde{S}^{(1)'\,E}}(\lambda)\frac{J_{\tilde{S}^{(1)'\,E}_s}}{s-1}-Z_{\tilde{S}^{(2)'\,E}}(\lambda)\frac{J_{\tilde{S}^{(2)'\,E}_s}}{s+2})(-\overrightarrow{\partial}_z)^{k+1}  \Bigg)^{-1}\\
				&\quad\sum_{k_1 = 0}^{s_1-1}{s_1\choose k_1}{s_1+1\choose k_1+2} (-1)^{s_1-1}
				(-\overrightarrow{\partial}_z)^{s_1-k_1}\frac{\Laplace^{-1}}{\lambda^{s_1+2}}Z_{M}(\lambda)J_{\bar M^{'E}_{s_1}}(-\overrightarrow{\partial}_z)^{k_1+1}  \\
				&\quad\Bigg(I+\frac{1}{N}\frac{Z_{\tilde{O}^\lambda_1}(\lambda)}{\lambda^3}\Laplace^{-1}J_{\tilde{O}^{\lambda\,E}_1}\\
				&-\frac{1}{N}\sum_{k_2=0}^{{s_2}-1}{{s_2}\choose k_2}{{s_2}\choose k_2+1}(-\overrightarrow{\partial}_z)^{{s_2}-k_2}\frac{\Laplace^{-1}}{\lambda^{s_2+2}}(-Z_{S^{(1)'\,E}}(\lambda)J_{S^{(1)'\,E}_{s_2}}+Z_{S^{(2)'\,E}}(\lambda)J_{S^{(2)'\,E}_{s_2}}+Z_{\tilde{S}^{(1)'\,E}}(\lambda)J_{\tilde{S}^{(1)'\,E}_{s_2}}-Z_{\tilde{S}^{(2)'\,E}}(\lambda)J_{\tilde{S}^{(2)'\,E}_{s_2}})(-\overrightarrow{\partial}_z)^{k_2} \Bigg)^{-1}\\
				&\quad \sum_{k_3 = 0}^{s_3-1}{s_3+1\choose k_3}{s_3\choose k_3+1} (-1)^{s_3-1}
				(-\overrightarrow{\partial}_z)^{s_3-k_3}\frac{\Laplace^{-1}}{\lambda^{s_3+2}}Z_{M}(\lambda)\bar{J}_{M^{'E}_{s_3}} (-\overrightarrow{\partial}_z)^{k_3}\Bigg]\\
				&+\frac{1}{2}\log\Det\Bigg[I\\
				&+\frac{1}{N^2}\left(I+\frac{1}{N}6\sum_{k=0}^{s-2}{s\choose k}{s\choose k+2}(-\overrightarrow{\partial}_z)^{s-k-1}\frac{\Laplace^{-1}}{\lambda^{s+2}}(Z_{S^{(1)'\,E}}(\lambda)\frac{J_{S^{(1)'\,E}_s}}{s-1}+Z_{S^{(2)'\,E}}(\lambda)\frac{J_{S^{(2)'\,E}_s}}{s+2}+Z_{\tilde{S}^{(1)'\,E}}(\lambda)\frac{J_{\tilde{S}^{(1)'\,E}_s}}{s-1}+Z_{\tilde{S}^{(2)'\,E}}(\lambda)\frac{J_{\tilde{S}^{(2)'\,E}_s}}{s+2})(-\overrightarrow{\partial}_z)^{k+1}\right)^{-1}\\
				&\quad \sum_{k_1 = 0}^{s_1-1}{s_1\choose k_1}{s_1+1\choose k_1+2} 
				(-\overrightarrow{\partial}_z)^{s_1-k_1}\frac{\Laplace^{-1}}{\lambda^{s_1+2}}Z_{M}(\lambda)\bar{J}_{M^{'E}_{s_1}}(-\overrightarrow{\partial}_z)^{k_1+1}   \\
				&\quad\Bigg(I-\frac{1}{N}\frac{Z_{\tilde{O}^\lambda_1}(\lambda)}{\lambda^3}\Laplace^{-1}J_{\tilde{O}^{\lambda\,E}_1}\\
				&-\frac{1}{N}\sum_{{k_2}=0}^{{s_2}-1}{{s_2}\choose {k_2}}{{s_2}\choose {k_2}+1}(-\overrightarrow{\partial}_z)^{{s_2}-{k_2}}\frac{\Laplace^{-1}}{\lambda^{s_2+2}}(-Z_{S^{(1)'\,E}}(\lambda)J_{S^{(1)'\,E}_{s_2}}+Z_{S^{(2)'\,E}}(\lambda)J_{S^{(2)'\,E}_{s_2}}-Z_{\tilde{S}^{(1)'\,E}}(\lambda)J_{\tilde{S}^{(1)'\,E}_{s_2}}+Z_{\tilde{S}^{(2)'\,E}}(\lambda)J_{\tilde{S}^{(2)'\,E}_{s_2}})(-\overrightarrow{\partial}_z)^{{k_2}}\Bigg)^{-1}\\
				&\quad  \sum_{k_3 = 0}^{s_3-1}{s_3+1\choose k_3}{s_3\choose k_3+1} 
				(-\overrightarrow{\partial}_z)^{s_3-k_3}\frac{\Laplace^{-1}}{\lambda^{s_3+2}}Z_{M}(\lambda)J_{\bar M^{'E}_{s_3}} (-\overrightarrow{\partial}_z)^{k_3}\Bigg]\,.
			\end{aligned}$
		}
	\end{equation}

\section{An equality of determinants \label{appM}}

We prove the equality of the two functional determinants in Eq.~\eqref{2det} arising from fermionic integration. Their expansion reads 
	\begin{align}
		\label{primodet}
		&\frac{1}{2}\log\Det \Big(\mathcal{I}+\frac{1}{4} i\partial_+i\square^{-1}\bar{J}_{M_{s_1}}\otimes\mathcal{G}_{s_1-1}^{(1,2)}(-1)^{s_1-1}i\square^{-1} J_{\bar M_{s_3}}\otimes\mathcal{G}_{s_3-1}^{(2,1)}(-1)^{s_3-1} \Big)\nonumber\\
		&=\frac{N^2-1}{2}\sum_{l=1}^{\infty}\frac{(-1)^{l+1}}{l}\frac{1}{2^{2l}}\int d^4u_1\ldots d^4u_ld^4v_1\ldots d^4v_l \sum_{s_1}\ldots \sum_{s_l}\sum_{s'_1}\ldots\sum_{s'_l} \nonumber\\
		&\quad(-1)^{\sum_{i=1}^l s_i+s'_i}\nonumber\\ 
		&\quad i\partial_{v_1^+}i\Box^{-1}(v_1-u_1)\mathcal{G}_{s_1-1}^{(1,2)}(\overleftarrow{\partial}_{u_1^+},\overrightarrow{\partial}_{u_1^+})\otimes \bar{J}_{M_{s_1}}(u_1)\nonumber\\
		&\quad i\Box^{-1}(u_1-v_2) \mathcal{G}_{s'_2-1}^{(2,1)}(\overleftarrow{\partial}_{v_2^+},\overrightarrow{\partial}_{v_2^+})\otimes J_{\bar M_{s'_2}}(v_2)\nonumber\\
		&\quad\ldots i\partial_{v_l^+}i\Box^{-1}(v_l-u_l)\mathcal{G}_{s_l-1}^{(1,2)}(\overleftarrow{\partial}_{u_l^+},\overrightarrow{\partial}_{u_l^+})\otimes \bar{J}_{M_{s_l}}(u_l)\nonumber\\
		&\quad i\Box^{-1}(u_l-v_1) \mathcal{G}_{s'_1-1}^{(2,1)}(\overleftarrow{\partial}_{v_1^+},\overrightarrow{\partial}_{v_1^+})\otimes J_{\bar M_{s'_1}}(v_1)
	\end{align}
	and
	\begin{align}
		&\frac{1}{2}\log\Det \Big(\mathcal{I}  + \frac{1}{4}i\partial_+i\square^{-1}J_{\bar M_{s_1}}\otimes\mathcal{G}_{s_1-1}^{(1,2)} i\square^{-1}\bar{J}_{M_{s_3}}\otimes\mathcal{G}_{s_3-1}^{(2,1)}\Big)\nonumber\\
		&=\frac{N^2-1}{2}\sum_{l=1}^{\infty}\frac{(-1)^{l+1}}{l}\frac{1}{2^{2l}}\int d^4u_1\ldots d^4u_ld^4v_1\ldots d^4v_l \sum_{s_1}\ldots \sum_{s_l}\sum_{s'_1}\ldots\sum_{s'_l} \nonumber\\
		&\quad i\partial_{u_1^+}i\Box^{-1}(u_1-v_1)
		\mathcal{G}_{s'_1-1}^{(1,2)}(\overleftarrow{\partial}_{v_1^+},\overrightarrow{\partial}_{v_1^+})\otimes J_{\bar M_{s'_1}}(v_1)\nonumber\\
		&\quad i\Box^{-1}(v_1-u_2) \mathcal{G}_{s_2-1}^{(2,1)}(\overleftarrow{\partial}_{u_2^+},\overrightarrow{\partial}_{u_2^+})\otimes \bar{J}_{M_{s_2}}(u_2)\nonumber\\
		&\quad\ldots i\partial_{u_l^+}i\Box^{-1}(u_l-v_l)\mathcal{G}_{s'_l-1}^{(1,2)}(\overleftarrow{\partial}_{v_l^+},\overrightarrow{\partial}_{v_l^+})\otimes J_{\bar M_{s'_l}}(v_l)\nonumber\\
		&\quad i\Box^{-1}(v_l-u_1) \mathcal{G}_{s_1-1}^{(2,1)}(\overleftarrow{\partial}_{u_1^+},\overrightarrow{\partial}_{u_1^+})\otimes \bar{J}_{M_{s_1}}(u_1) \,.
	\end{align}
In Eq. \eqref{primodet} employing the property of the Jacobi polynomials
\begin{align}
	&f(x)\mathcal{G}_{s-1}^{(\alpha,\beta)}(\overleftarrow{\partial}_+,\overrightarrow{\partial}_+)g(x)\nonumber\\
	&= (-1)^{s-1}g(x)\mathcal{G}_{s-1}^{(\beta,\alpha)}(\overleftarrow{\partial}_+,\overrightarrow{\partial}_+)f(x)
\end{align}
and the cyclicity of the trace, we obtain
\begin{equation}
	\resizebox{0.67\textwidth}{!}{
		$\begin{aligned}
			&(-1)^{\sum_{i=1}^l s_i+s'_i}\\ & i\partial_{v_1^+}i\Box^{-1}(v_1-u_1)\mathcal{G}_{s_1-1}^{(1,2)}(\overleftarrow{\partial}_{u_1^+},\overrightarrow{\partial}_{u_1^+})\otimes \bar{J}_{M_{s_1}}(u_1)\\
			& i\Box^{-1}(u_1-v_2) \mathcal{G}_{s'_2-1}^{(2,1)}(\overleftarrow{\partial}_{v_2^+},\overrightarrow{\partial}_{v_2^+})\otimes J_{\bar M_{s'_2}}(v_2)\\
			&\ldots i\partial_{v_l^+}i\Box^{-1}(v_l-u_l)\mathcal{G}_{s_l-1}^{(1,2)}(\overleftarrow{\partial}_{u_l^+},\overrightarrow{\partial}_{u_l^+})\otimes \bar{J}_{M_{s_l}}(u_l) \\
			& i\Box^{-1}(u_l-v_1) \mathcal{G}_{s'_1-1}^{(2,1)}(\overleftarrow{\partial}_{v_1^+},\overrightarrow{\partial}_{v_1^+})\otimes J_{\bar M_{s'_1}}(v_1)\\
			&=\\
			& i\partial_{u_1^+}i\Box^{-1}(u_1-v_1)\mathcal{G}_{s'_1-1}^{(1,2)}(\overleftarrow{\partial}_{v_1^+},\overrightarrow{\partial}_{v_1^+})\otimes J_{\bar M_{s'_1}}(v_1)\\
			&
				i\Box^{-1}(v_1-u_l) \mathcal{G}_{s_l-1}^{(2,1)}(\overleftarrow{\partial}_{u_l^+},\overrightarrow{\partial}_{u_l^+})\otimes\bar{J}_{M_{s_l}}(u_l)\\
			& i\partial_{u_l^+}i\Box^{-1}(u_l-v_l)\mathcal{G}_{s'_l-1}^{(1,2)}(\overleftarrow{\partial}_{v_l^+},\overrightarrow{\partial}_{v_l^+})\otimes J_{\bar M_{s'_l}}(v_l)\\
			&
			i\Box^{-1}(v_l-u_{l-1}) \mathcal{G}_{s_{l-1}-1}^{(2,1)}(\overleftarrow{\partial}_{u_{l-1}^+},\overrightarrow{\partial}_{u_{l-1}^+})\\
			&\otimes \bar{J}_{M_{s_{l-1}}}(u_{l-1})\\
			&\ldots i\partial_{u_2^+}i\Box^{-1}(u_2-v_2)\mathcal{G}_{s'_2-1}^{(1,2)}(\overleftarrow{\partial}_{v_2^+},\overrightarrow{\partial}_{v_2^+})\otimes J_{\bar M_{s'_2}}(v_2)\\
			& i\Box^{-1}(v_2-u_1) \mathcal{G}_{s_1-1}^{(2,1)}(\overleftarrow{\partial}_{u_1^+},\overrightarrow{\partial}_{u_1^+})\otimes\bar{J}_{M_{s_1}}(u_1)\,,
		\end{aligned}$
	} 
\end{equation}
where we have employed
\begin{equation}
	\label{uvbox}
	\partial_{v^+}\Box^{-1}(v-u)=-\partial_{u^+}\Box^{-1}(u-v)
\end{equation}
and cyclically permuted the anticommuting sources, so that the factors of $(-1)^n$ arising from Eq. \eqref{uvbox} and the permutation of the anticommuting sources cancel out. Relabeling the variables $x_{i}s_{i}k_{i}\rightarrow x_{l-i+2}s_{l-i+2}k_{l-i+2}$ for $2 \leq i \leq l$ and keeping $x_1s_1k_1$ fixed with $x=u,v$, we obtain
\begin{equation}
	\resizebox{0.67\textwidth}{!}{
		$\begin{aligned}
			& i\partial_{u_1^+}i\Box^{-1}(u_1-v_1)\mathcal{G}_{s'_1-1}^{(1,2)}\otimes J_{\bar M_{s'_1}}(v_1)(\overleftarrow{\partial}_{v_1^+},\overrightarrow{\partial}_{v_1^+})\\
			&
			i\Box^{-1}(v_1-u_l) \mathcal{G}_{s_l-1}^{(2,1)}(\overleftarrow{\partial}_{u_l^+},\overrightarrow{\partial}_{u_l^+})\otimes\bar{J}_{M_{s_l}}(u_l)\\
			& i\partial_{u_l^+}i\Box^{-1}(u_l-v_l)\mathcal{G}_{s'_l-1}^{(1,2)}(\overleftarrow{\partial}_{v_l^+},\overrightarrow{\partial}_{v_l^+})\otimes J_{\bar M_{s'_l}}(v_l)\\
			&
			i\Box^{-1}(v_l-u_{l-1}) \mathcal{G}_{s_{l-1}-1}^{(2,1)}(\overleftarrow{\partial}_{u_{l-1}^+},\overrightarrow{\partial}_{u_{l-1}^+})\\
			&\otimes \bar{J}_{M_{s_{l-1}}}(u_{l-1})\\
			&\ldots i\partial_{u_2^+}i\Box^{-1}(u_2-v_2)\mathcal{G}_{s'_2-1}^{(1,2)}(\overleftarrow{\partial}_{v_2^+},\overrightarrow{\partial}_{v_2^+})\otimes J_{\bar M_{s'_2}}(v_2)\\
			& i\Box^{-1}(v_2-u_1) \mathcal{G}_{s_1-1}^{(2,1)}(\overleftarrow{\partial}_{u_1^+},\overrightarrow{\partial}_{u_1^+})\otimes\bar{J}_{M_{s_1}}(u_1)\\
			&=\\
			& i\partial_{u_1^+}i\Box^{-1}(u_1-v_1)
			\mathcal{G}_{s'_1-1}^{(1,2)}(\overleftarrow{\partial}_{v_1^+},\overrightarrow{\partial}_{v_1^+})\otimes J_{\bar M_{s'_1}}(v_1)\\
			&
			i\Box^{-1}(v_1-u_2) \mathcal{G}_{s_2-1}^{(2,1)}(\overleftarrow{\partial}_{u_2^+},\overrightarrow{\partial}_{u_2^+})\otimes \bar{J}_{M_{s_2}}(u_2)\\
			&\ldots i\partial_{u_l^+}i\Box^{-1}(u_l-v_l)\mathcal{G}_{s'_l-1}^{(1,2)}(\overleftarrow{\partial}_{v_l^+},\overrightarrow{\partial}_{v_l^+})\otimes J_{\bar M_{s'_l}}(v_l)\\
			& i\Box^{-1}(v_l-u_1) \mathcal{G}_{s_1-1}^{(2,1)}(\overleftarrow{\partial}_{u_1^+},\overrightarrow{\partial}_{u_1^+})\otimes \bar{J}_{M_{s_1}}(u_1)\,.
		\end{aligned}$
	} 
\end{equation}
It follows
	\begin{align}
		&\log\Det \Big(\mathcal{I}+\frac{1}{4} i\partial_+i\square^{-1}\bar{J}_{M_{s_1}}\otimes\mathcal{G}_{s_1-1}^{(1,2)}(-1)^{s_1-1}i\square^{-1} J_{\bar M_{s_3}}\otimes\mathcal{G}_{s_3-1}^{(2,1)}(-1)^{s_3-1} \Big)\nonumber\\
		&=\log\Det \Big(\mathcal{I}  + \frac{1}{4}i\partial_+i\square^{-1}J_{\bar M_{s_1}}\otimes\mathcal{G}_{s_1-1}^{(1,2)} i\square^{-1}\bar{J}_{M_{s_3}}\otimes\mathcal{G}_{s_3-1}^{(2,1)}\Big)\,.
	\end{align}

 \bibliographystyle{JHEP}
\bibliography{mybib} 

\providecommand{\href}[2]{#2}\begingroup\raggedright\begin{thebibliography}{10}

\bibitem{BPSpaper2}
M.~Bochicchio, M.~Papinutto and F.~Scardino, \emph{{UV asymptotics of n-point
  correlators of twist-2 operators in SU($N$) Yang-Mills theory}},
  \href{http://dx.doi.org/10.1103/PhysRevD.108.054023}{\emph{Phys. Rev. D} {\bf
  108} (2023) 054023}, [\href{https://arxiv.org/abs/2208.14382}{{\tt
  2208.14382}}].

\bibitem{BPS1}
M.~Bochicchio, M.~Papinutto and F.~Scardino, \emph{{n-point correlators of
  twist-2 operators in SU($N$) Yang-Mills theory to the lowest perturbative
  order}}, \href{http://dx.doi.org/10.1007/JHEP08(2021)142}{\emph{JHEP} {\bf
  08} (2021) 142}, [\href{https://arxiv.org/abs/2104.13163}{{\tt 2104.13163}}].

\bibitem{tHooft:1973alw}
G.~'t~Hooft, \emph{{A Planar Diagram Theory for Strong Interactions}},
  \href{http://dx.doi.org/10.1016/0550-3213(74)90154-0}{\emph{Nucl. Phys. B}
  {\bf 72} (1974) 461}.

\bibitem{MB1}
M.~Bochicchio, \emph{{On the geometry of operator mixing in massless QCD-like
  theories}},
  \href{http://dx.doi.org/10.1140/epjc/s10052-021-09543-5}{\emph{Eur. Phys. J.
  C} {\bf 81} (2021) 749}, [\href{https://arxiv.org/abs/2103.15527}{{\tt
  2103.15527}}].

\bibitem{BPSL}
M.~Bochicchio, M.~Papinutto and F.~Scardino, \emph{{On the structure of the
  large-$N$ expansion in SU($N$) Yang-Mills theory}},
  \href{https://arxiv.org/abs/2401.09312}{{\tt 2401.09312}}.

\bibitem{QCD24}
M.~Bochicchio, M.~Papinutto and F.~Scardino, \emph{{Topology of the large-$N$
  expansion in SU($N$) Yang-Mills theory and spin-statistics theorem}},
  \href{http://dx.doi.org/10.1051/epjconf/202431400025}{\emph{EPJ Web Conf.}
  {\bf 314} (2024) 00025}.

\bibitem{Witten:1979kh}
E.~Witten, \emph{{Baryons in the 1/N Expansion}},
  \href{http://dx.doi.org/10.1016/0550-3213(79)90232-3}{\emph{Nucl. Phys. B}
  {\bf 160} (1979) 57--115}.

\bibitem{1987gauge}
A.~M. Polyakov, \emph{Gauge Fields and Strings}, vol.~3 of \emph{Contemporary
  concepts in physics}.
\newblock Taylor \& Francis, 1987.

\bibitem{Bochicchio:2013eda}
M.~Bochicchio, \emph{{Glueball and meson propagators of any spin in large-$N$
  {QCD}}}, \href{http://dx.doi.org/10.1016/j.nuclphysb.2013.07.023}{\emph{Nucl.
  Phys. B} {\bf 875} (2013) 621--649},
  [\href{https://arxiv.org/abs/1305.0273}{{\tt 1305.0273}}].

\bibitem{Bochicchio:2023ols}
M.~Bochicchio, \emph{{Higher-Spin Currents, Operator Mixing and UV Asymptotics
  in Large-$N$ QCD-like Theories}},
  \href{http://dx.doi.org/10.3390/universe9020057}{\emph{Universe} {\bf 9}
  (2023) 57}.

\bibitem{Bochicchio:2016toi}
M.~Bochicchio, \emph{{An asymptotic solution of Large-$N$ QCD, for the glueball
  and meson spectrum and the collinear S-matrix}},
  \href{http://dx.doi.org/10.1063/1.4949387}{\emph{AIP Conf. Proc.} {\bf 1735}
  (2016) 030004}.

\bibitem{Veneziano:1976wm}
G.~Veneziano, \emph{{Some Aspects of a Unified Approach to Gauge, Dual and
  Gribov Theories}},
  \href{http://dx.doi.org/10.1016/0550-3213(76)90412-0}{\emph{Nucl. Phys. B}
  {\bf 117} (1976) 519--545}.

\bibitem{Migdal:1977nu}
A.~A. Migdal, \emph{{Multicolor QCD as Dual Resonance Theory}},
  \href{http://dx.doi.org/10.1016/0003-4916(77)90181-6}{\emph{Annals Phys.}
  {\bf 109} (1977) 365}.

\bibitem{zinn1993quantum}
J.~Zinn-Justin, \emph{{Quantum field theory and critical phenomena}}.
\newblock Oxford University Press, 2021.

\bibitem{Brink:1976bc}
L.~Brink, J.~H. Schwarz and J.~Scherk, \emph{{Supersymmetric Yang-Mills
  Theories}}, \href{http://dx.doi.org/10.1016/0550-3213(77)90328-5}{\emph{Nucl.
  Phys. B} {\bf 121} (1977) 77--92}.

\bibitem{Belitsky:2004sc}
A.~V. Belitsky, S.~E. Derkachov, G.~P. Korchemsky and A.~N. Manashov,
  \emph{{Dilatation operator in (super-)Yang-Mills theories on the
  light-cone}},
  \href{http://dx.doi.org/10.1016/j.nuclphysb.2004.11.034}{\emph{Nucl. Phys. B}
  {\bf 708} (2005) 115--193}, [\href{https://arxiv.org/abs/hep-th/0409120}{{\tt
  hep-th/0409120}}].

\bibitem{Belitsky:2003sh}
A.~V. Belitsky, S.~E. Derkachov, G.~P. Korchemsky and A.~N. Manashov,
  \emph{{Superconformal operators in N=4 superYang-Mills theory}},
  \href{http://dx.doi.org/10.1103/PhysRevD.70.045021}{\emph{Phys. Rev. D} {\bf
  70} (2004) 045021}, [\href{https://arxiv.org/abs/hep-th/0311104}{{\tt
  hep-th/0311104}}].

\bibitem{Belitsky:1998gu}
A.~V. Belitsky, D.~Mueller and A.~Schafer, \emph{{Implications of N=1
  supersymmetry for QCD conformal operators}},
  \href{http://dx.doi.org/10.1016/S0370-2693(99)00146-X}{\emph{Phys. Lett. B}
  {\bf 450} (1999) 126--135}, [\href{https://arxiv.org/abs/hep-ph/9811484}{{\tt
  hep-ph/9811484}}].

\bibitem{SS1}
G.~Santoni and F.~Scardino, \emph{{Superfield twist-$2$ operators in
  $\mathcal{N} = 1$ SCFTs and their renormalization-group improved generating
  functional in $\mathcal{N} = 1$ SYM theory}},
  \href{https://arxiv.org/abs/2410.01435}{{\tt 2410.01435}}.

\bibitem{S1}
F.~Scardino, \emph{{Nonresonant renormalization scheme for twist-2 operators in
  SU($N$) Yang\textendash{}Mills theory}},
  \href{http://dx.doi.org/10.1140/epjc/s10052-024-13590-z}{\emph{Eur. Phys. J.
  C} {\bf 84} (2024) 1229}, [\href{https://arxiv.org/abs/2410.15366}{{\tt
  2410.15366}}].

\bibitem{Braun:2003rp}
V.~M. Braun, G.~P. Korchemsky and D.~Mueller, \emph{{The uses of conformal
  symmetry in QCD}},
  \href{http://dx.doi.org/10.1016/S0146-6410(03)90004-4}{\emph{Prog. Part.
  Nucl. Phys.} {\bf 51} (2003) 311--398},
  [\href{https://arxiv.org/abs/hep-ph/0306057}{{\tt hep-ph/0306057}}].

\bibitem{szego1959orthogonal}
G.~Szeg{\"o}, \emph{Orthogonal Polynomials}, vol.~23 of \emph{American
  Mathematical Society colloquium publications}.
\newblock American Mathematical Society, 1959.

\bibitem{makeenko}
Y.~M. Makeenko, \emph{On conformal operators in quantum chromodynamics},  tech.
  rep., USSR, 1980.

\bibitem{Belitsky:2007jp}
A.~V. Belitsky, J.~Henn, C.~Jarczak, D.~Mueller and E.~Sokatchev,
  \emph{{Anomalous dimensions of leading twist conformal operators}},
  \href{http://dx.doi.org/10.1103/PhysRevD.77.045029}{\emph{Phys. Rev. D} {\bf
  77} (2008) 045029}, [\href{https://arxiv.org/abs/0707.2936}{{\tt
  0707.2936}}].

\bibitem{Beisert:2004fv}
N.~Beisert, G.~Ferretti, R.~Heise and K.~Zarembo, \emph{{One-loop QCD spin
  chain and its spectrum}},
  \href{http://dx.doi.org/10.1016/j.nuclphysb.2005.04.004}{\emph{Nucl. Phys. B}
  {\bf 717} (2005) 137--189}, [\href{https://arxiv.org/abs/hep-th/0412029}{{\tt
  hep-th/0412029}}].

\bibitem{Belitsky:1998gc}
A.~V. Belitsky and D.~Mueller, \emph{{Broken conformal invariance and spectrum
  of anomalous dimensions in QCD}},
  \href{http://dx.doi.org/10.1016/S0550-3213(98)00677-4}{\emph{Nucl. Phys. B}
  {\bf 537} (1999) 397--442}, [\href{https://arxiv.org/abs/hep-ph/9804379}{{\tt
  hep-ph/9804379}}].

\bibitem{Kazakov:2012ar}
V.~Kazakov and E.~Sobko, \emph{{Three-point correlators of twist-2 operators in
  N=4 SYM at Born approximation}},
  \href{http://dx.doi.org/10.1007/JHEP06(2013)061}{\emph{JHEP} {\bf 06} (2013)
  061}, [\href{https://arxiv.org/abs/1212.6563}{{\tt 1212.6563}}].

\end{thebibliography}\endgroup
\end{document}